\documentclass[preprint,3p,twocolumn,times]{elsarticle}
\usepackage{pifont}
\usepackage{color}
\usepackage{subfigure}
\usepackage{epsfig}
\usepackage{upgreek}

\usepackage{amssymb}
\usepackage{lineno}
\usepackage{hyperref}
\hypersetup{
 colorlinks=true,     
 urlcolor=blue,       
 linkcolor=blue,     
 citecolor=blue,     
 plainpages=false,
 breaklinks=true,
 bookmarksnumbered=true,
 bookmarksopen=true,
 pdfkeywords={},
 pdfpagemode=UseOutlines  
 }

\biboptions{sort&compress}

\journal{arXiv}

\begin{document}

\begin{frontmatter}



\title{Focal-plane detector system for the KATRIN experiment
}

\author[uw]{J.\,F.~Amsbaugh}
\author[mitt]{J.~Barrett}
\author[kitipe]{A.~Beglarian}
\author[kitipe]{T.~Bergmann}
\author[uw]{H.~Bichsel}
\author[uw]{L.\,I.~Bodine}
\author[mainz]{J.~Bonn\corref{deceased}}
\author[uw]{N.\,M.~Boyd}
\author[uw]{T.\,H.~Burritt}
\author[setif]{Z.~Chaoui}
\author[kitipe]{S.~Chilingaryan}
\author[unc]{T.\,J.~Corona}
\author[uw]{P.\,J.~Doe}
\author[uw]{J.\,A.~Dunmore\fnref{jessica}}
\author[uw]{S.~Enomoto}
\author[mitt]{J.\,A.~Formaggio}
\author[unc]{F.\,M.~Fr\"{a}nkle\fnref{florian}}
\author[mitt]{D.~Furse}
\author[kitipe]{H.~Gemmeke}
\author[kitiekp,wigner]{F.~Gl\"{u}ck}
\author[kitiekp]{F.~Harms}
\author[uw]{G.\,C.~Harper}
\author[kitipe]{J.~Hartmann}
\author[unc]{M.\,A.~Howe}
\author[mitt]{A.~Kaboth}
\author[mitt]{J.~Kelsey}
\author[kitik]{M.~Knauer}
\author[kitipe]{A.~Kopmann}
\author[uw]{M.\,L.~Leber}
\author[uw]{E.\,L.~Martin}
\author[dares]{K.\,J.~Middleman}
\author[uw]{A.\,W.~Myers\fnref{brent}}
\author[mitt]{N.\,S.~Oblath}
\author[uw]{D.\,S.~Parno\corref{Diana}}
\author[uw]{D.\,A.~Peterson}
\author[kitipe]{L.~Petzold}
\author[unc]{D.\,G.~Phillips II}
\author[kitiekp]{P.~Renschler}
\author[uw]{R.\,G.\,H.~Robertson}
\author[kitiekp]{J.~Schwarz}
\author[kitik]{M.~Steidl}
\author[kitipe]{D.~Tcherniakhovski}
\author[kitik]{T.~Th\"{u}mmler}
\author[uw]{T.\,D.~Van Wechel}
\author[uw]{B.\,A.~VanDevender\fnref{brent}}
\author[muenster]{S.~V\"{o}cking}
\author[uw]{B.\,L.~Wall}
\author[uw,unc]{K.\,L.~Wierman}
\author[unc,ornl]{J.\,F.~Wilkerson}
\author[kitipe]{S.~W\"{u}stling}

\address[uw]{Center for Experimental Nuclear Physics and Astrophysics, and Department of Physics, University of Washington, Seattle, WA 98195, USA}
\address[mitt]{Laboratory for Nuclear Science, Massachusetts Institute of Technology, Cambridge, MA 02139, USA}
\address[kitipe]{Institute for Data Processing and Electronics, Karlsruhe Institute of Technology, 76344 Eggenstein-Leopoldshafen, Germany}
\address[mainz]{Institute of Physics, Johannes Gutenberg-Universit\"{a}t Mainz, 55099 Mainz, Germany}
\address[setif]{Laboratory of Optoelectronics and Devices, University of Setif, UFA Setif, Setif 19000, Algeria}
\address[unc]{Department of Physics, University of North Carolina, Chapel Hill, NC 27599, USA and \\
						Triangle Universities Nuclear Laboratory, Durham, NC 27708, USA}
\address[kitiekp]{Institute for Experimental Nuclear Physics, Karlsruhe Institute of Technology, 76344 Eggenstein-Leopoldshafen, Germany}
\address[wigner]{Wigner Research Center for Physics, POB 49, 1525 Budapest, Hungary}
\address[kitik]{Institute for Nuclear Physics, Karlsruhe Institute of Technology, 76344 Eggenstein-Leopoldshafen, Germany}
\address[dares]{ASTeC Vacuum Science Group, STFC Daresbury Laboratory, Warrington, Cheshire WA4 4AD, UK}
\address[muenster]{Institute of Nuclear Physics, Westf\"{a}lische Wilhelms-Universit\"{a}t M\"{u}nster, 48149  M\"{u}nster, Germany}
\address[ornl]{Oak Ridge National Laboratory, Oak Ridge, TN 37831, USA}

\fntext[jessica]{Present address: Dept.\,of Physics, University of Texas at El Paso, El Paso, TX, USA}
\fntext[florian]{Present address: Inst.\ for Nuclear Physics, Karlsruhe Institute of Technology, Karlsruhe, Germany}
\fntext[brent]{Present address: Pacific Northwest National Lab, Richland, WA, USA}

\cortext[Diana]{Corresponding author.  dparno@uw.edu}
\cortext[deceased]{Deceased}

\begin{abstract}
The focal-plane detector system for the KArlsruhe TRItium Neutrino (KATRIN) experiment consists of a multi-pixel silicon p-i-n-diode array, custom readout electronics, two superconducting solenoid magnets, an ultra high-vacuum system, a high-vacuum system, calibration and monitoring devices, a scintillating veto, and a custom data-acquisition system.  It is designed to detect the low-energy electrons selected by the KATRIN main spectrometer. We describe the system and summarize its performance after its final installation. 
\end{abstract}


\end{frontmatter}

\tableofcontents


\section{Introduction: The KATRIN Experiment}
\label{sec:intro}

The discovery of neutrino flavor oscillation~\cite{Fukuda:SuperK1998, Ahmad:SNO2002} showed that the neutrino flavor eigenstates $\nu_{e}, \nu_{\mu},\nu_{\tau}$ are not states of fixed mass; instead, each is a coherent superposition of mass eigenstates $\nu_{1}, \nu_{2}, \nu_{3}$. However, these and subsequent oscillation experiments are not sensitive to the absolute value of the neutrino mass, indicating only that at least two neutrinos have mass, and at least one of them has a mass $\geq 48$~meV~\cite{Schwetz:NeutrinoParam2011}. Neutrino mass is considered to provide unique early insight into electroweak physics beyond the standard model, and plays a role in the evolution of large-scale structure in the universe.   A laboratory determination of the mass scale would constrain cosmological models.  Recently reported results from surveys of the cosmic microwave background, including the WMAP~\cite{wmap:2013} and Planck missions~\cite{Ade:2013zuv}, confirm that observational cosmology is indeed sensitive to the sum of the masses $\sum_j m_j$, but that the limit or value obtained depends on the types of data and assumptions included in the analysis.  With very conservative uncertainty estimates, one may obtain a limit of  $\sum_j m_j \leq 1.3$~eV at the 95\% confidence level~\cite{PDG12}.

\begin{figure*}[tb]
\begin{center}
\includegraphics[width=\textwidth]{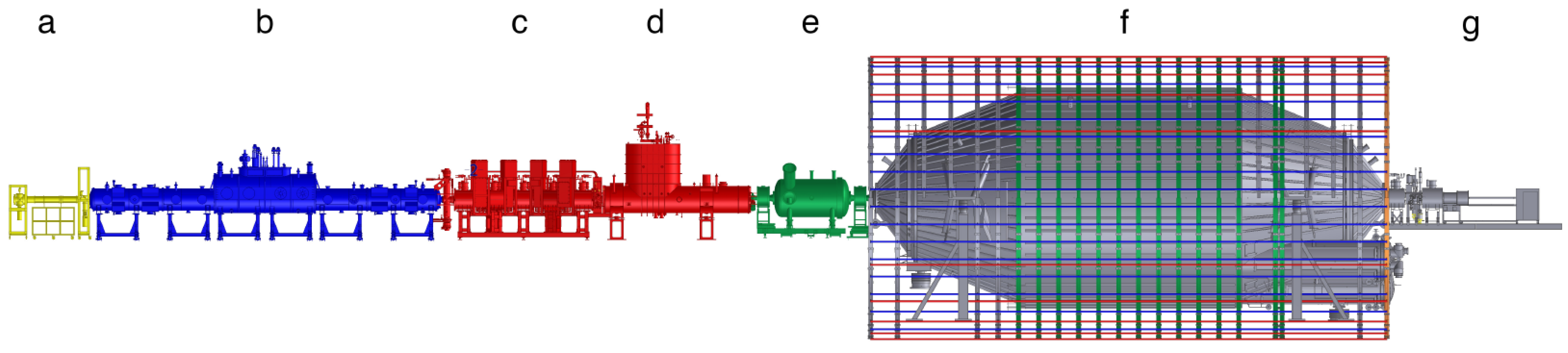}
\end{center}
\caption{\label{overallkatfig}Schematic overview of the 70-m KATRIN experimental beamline: (a) rear section, (b) tritium source, (c) differential-pumping section, (d) cryogenic-pumping section, (e) pre-spectrometer, (f) main spectrometer in air-coil framework, (g) focal-plane detector system.}
\end{figure*}

The KArlsruhe TRItium Neutrino experiment~\cite{katrin_dr2004}, KATRIN, will make a model-independent measurement of the mass $m_{\overline{\nu}}$ of the electron antineutrino in the quasi-degenerate regime ($m_1 \approx m_2 \approx m_3$). It continues a series of tritium-based experiments, including Mainz~\cite{MainzFinal2005} and Troitsk~\cite{TroitskFinal2003, Troitsk:Reanalysis2011}, which established the present model-independent limit of $m_{\overline{\nu}} <2$~eV~\cite{PDG12}. 

Tritium beta decay, $^{3}\mathrm{H}\longrightarrow {^{3}\mathrm{He}^{+}} + e^{-} + \overline{\nu}_{e}$, has an endpoint energy of 18.6~keV and a half-life of 12.3~years.  The shape of the electron energy spectrum near the endpoint is sensitive to the neutrino mass. The KATRIN experiment will use the Magnetic Adiabatic Collimation-Electrostatic (MAC-E) filter technique~\cite{Beamson:MacEfilter1980}, in which multiple integrating measurements with varying thresholds are combined to map the spectral shape near the endpoint. KATRIN is currently under construction at the Karlsruhe Institute of Technology in Karlsruhe, Germany and is expected to achieve a sensitivity of 200~meV at the 90\% confidence level. 

Figure~\ref{overallkatfig} is a schematic of KATRIN's overall layout. A $10^{11}$-Bq, windowless, gaseous tritium source provides beta electrons. An electron gun for calibration is located upstream in the rear section. Downstream of the source, an extensive transport section removes errant tritium molecules via differential and cryogenic pumping systems while guiding beta electrons adiabatically to the low-resolution pre-spectrometer~\cite{prall:2012}. The pre-spectrometer reduces the electron flux by seven orders of magnitude prior to entry into the main spectrometer, which further filters out lower-energy electrons with a designed energy resolution of 0.93~eV. Beta electrons that pass through these two MAC-E filters are magnetically guided to the focal-plane detector (FPD). 

\begin{figure}
\begin{center}
\includegraphics[width=\columnwidth]{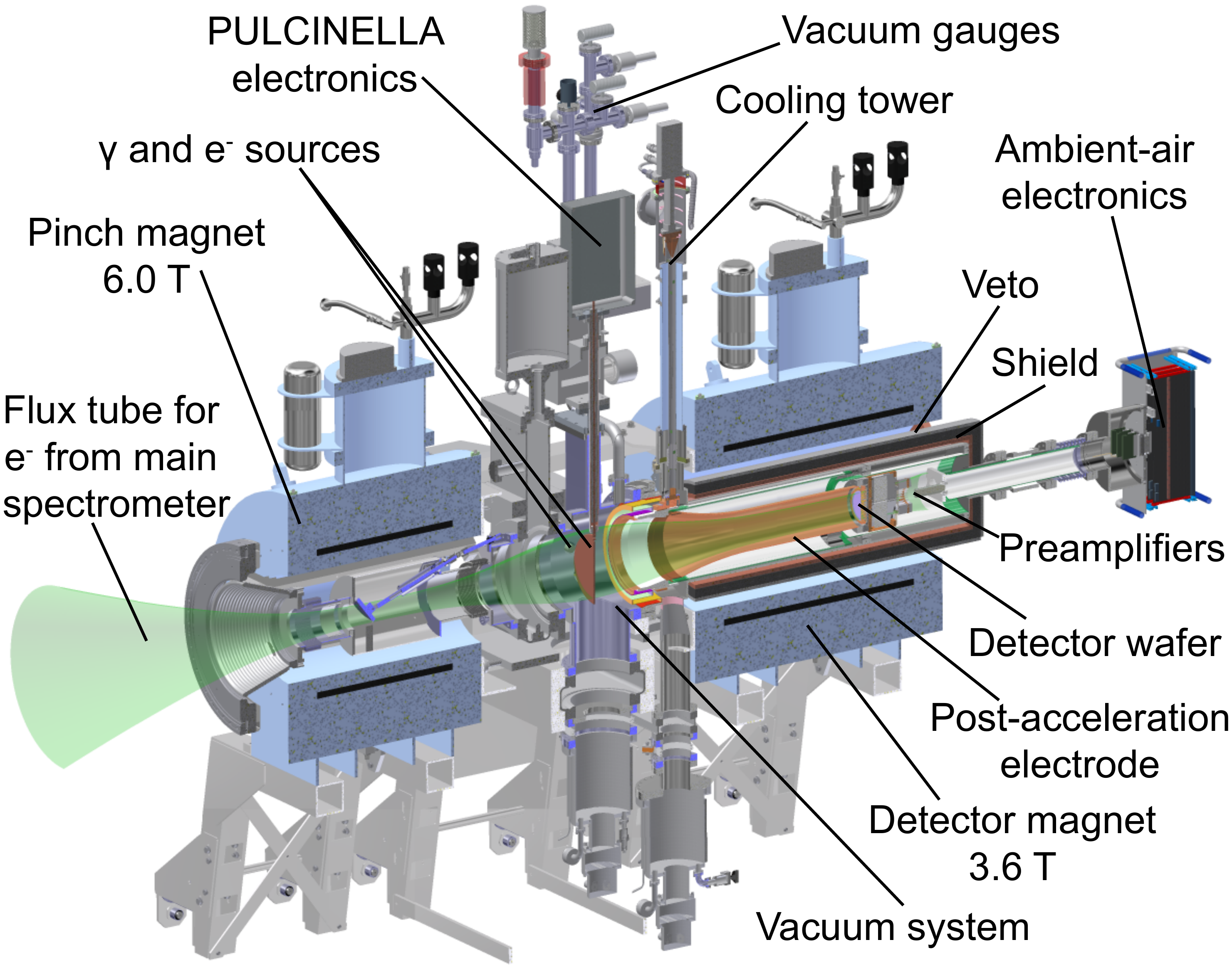}
\end{center}
\caption{\label{detsys}The primary components of the FPD system. The main spectrometer is positioned at bottom left; the data acquisition system is located beyond the top right corner.}
\end{figure}

The focus of this paper is the FPD system, shown in Fig.~\ref{detsys}. This system was constructed and commissioned at the University of Washington in Seattle, USA, prior to installation in Karlsruhe in summer 2011. In Sec.~\ref{sec:apparatus}, we provide a detailed description of the elements of this system, including design constraints (Sec.~\ref{sec:design_constraints}) and upgrades undertaken in Karlsruhe. Section~\ref{sec:performance} presents detector performance results, primarily drawn from the commissioning of the system in Karlsruhe. In Sec.~\ref{sec:det_sys_environment}, we discuss the backgrounds and consider the way in which the FPD system affects the overall KATRIN measurement via its resolution and backgrounds. 

\section{Apparatus}
\label{sec:apparatus}

In the KATRIN experiment, beta electrons produced near the endpoint energy of tritium decay will pass the energy threshold set by the main spectrometer and enter the FPD system (Fig.~\ref{detsys}) at the spectrometer exit. This loosely collimated ``beam'' of beta electrons travels within a flux tube defined by the local magnetic field, which in the FPD system is provided by two warm-bore superconducting solenoids (Sec.~\ref{sec:apparatus:magnet}). Beta electrons move through the bore of the high-field pinch magnet and past a gate valve that separates the main-spectrometer vacuum from the FPD-system vacuum system (Sec.~\ref{sec:apparatus:vacuum}), before entering the bore of the lower-field detector magnet and striking the multi-pixel silicon p-i-n-diode detector (Sec.~\ref{sec:apparatus:detector}). In this final stage of the electron trajectory, the flux tube is contained within a post-acceleration electrode that permits increasing the electron energy to a range with a more favorable background rate.

The first readout stage for detector signals consists of preamplifiers mounted directly onto feedthrough pins on the detector flange; preamplifier signals proceed along coaxial cable to the second stage, mounted outside the vacuum system (Sec.~\ref{sec:apparatus:fpd_electronics}). A liquid-nitrogen thermosiphon (Sec.~\ref{sec:apparatus:thermosiphon}) cools the detector and preamplifiers through the post-acceleration electrode. A shield and a veto system (Sec.~\ref{sec:apparatus:veto}) line the bore of the detector magnet, reducing backgrounds in the detector. An electron source and a $\gamma$ emitter (Sec.~\ref{sec:apparatus:calibration}), located between the two magnets, serve as calibration sources.   

FPD and veto data are recorded in a data-acquisition system with a graphical user interface (Sec.~\ref{sec:apparatus:daq}) while other hardware elements of the system are monitored and controlled via a separate slow-controls system (Sec.~\ref{sec:apparatus:slowcontrols}). The demands of the KATRIN experiment require an extensive data-management system, described in Sec.~\ref{sec:apparatus:datamanagement}. 

\subsection{Design Constraints}
\label{sec:design_constraints}

A number of strict requirements guided the FPD-system design. For example, the pinch magnet helps to complete KATRIN's primary MAC-E filter, and the quality of the FPD-system vacuum affects the performance of the main spectrometer. Here, we summarize the most important constraints on the system design, which fall into two categories: electromagnetic (Sec.~\ref{sec:em_constraints}) and vacuum (Sec.~\ref{sec:mainspec_adapt}).

\subsubsection{Electromagnetic Constraints}
\label{sec:em_constraints}

The KATRIN spectrometers act as integrating high-pass filters. Superconducting magnets guide electrons adiabatically along magnetic field lines, producing a flux that encounters a high electrostatic potential. Electrons with energies below the threshold are reflected toward the source, while higher-energy electrons pass through the analyzing plane at the spectrometer mid-plane, are re-accelerated, and are focused onto the detector. A ceramic break, located at the downstream exit of the main spectrometer, isolates the FPD system from the retarding potential of the MAC-E filter.

\begin{figure*}[tbp]
  \centering
    \subfigure[]{
    \includegraphics[width=0.45\textwidth]{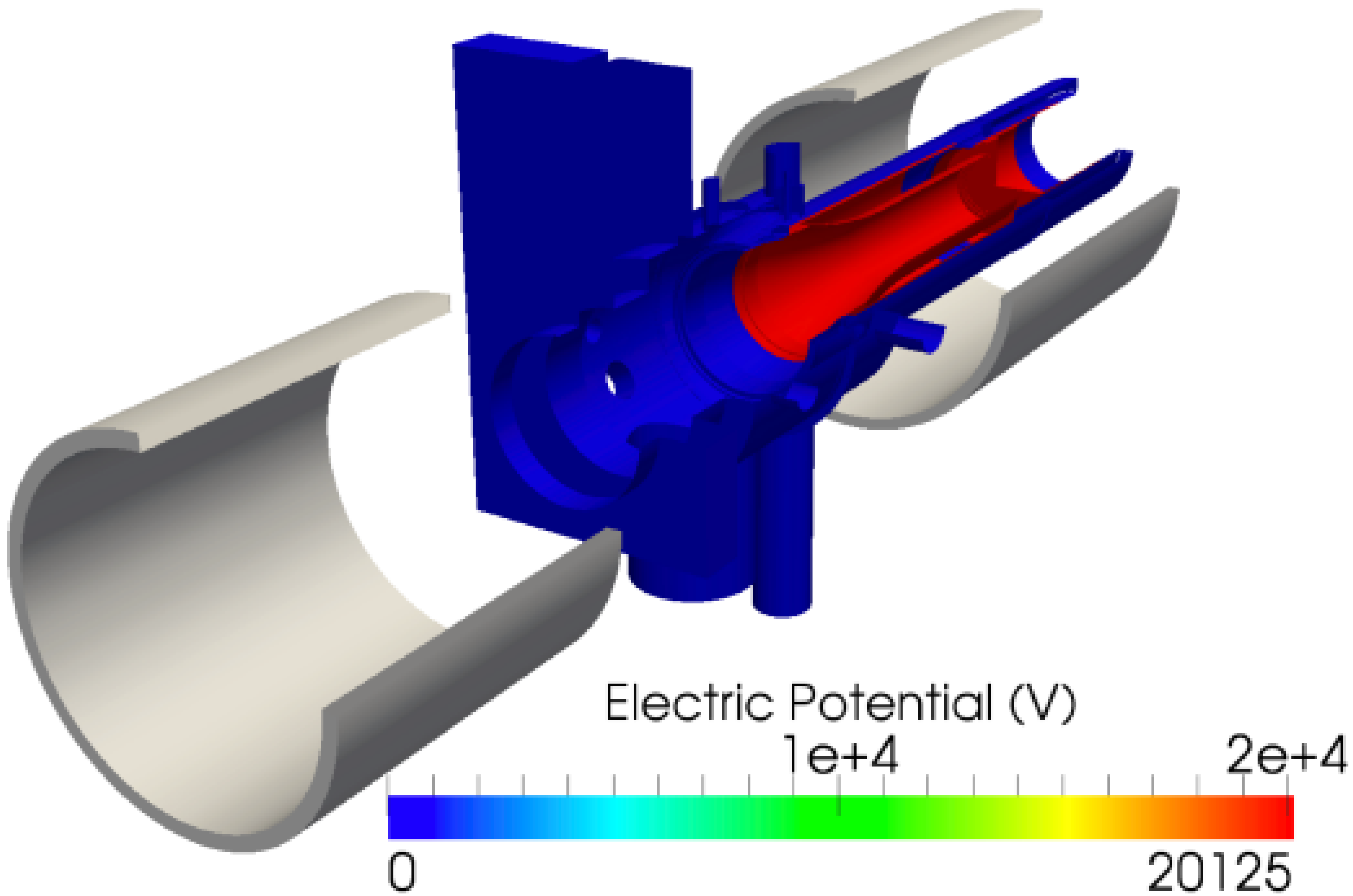}
    \label{fig:DetectorRegionPotential}
  }
    \subfigure[]{
    \includegraphics[width=0.45\textwidth]{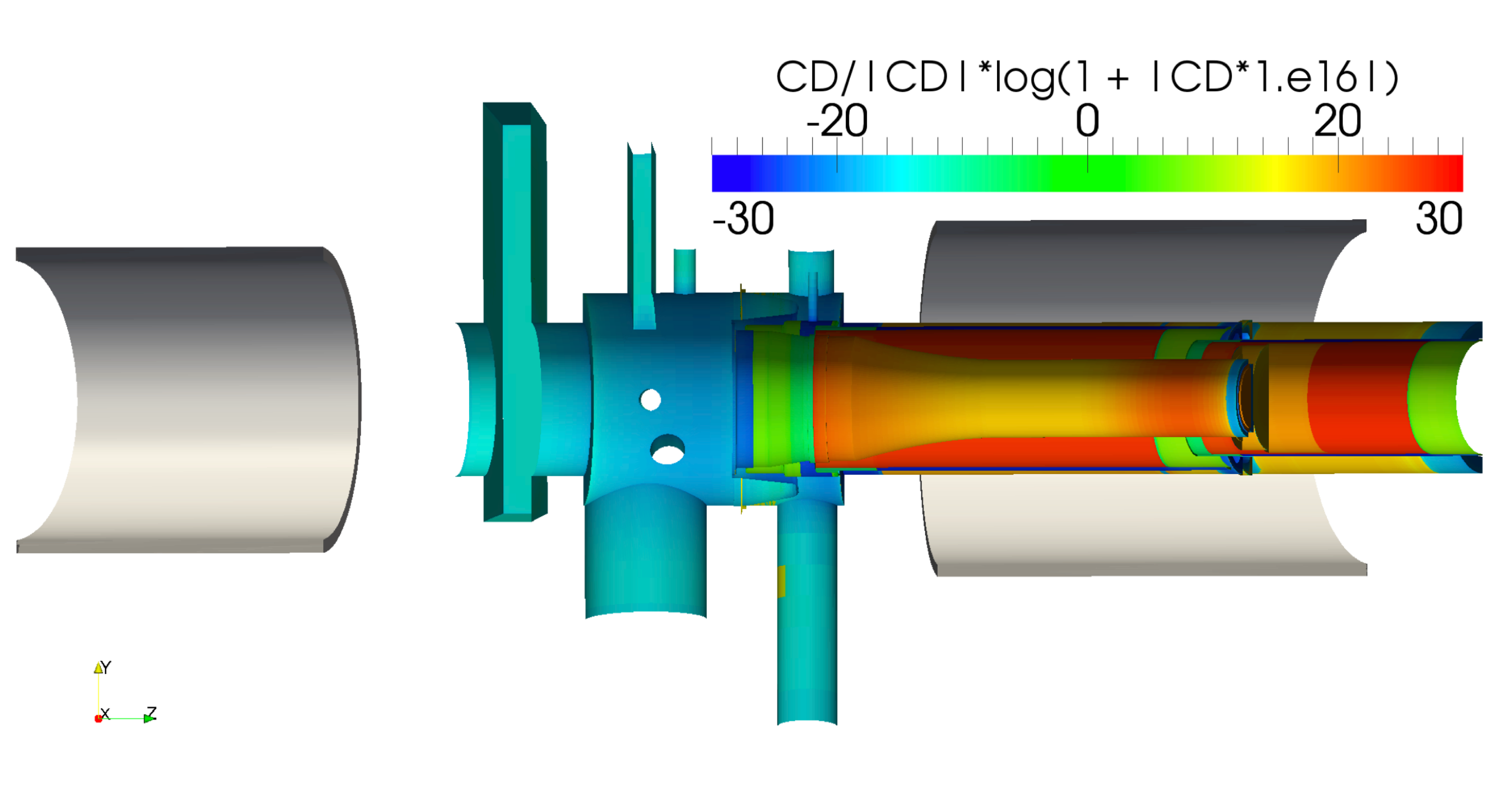}
    \label{fig:DetectorRegionCD}
  }
   \subfigure[]{
    \includegraphics[width=0.45\textwidth]{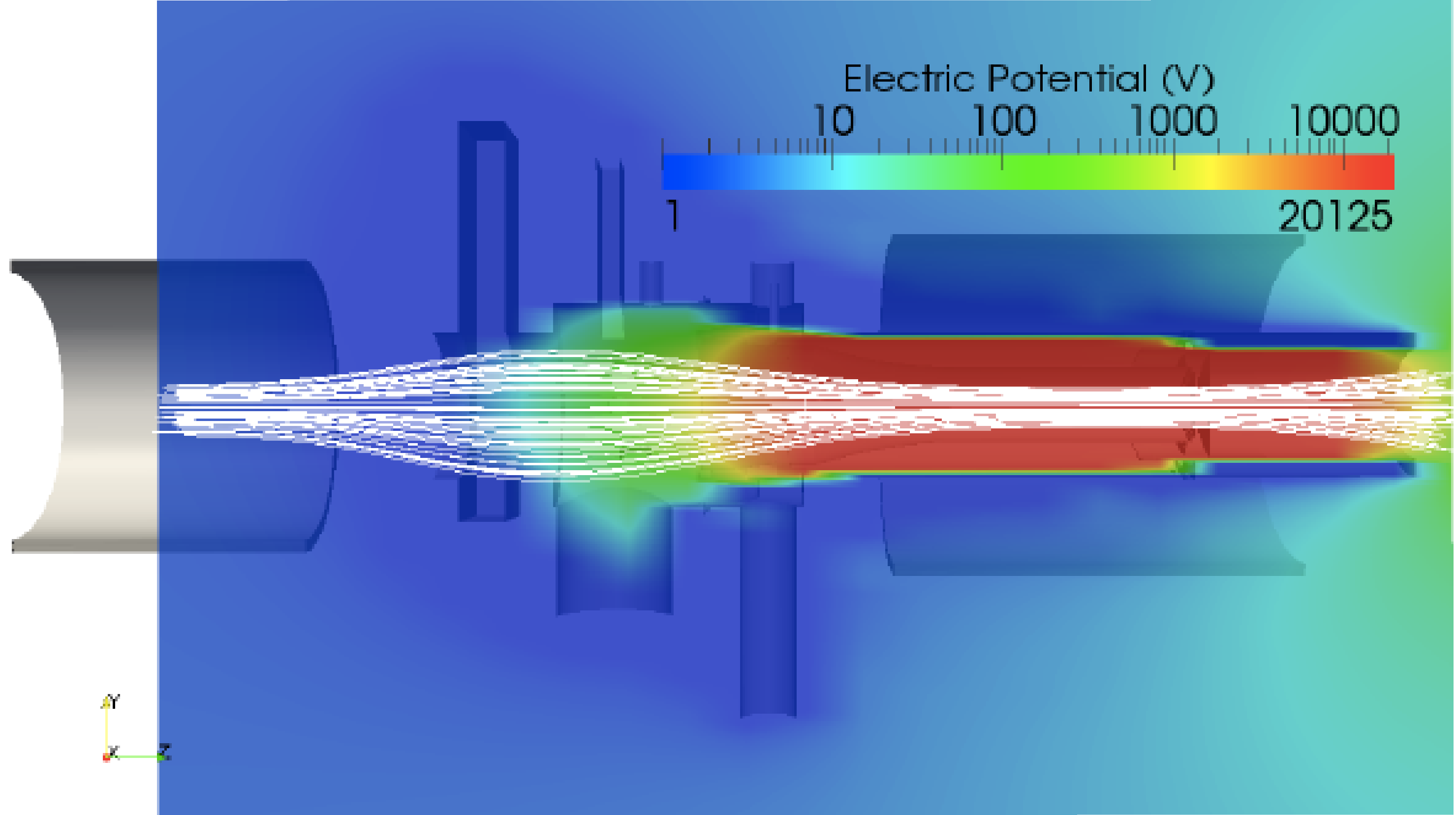}
    \label{fig:DetectorPhiAndBField}
	}
   \subfigure[]{
    \includegraphics[width=0.45\textwidth]{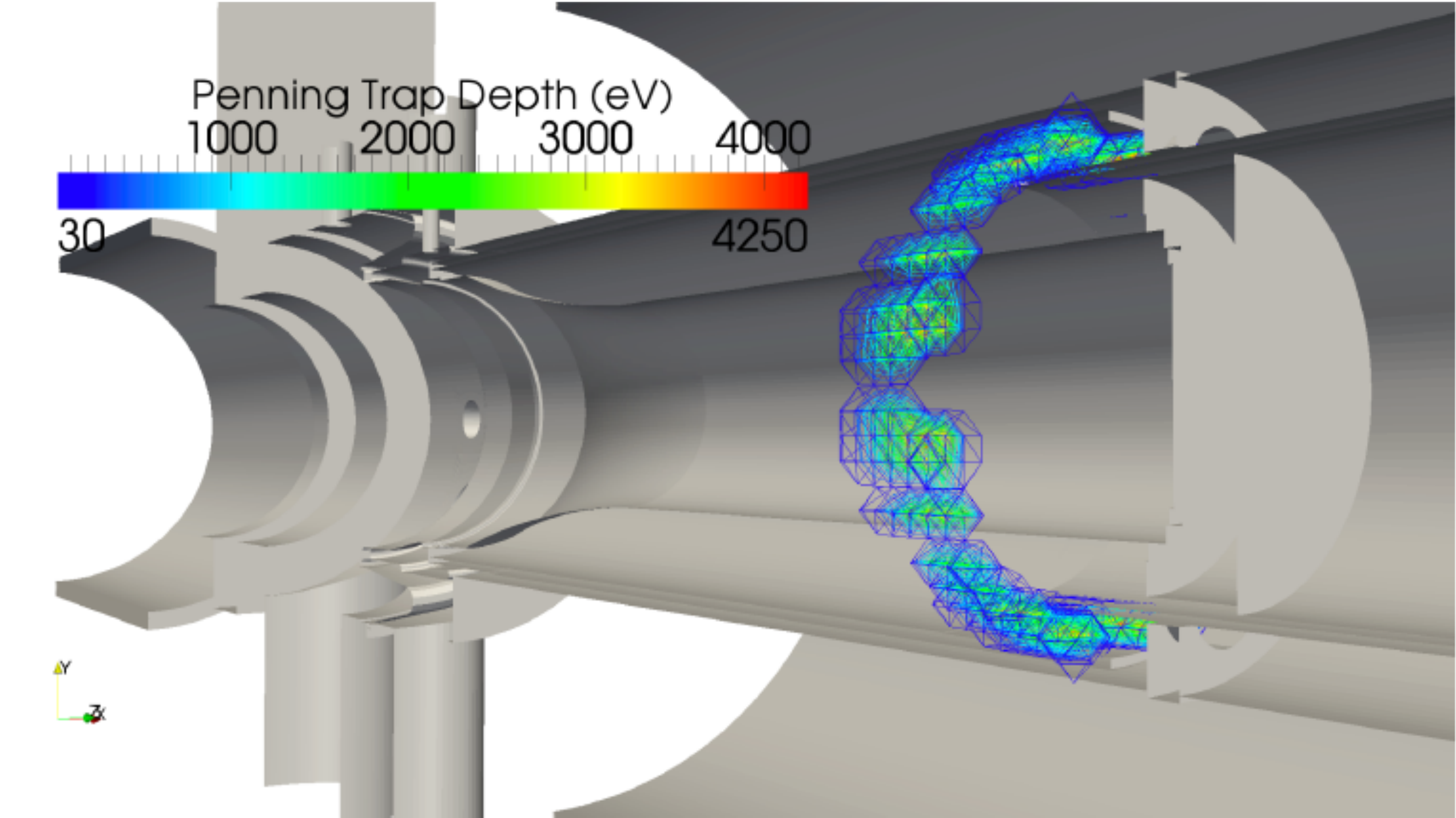}
    \label{fig:DetectorRegionPTMap}
	}
    \caption[]
    {(Color online.) \subref{fig:DetectorRegionPotential} Magnet (gray) and electrode configuration for the KATRIN detector region. \subref{fig:DetectorRegionCD} A cross section of the computed charge density (CD) profile. \subref{fig:DetectorPhiAndBField} Electrical potential (color) and magnetic field lines (white). \subref{fig:DetectorRegionPTMap} Result of an automated Penning trap search; the identified trap is outside the flux tube region. The pinch magnet is shown at left in all images; the detector is within the bore of the magnet at right. These visualizations were generated with ParaView~\cite{paraview:2007}.}
	\label{fig:PenningTrapSearch}
\end{figure*}

The FPD-system pinch magnet forms a part of the MAC-E filter for the main spectrometer, and provides the largest magnetic guiding field (6~T) in the entire experiment. This magnetic-field maximum is located downstream of the spectrometers, rather than upstream, to reduce background. With the maximum field at the detector end, electrons reflected by the spectrometer will preferentially escape toward the source. 

The transverse extent of the electron flux, and therefore the detector diameter necessary to image the entire analyzing plane at the center of the main spectrometer, is determined by the source diameter and by the magnetic-field strengths at the source and at the FPD. While a small detector wafer is desirable, a shallow angle of incidence is not, as it would increase backscattering. Our design therefore takes a moderate field strength, 3.3~T, as the nominal operating condition at the detector location, permitting a trade-off between backscattering and detector background. This field is provided by a second magnet, the detector magnet, about 1~m downstream of the pinch magnet; the physical separation between magnets provides space for a calibration system (Sec.~\ref{sec:apparatus:calibration}) that is independent of the main spectrometer. 

With a nominal field of 3.3~T at the detector location, a detector diameter of 90~mm is required to cover the 191-T~cm$^2$ flux tube available from the source, plus a margin of 10\%; with a higher magnetic field at the detector, the source will make a smaller image on the detector. In order to ensure that the flux tube is fully contained within the vacuum system, the field axis of each magnet must lie within 5~mm of the axis of its cryostat bore.

Background in the 18.6-keV region of the endpoint beta-electron signal can be reduced by means of a post-acceleration electrode, whose potential boosts the signal (and thus the region of interest) to an energy where the FPD-system background is lower. It also allows various calibrations with a low retarding potential, such as the in-situ measurement of low-energy background electrons from the main spectrometer. This electrode fits inside the detector-magnet bore and contains the entire electron flux tube in order to avoid scattering. The detector and most of the readout electronics float at post-acceleration potential.

An additional source of background for an apparatus like the FPD system, which employs strong electrostatic and magnetostatic fields in vacuum, occurs when charged particles confined within a Penning trap (a potential well along a magnetic field line) ionize residual gas molecules, producing background particles, allowing the formation of unstable plasmas, and possibly causing electrical breakdowns~\cite{Penning:1936,Haefer:1953,Hara:1989,Picard:MACE1992,Belesev:Troitsk1995,byrne:2005, byrne:2014}.  Penning traps are a geometry-dependent phenomenon, related to the configurations of both magnets and electrodes. To help avoid these traps, the high-voltage region is surrounded by fused-silica insulation within the vacuum system. To facilitate assembly and allow pumping of the interior spaces, this insulation is not one long tube, but a series of three shorter tubes with different diameters. Each silica tube is lined with a pair of stainless-steel sleeves, defining the inner and outer surface potentials. Inner surfaces are held at the post-acceleration potential while the outer surfaces are held at ground potential, confining most of the electric field within the insulator and reducing the probability of breakdowns. 

Further efforts to eliminate Penning traps must rely on field simulations, since the fields can quickly become complex for even a simple electrode-and-magnet configuration and since it is difficult (and dangerous to the detector) to experimentally locate Penning traps.  We have developed simulation software to perform this task using the Robin Hood method~\cite{formaggio-RobinHood:2012}. Our model of KATRIN's FPD system (Fig.~\ref{fig:DetectorRegionPotential}) consists of $\sim450,000$ sub-elements.  The computation of the charge-density profile (Fig.~\ref{fig:DetectorRegionCD}) took $\sim 10,000$ real-time hours on the Topsail cluster at the University of North Carolina, which is comprised of 4160 2.3-GHz EM64T central processing units. To map Penning traps, the algorithm first chooses sampling points within our region of interest and identifies the intersecting magnetic field line.  The program then traverses each such field line, sampling the electric potential along the way (Fig.~\ref{fig:DetectorPhiAndBField}). If a potential well is found, the surrounding voxel is assigned the value of the well depth in eV (Fig.~\ref{fig:DetectorRegionPTMap}).  In the final search for Penning traps, we sampled the detector region with $2 \times 2 \times 2$-cm$^3$ voxels ($\sim 15,000$ vertices), taking the equivalent of $\sim 2,000$ central-processing-unit~hours to compute.  The resulting Penning-trap map (Fig.~\ref{fig:DetectorRegionPTMap}) demonstrates that, within the sensitive region of the FPD system, there are no Penning traps that would contribute to the background via Penning discharge.  A very weak Penning trap exists between the first and second silica tubes in the high-vacuum system, and is probably responsible for the limitation of the post-acceleration potential to 12~kV without breakdown.

\subsubsection{Vacuum Constraints}
\label{sec:mainspec_adapt}

Since the FPD-system beam pipe couples to the KATRIN main spectrometer, it must be maintained at a pressure on the order of $10^{-9}$ mbar or less. We modeled the pressure profile throughout the FPD-system vacuum chamber using the Monte-Carlo-based MOLFLOW program~\cite{Kersevan:MOLFLOW}, starting from a conservative estimate of $1 \times 10^{-11}$~mbar\,l\,s$^{-1}$\,cm$^{-2}$ for outgassing from the interior of the vacuum system at room temperature. Non-evaporable getter pumps cannot be used in the FPD system due to their radioactivity, so cryo\-pumps alone must maintain the vacuum. Under these conditions, our model predicts a base pressure of $6 \times 10^{-11}$~mbar at room temperature, dominated by hydrogen from the stainless steel walls of the vacuum vessel. Section~\ref{sec:apparatus:vacuum} reports the pressures attained in the apparatus as built.

An all-metal gate valve with DN250 flanges is placed downstream of the pinch magnet and protects the main-spectrometer and FPD vacuum systems from each other. The pneumatically operated valve closes automatically if the pressure in either system exceeds $10^{-8}$~mbar. A custom-built valve along the beam pipe connecting the main spectrometer to the FPD system allows the temporary separation of the two vacuum systems until the beam pipe can be sealed with an all-metal flange. This measure permits the systems to be decoupled for spectrometer bakeout.

\subsection{Magnet System}
\label{sec:apparatus:magnet}

\begin{figure}[tbp]
   \begin{center}
   \includegraphics[width=\columnwidth]{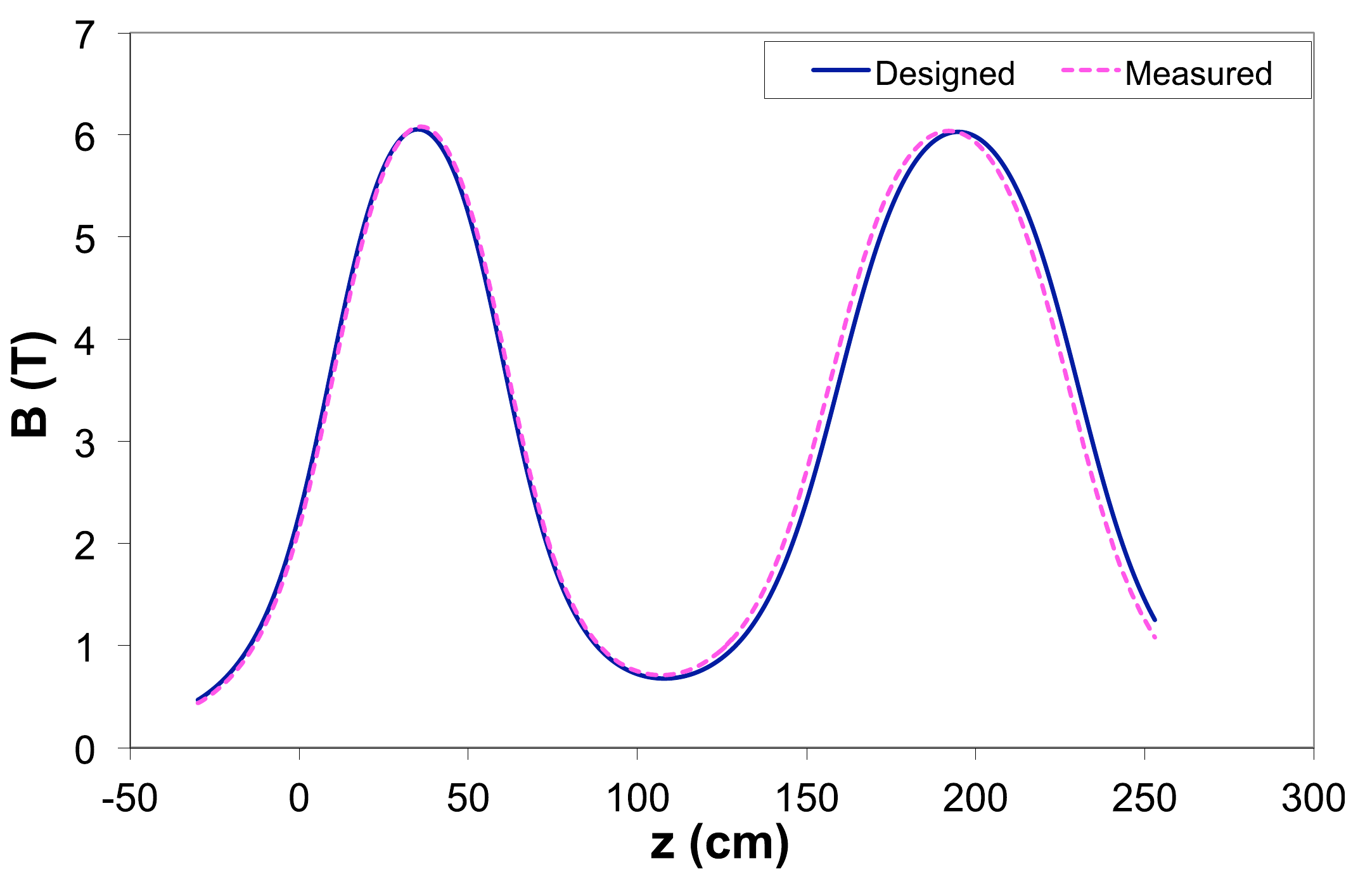}
   \end{center}
   \caption{Designed and measured axial magnetic on-axis fields as functions of position with both magnets at 6 T. The position is measured from the upstream face of the pinch magnet at $z=0$.}
\label{axialfieldscan}  
\end{figure}

The FPD system's two 6-T superconducting solenoids, in separate warm-bore cryostats, were designed and manufactured by Cryomagnetics, Inc, of Oak Ridge, Tennessee. Figure~\ref{axialfieldscan} shows the designed and measured axial fields as functions of axial position. The cryostats are made of stainless steel to minimize their radioactivity, and a vacuum system (Sec.~\ref{sec:apparatus:vacuum}) and calibration equipment (Sec.~\ref{sec:apparatus:eandgammasource}) are housed between the cryostats and within their bores. The magnetic center of each coil is within 2~mm of the center of the bore, as demonstrated by on-axis magnetic field mapping during acceptance tests. 

\begin{figure}[tbp]
   \begin{center}
   \includegraphics[width=\columnwidth]{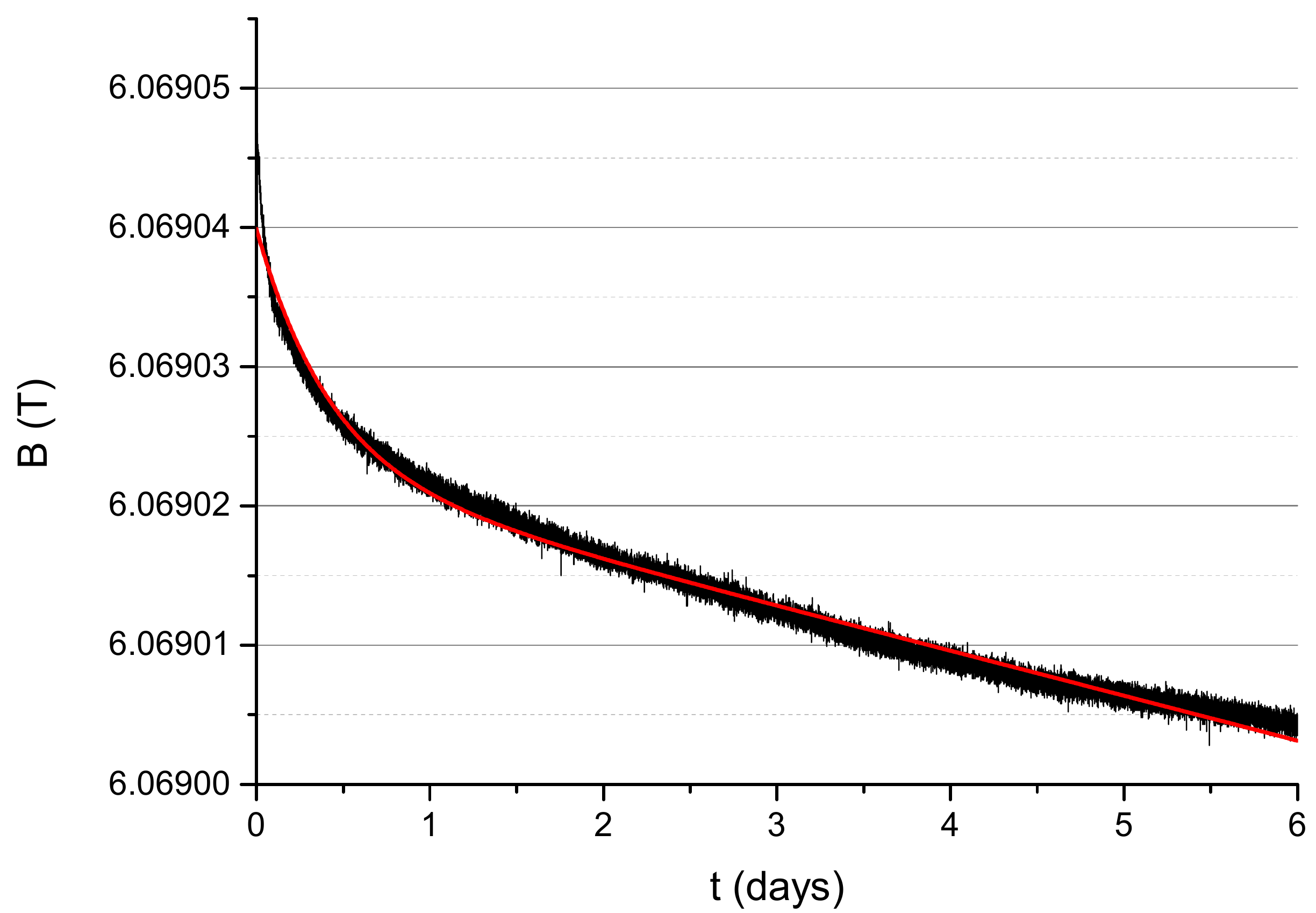}
   \end{center}
   \caption{(Color online.) Sample nuclear-magnetic-resonance probe data from the center of the pinch magnet, showing drift of the on-axis field over several days after the magnet initially entered persistent mode. The fit (thin red line) is to Eq.~\ref{eq:Bfield_decay}. The noise reflects the high axial field gradient.}
\label{fieldstability}  
\end{figure}

Table~\ref{magnet_characteristics} gives the characteristics of the two magnets. Each coil consists of twisted multifilamentary NbTi/Cu wire and a bare NbTi persistent switch that allows operation in either persistent or driven mode. The coils are immersed in liquid helium and operate at a nominal temperature of 4.2~K. Each cryostat is equipped with a Cryomech PT410 pulse-tube cooler for helium recondensing and shield cooling.   A heating element regulates the pressure at 15.1~psia (104.1~kPa) and, under normal operating conditions, there is no measurable loss of cryogens.  Figure~\ref{fieldstability} shows a measurement of the field drift after the pinch magnet enters persistent mode, measured by nuclear magnetic resonance and fit to the expression~\cite{gil:2012}

\begin{equation}
\label{eq:Bfield_decay}
B(t) = B_0 \left[ (1 + \alpha) e^{-t/\tau_1} - \alpha e^{-t/\tau_2} \right].	
\end{equation}  

\noindent Here, $B(t)$ is the magnetic flux density over time, with $B_0 = 6.069$~T the initial magnetic flux density. The time constants $\tau_1 = 1.879(2)\times10^6$~days and $\tau_2 = 0.423(1)$~days parameterize the decays of the transport and screening currents, with $\alpha = -2.863(4)\times10^{-6}$ the mutually inductive coupling coefficient between the two current types. Figure~\ref{fieldstability} shows that Eq.~\ref{eq:Bfield_decay} overestimates the magnetic-field drift at large times $t$. Nonetheless, the magnetic field drift extrapolated over the course of a month is less than 20~ppm, well within the KATRIN design goal of 0.1\%. Other magnetic-field measurements, taken over longer durations, showed slightly smaller field drifts at large $t$.

With both magnets at 6~T, the attractive force between them is 54~kN.  Aluminum bars separate the cryostats, and Cryomagnetics implemented additional internal bracing.  

\begin{table}
\begin{center}
	\begin{tabular}{lcc}
	\hline
	Characteristic & Pinch	& Detector \\ \hline \hline
	Coil length & 500 mm & 700 mm \\
	Cryostat length & 711 mm & 910 mm \\ \hline
	Cryostat inner diameter & 346 mm & 448 mm \\
	Coil inner diameter & 454 mm & 540 mm \\
	Coil outer diameter & 498 mm & 680 mm \\ \hline
	Cryostat He capacity & 65 L & 75 L \\ \hline
	Inductance & 432 H & 647 H \\ \hline
	Maximum field & 6 T & 6 T \\
	Nominal field setting & 6 T  & 3.6 T  \\
	& (87.15 A) & (56.15 A) \\ \hline
	\end{tabular}
\end{center}
	\caption{Magnet-system characteristics. Magnetic field values are quoted for the center of each coil.}
	\label{magnet_characteristics}
\end{table}

\subsection{Vacuum System}
\label{sec:apparatus:vacuum}

\begin{figure}
\begin{center}
\includegraphics[width=\columnwidth]{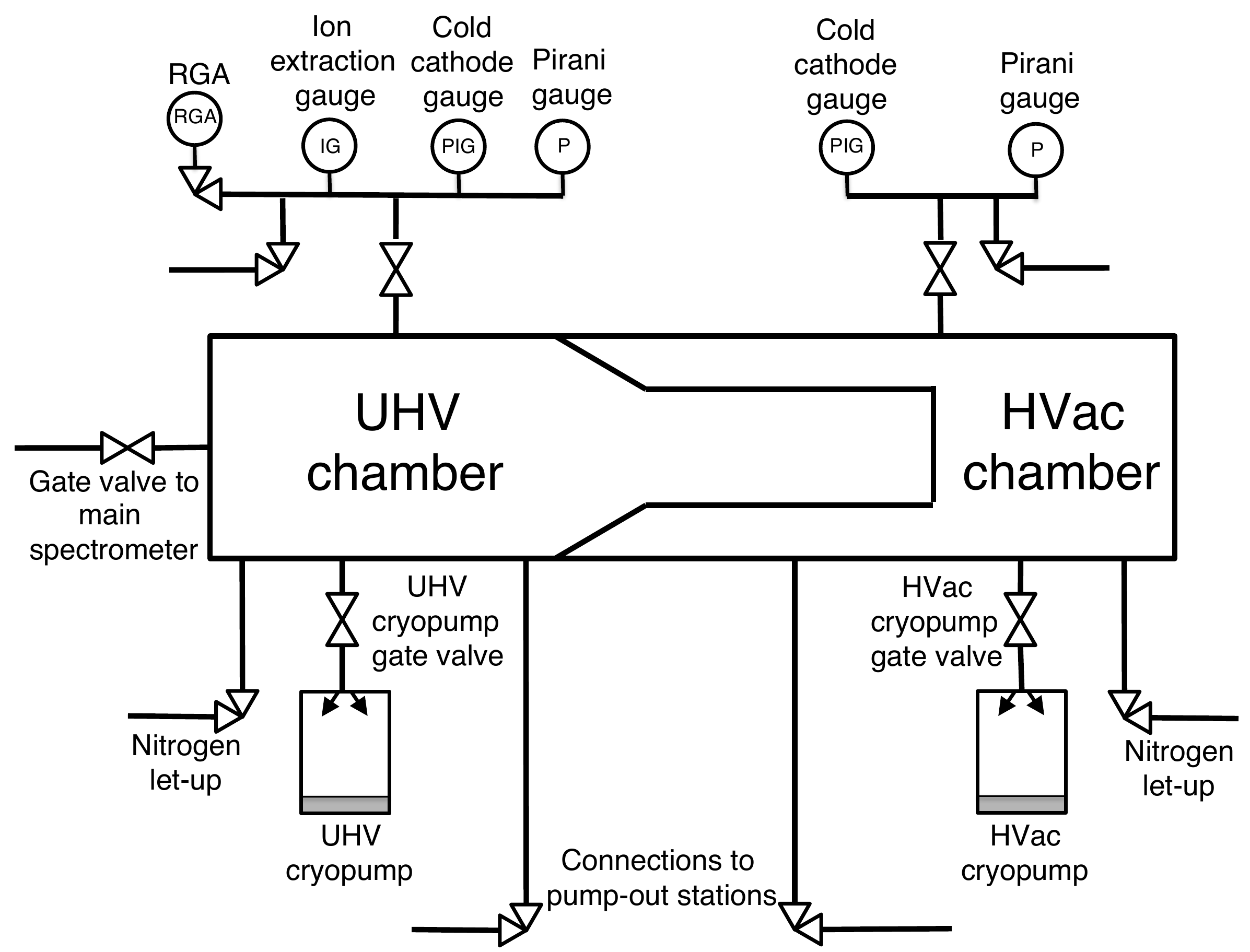}
\end{center}
\caption{\label{vacsys_schematic}Schematic layout of the FPD-system vacuum components in steady-state running.} 
\end{figure}

The FPD system incorporates two independent vacuum chambers. The ultra high-vacuum (UHV) chamber couples to the main spectrometer and houses the FPD wafer (Sec.~\ref{sec:apparatus:detector}). The high-vacuum (HVac) chamber houses the front-end electronics (Sec.~\ref{sec:apparatus:fpd_electronics}), whose outgassing rates are too high to permit UHV pressures. Figure~\ref{vacsys_schematic} is a schematic diagram of the FPD vacuum system. 

A set of four sensors -- a convection-enhanced Pirani (CEP) sensor, a cold-cathode sensor (CC), an extractor ion gauge, and a residual-gas analyzer (RGA) -- monitors the pressure and gas composition in the UHV chamber. A CEP sensor and a CC sensor monitor the pressure in the HVac chamber. The sensors are mounted on extensions approximately 1.5~m from the main body of the vacuum vessel, away from the high-magnetic-field region. 

The trumpet-shaped, copper post-acceleration electrode, manufactured by the Beverly Microwave Division of Communications and Power Industries, forms part of the boundary between the UHV (inside) and HVac (outside) chambers and has been demonstrated to maintain a potential of up to 12~kV relative to ground. The detector and electronics are mounted at its downstream end. Active cooling for the detector and electronics (Sec.~\ref{sec:apparatus:thermosiphon}) is applied through a ceramic insulator brazed to the electrode.

The UHV and HVac chambers are initially evacuated using dedicated pump-out stations attached by metal bellows. Each station consists of a turbomolecular pump backed by a dry scroll pump. During the initial evacuation, each pump-out station is monitored by a CEP sensor and a CC sensor, with an additional RGA for the UHV pumpout station. A high-temperature bakeout at 150~$^{\circ}$C is performed before the detector wafer is installed; after this installation, the system undergoes an additional low-temperature bakeout at 65~$^{\circ}$C.  After the initial evacuation and final bakeout, the two pump-out stations are removed. Vacuum in both chambers is then maintained by cryo\-pumps, which are less sensitive to magnetic fields. All-metal gate valves allow isolation of the cryo\-pumps from their vacuum chambers. 

During the Karlsruhe commissioning described in this work, base pressures of $2.8 \times 10^{-9}$~mbar (UHV) and $3.0 \times 10^{-6}$~mbar (HVac) were achieved with a cooled system (Sec.~\ref{sec:apparatus:thermosiphon}) after bakeout and installation of the electronics. The UHV pressure was later improved to $1.7 \times 10^{-9}$~mbar after the repair of a known air leak.

Vacuum-system maintenance, including regeneration of the cryo\-pumps, can easily be scheduled for general maintenance periods so as not to affect regular data-taking.

\subsection{Focal-Plane Detector}
\label{sec:apparatus:detector}

\begin{figure}[tbp]
   \begin{center}
   \includegraphics[width=\columnwidth]{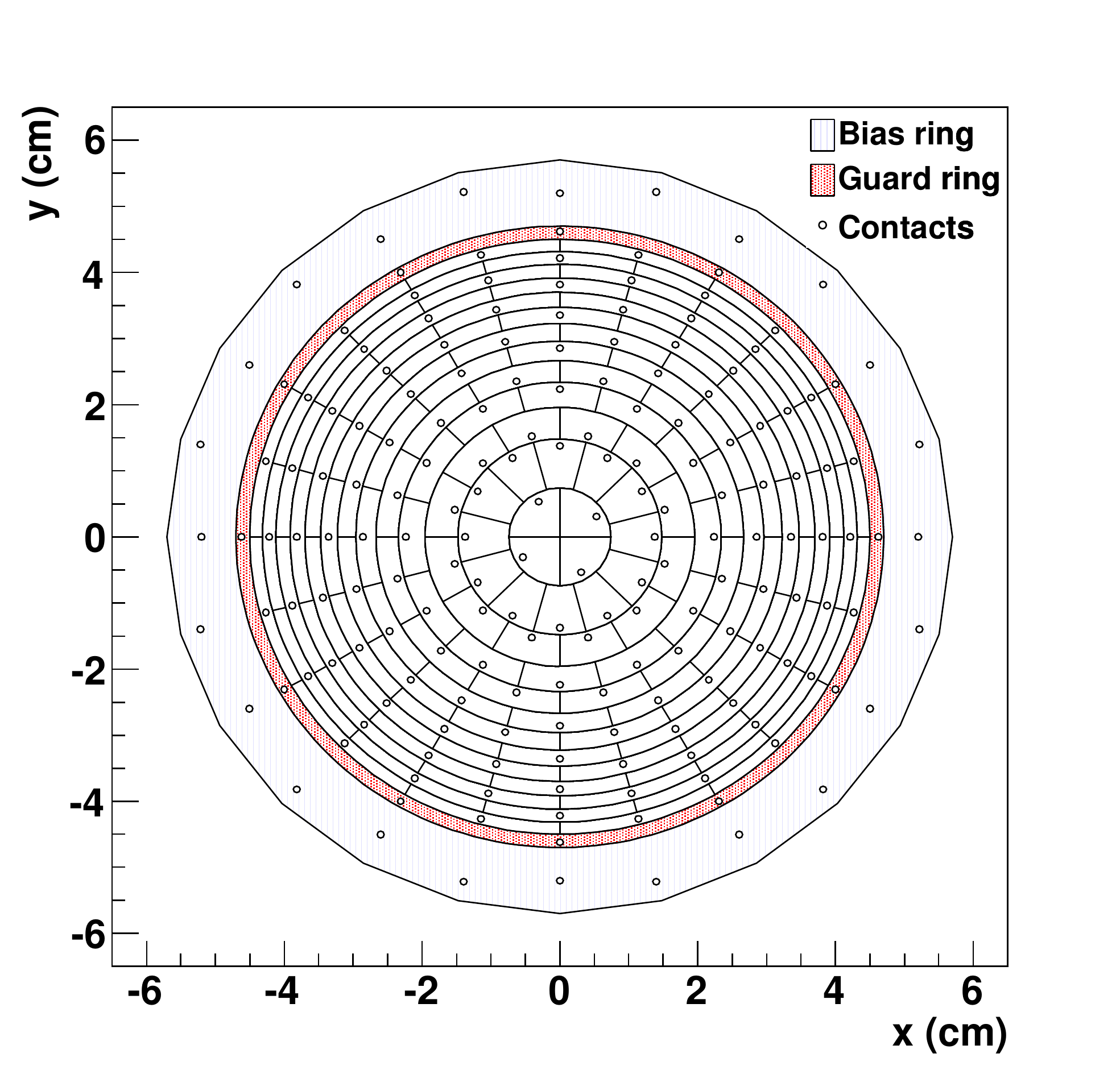}
   \end{center}
   \caption{FPD dartboard pixel pattern, surrounded by a guard ring and bias ring. Open circles show the points where electrical contacts are made.}
\label{FPD_back}  
\end{figure}

Beta electrons from the main spectrometer strike the shallowly ion-implanted, $n\tiny{++}$ ohmic face of a monolithic 148-pixel p-i-n-diode array on a single silicon wafer. This detector is made via a double-sided process~\cite{Canberra03} on a wafer that is 503 $\mu$m thick with a 125-mm diameter. Manufactured by Canberra Belgium according to the pattern shown in Fig.~\ref{FPD_back}, it has a sensitive area 90 mm in diameter, surrounded by a 2-mm guard ring and a 15.5-mm bias ring. Its crystal orientation is $\langle 111 \rangle$. The resistivity is 8~k$\Omega$-cm. A non-oxidizing TiN coating on the pixelated side facilitates electrical connections; this coating wraps around the edges to the insensitive, outer areas of the unsegmented front face to allow the application of a 120-V bias from the pixel side. The specified dead-layer thickness for all wafers, assuming a complete absence of charge collection in the dead layer, was 100~nm. A comparison of simulated spectra to photoelectron data taken in Seattle revealed this assumption to be inaccurate: some charge deposited in the dead layer is collected as part of the measured pulse. With this more realistic definition, the dead-layer thickness was found to be $155.4 \pm 0.5_{\mathrm{stat}} \pm 0.2_{\mathrm{sys}}$~nm with 46\% charge collection~\cite{wall:2013}. The precision of this measurement, and of the simulation package developed for electron interaction in our detector~\cite{Renschler_thesis}, allows our analysis tools to compensate for the deviation from the original specification.

The segmented back face has 148 ion-implanted, p-type pixels, which are separated by 50-$\mu$m boundaries with a specified pixel-to-pixel resistance larger than 1~G$\Omega$. Each pixel has an area of 44~mm$^2$ and a design capacitance of 8.2~pF. The pixels are grouped into twelve concentric, equal-area rings of twelve pixels each; in the center, the bull's eye provides an additional four pixels. This arrangement allows later correction for radial electrical and magnetic inhomogeneities in the analyzing plane. 

\begin{figure}[tbp]
   \begin{center}
   \includegraphics[width=\columnwidth]{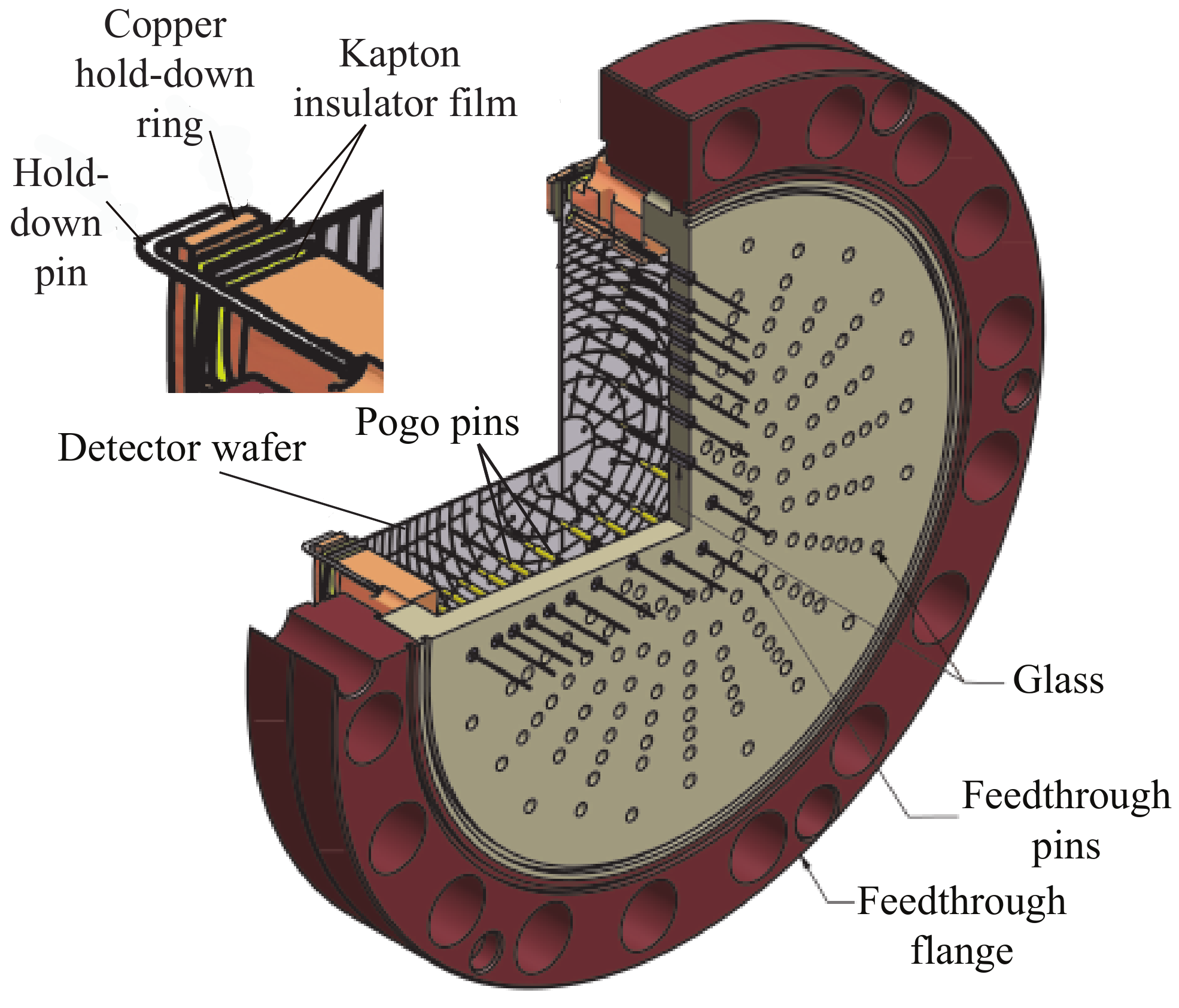}
   \end{center}
   \caption{Mounted detector wafer (back) with spring-loaded pogo pins and feedthrough flange. The inset is a close-up view of the detector compression scheme.}
\label{FPD_mount}  
\end{figure}

The electrical connection with each pixel is made by a spring-loaded Interconnect Devices pogo pin with a Ni-Ag barrel, Be-Cu plunger and a stainless-steel spring, all gold-plated. For optimal contact, the wafer must compress the pins by at least 0.38~mm, corresponding to a deformation of 0.24~mm at the wafer center. A novel mount~\cite{VanDevender:CARAC2011}, shown in Fig.~\ref{FPD_mount}, achieves this compression by holding the bare detector wafer to its mount with a copper hold-down ring and six hold-down pins; copper stops keep the wafer on-axis, and Kapton film insulates the wafer from its mount. The mount is bolted directly to the feedthrough flange. At room temperature, typical leakage currents are $0.6 \pm 0.1$~nA per pixel. There is no indication of additional stress-induced leakage currents for wafer deformations of up to 0.41~mm~\cite{VanDevender:CARAC2011}.

The detector flange separates the UHV and HVac regions and provides a low-capacitance, low-microphonic connection between pixels and preamplifiers. Mill-Max brass-alloy adapter receptacles connect the pogo pins to the feedthrough pins in the detector flange. On the other side of the flange, the feedthrough pins make direct contact with the preamplifier modules. Made by Ametek Hermetic Seal Corporation, the feedthrough is a custom array of 184 gold-plated Inconel X-750 pins 0.5~mm in diameter: 148 detector-pixel contacts, 12 guard-ring contacts and 24 bias-ring contacts spatially arranged to match both the wafer mask and the preamp contacts.  The pins are sealed with type-L21 borosilicate glass into an Inconel X-750 plate, which in turn is welded into a 304L stainless-steel Conflat-style flange. The potassium content of the glass is 3.6(1)\% by weight, and betas from the decay of $^{40}$K in the feedthrough seals are magnetically guided to the associated detector pixels. Simulations (Sec.~\ref{sec:sim_bg}) indicated that these betas constituted a significant portion of the observed background. After the commissioning period in Seattle, cylindrical copper shields, 3.2~mm in height with a 33.5-mm$^2$ cross section, were designed, machined and installed on the detector flange. Each cylinder fits over one of the 148~signal pins, and shields that pixel from betas emitted by the feedthrough seal.

\subsection{Readout Electronics}
\label{sec:apparatus:fpd_electronics}

\begin{table}
\begin{center}
	\begin{tabular}{lc}
	\hline
	Parameter & Specification	 \\ \hline \hline
	Noise level (30 kHz $-$ 30 MHz) & $1-2$ nV Hz$^{-1/2}$ \\ 
	Power dissipation in HVac section & $\sim 15$ W \\ 
	Operating temperature range & $-40$ to $+ 70$~$^{\circ}$C \\ 
	Maximum temperature & $100$~$^{\circ}$C \\ 
	Maximum total event rate (100 keV) & 100 kcps \\ 
	Full-scale energy range (variable gain) & $100-6000$ keV \\
	Leakage-current measurement range & $0-5$ nA \\ 
	Integral nonlinearity (fiber-optic) & $< 1\%$ \\ 
	Gain stability (fiber-optic) ($0-30^{\circ}$C) & $<1\%$ \\ \hline
	\end{tabular}
\end{center}
	\caption{Design specifications for front-end electronics and fiber-optic links.}
	\label{fpdelec_specs}
\end{table}

\begin{figure}
\begin{center}
\includegraphics[width=\columnwidth]{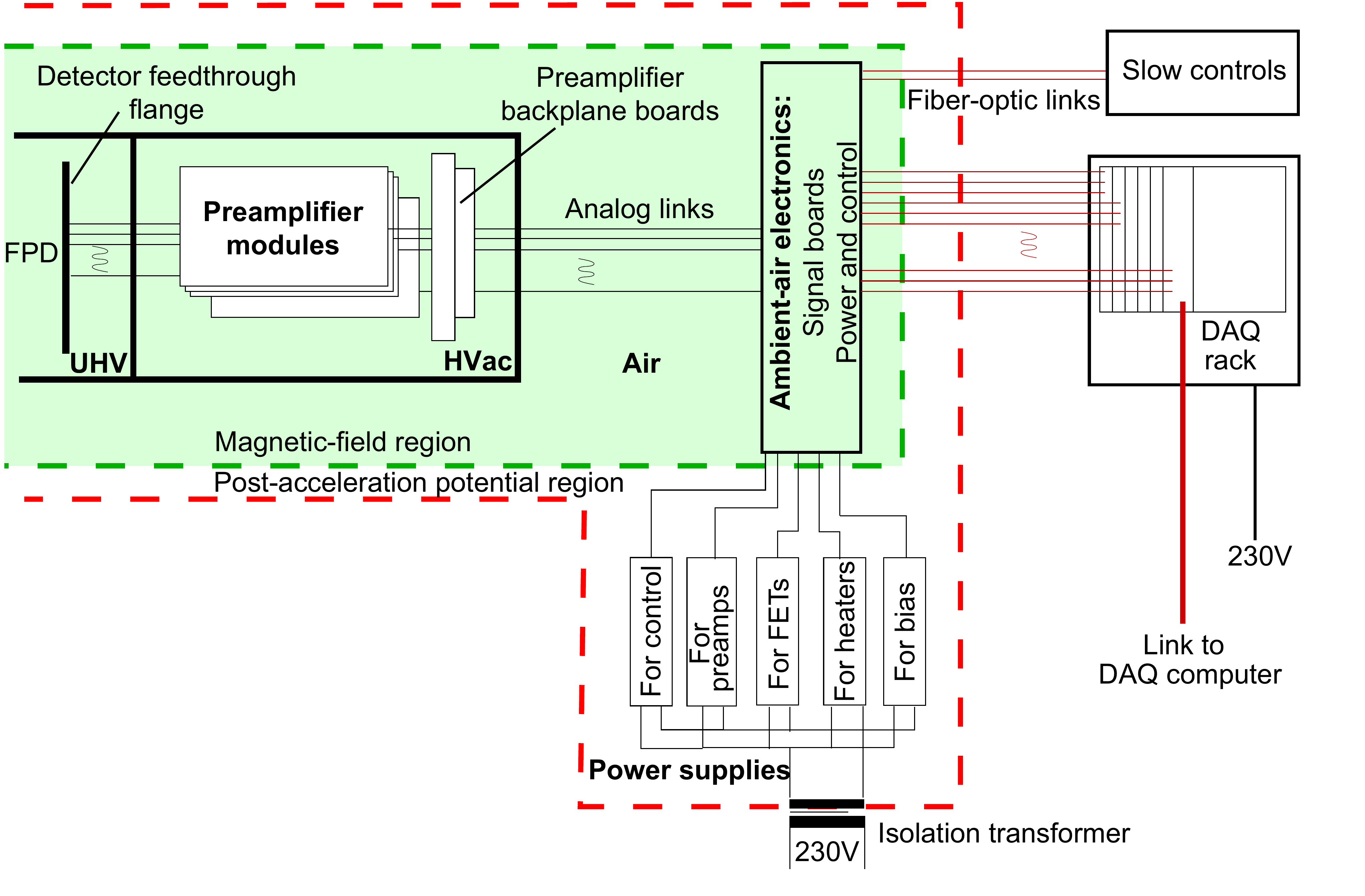}
\end{center}
\caption{\label{FPD_electronics_schematic}Schematic layout of the detector readout electronics. The shaded region is the high-magnetic-field area; the outer dashed line marks the region held at post-acceleration potential.}
\end{figure}

A custom-designed suite of electronics (Fig.~\ref{FPD_electronics_schematic}) reads out signals from the detector. To reduce noise, the preamplifier modules and service electronics are mounted directly on the electrical feedthroughs of the detector flange; all other electronics are mounted outside the HVac chamber and outside the radiation shield (Sec.~\ref{sec:apparatus:veto}). Plastic optical fibers carry analog signals between the detector electronics at post-acceleration potential and the data-acquisition (DAQ) system at ground potential. Table~\ref{fpdelec_specs} lists the design specifications. 

\begin{figure}
\begin{center}
\includegraphics[width=\columnwidth]{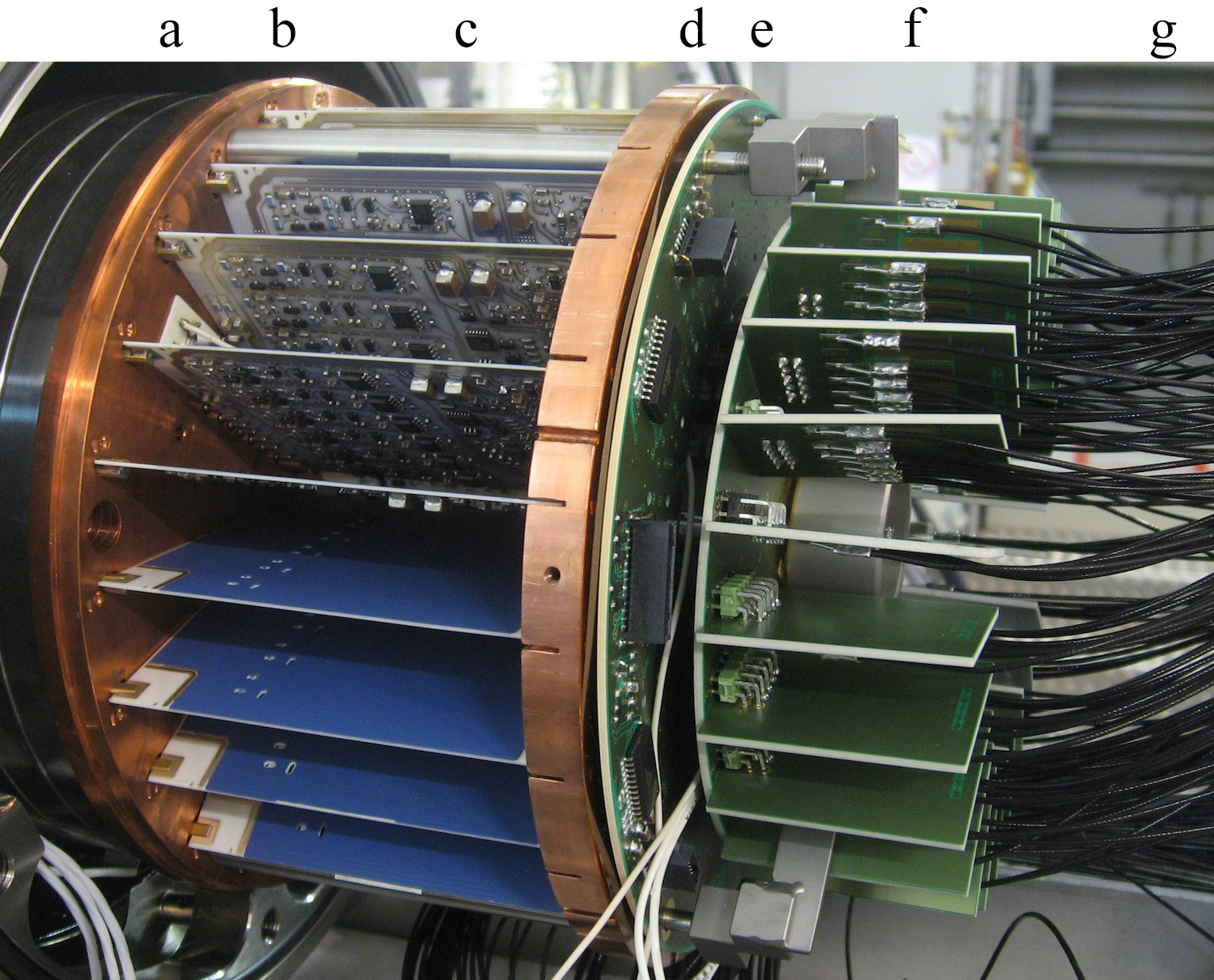}
\end{center}
\caption{\label{FrontEndElectronics}Front-end FPD electronics: (a) detector feedthrough flange, (b) copper mounting plate, (c) preamplifier modules, (d) copper support ring, (e) power-and-control distribution board, (f) signal distribution board, (g) coaxial signal cables.}
\end{figure}

\paragraph{Preamplifier Modules} On the downstream side of the detector flange, 24 charge-sensitive preamplifier modules are arranged in a radial pattern with $15^\circ$ angular spacing (Fig.~\ref{FrontEndElectronics}). Each module has either six or seven channels of charge-sensitive preamplifiers and serves pixels from alternating rings. 

The modules are built on 0.63-mm-thick Rubalit 710S aluminum-oxide ceramic boards manufactured by CeramTec. These boards are compatible both with high vacuum and with a specified heat dissipation of $0.6-1$~W per module; the measured heat dissipation in the system is 0.72~W per module. Mounting pins conduct heat to the detector flange, which is ultimately connected to an active cooling system (Sec.~\ref{sec:apparatus:thermosiphon}). Each module provides two feedthrough lines, each equipped with a noise filter, which supply the FPD with its guard-ring and bias potentials.

Only the first preamplification stage is included on each module, in order to limit power dissipation and allow sensitivity to 6-MeV background alpha particles from radon daughters. Each module houses multiplexers, switching devices, decoders, and current-to-voltage converter circuitry for test-pulse distribution and for temperature and leakage-current measurements. In Karlsruhe, the field-effect transistors (FETs) were replaced with low-noise AM radio front-end models (NXP/Philips BF862), which have improved the resolution. Single-ended signal transmission limits the number of feedthroughs. To avoid ground loops across the signal lines, differential transceivers are used outside the HVac chamber.

\paragraph{Preamplifier Backplane Boards} There are two backplane boards downstream of the preamplifier assembly (Fig.~\ref{FrontEndElectronics}). The ring-shaped board redistributes control and power lines onto miniature surface-mount connectors. The circular board distributes analog preamplifier output signals to the cable harness. Both boards have low heat dissipation and are made of Rexolite plastic.

\paragraph{Ambient-Air Electronics} 

Four 37-channel signal boards and one power-and-control board are located outside the radiation shield and HVac region, and float at the post-acceleration potential. The signal boards provide differential pick-up, additional signal amplification, variable-gain stages, and fiber-optic transmitters, which transmit the data from each channel via a separate plastic-optical-fiber link to the DAQ system. The power-and-control board includes power-conditioning circuits, overvoltage protection, variable-gain controls, and temperature readouts. The boards connect to the cable harness via 50-pin, vacuum-tight, sub-D type feedthroughs manufactured by the Ametek Hermetic Seal Corporation. 

The entire set of ambient-air boards is enclosed in an anti-corona housing and a Faraday cage for safety and for transient protection. Power for these boards is supplied through an insulated metal tube.

\paragraph{Data-Acquisition and Power-Supply Racks} The DAQ (Sec.~\ref{sec:apparatus:daq}) and power-supply racks are located 2.5~meters from the ambient-air electronics boards, outside the 7-mT magnetic-field contour where the power-supply fans can function reliably. Optical receiver boards in the analog backplane of the DAQ crate translate the optical signals to electrical signals for the digitizers. The power-supply rack floats at post-acceleration potential and is surrounded by a Faraday cage; the DAQ rack is held at ground potential.

\paragraph{Electronics Testing and Development} 
In order to maintain a continuous capability to test spare components, perform diagnostic tests on malfunctioning components, develop improved hardware and software, and train operational staff, we have built a complete test stand for the FPD electronics. The stand replicates the full signal chain, from a detector-wafer dummy, through the analog electronics inside a shielding can, to the DAQ hardware and software, including data storage, network access, and control software. The cable harness and analog fiber-optic links are of the same length as in the real FPD assembly.

\subsection{Cooling System}
\label{sec:apparatus:thermosiphon}

The detector and its readout electronics are cooled in order to reduce leakage currents and noise, and to conduct heat from the electronics. The problem of cooling is complicated by the location of these elements: in vacuum, in a strong magnetic field, and floating at the post-acceleration potential. Cooling is provided by a single-stage pulse-tube cryocooler (PT-60 UL / CP830 from Cryomech, Inc.), whose cold head must be mounted 1.2~m above the magnetic axis. In the prototype apparatus, a long copper cold finger and a set of flexible copper braids conducted heat to the cold head from a copper ring around the ceramic insulator surrounding the post-acceleration electrode (Sec.~\ref{sec:apparatus:vacuum}). This design proved inadequate for the heat load of $25-35$~W, however. A custom thermosiphon was designed, built and installed to conduct heat between the cooling ring and the cold head, replacing the copper rod and braids.

\begin{figure}
\begin{center}
\includegraphics[width=\columnwidth]{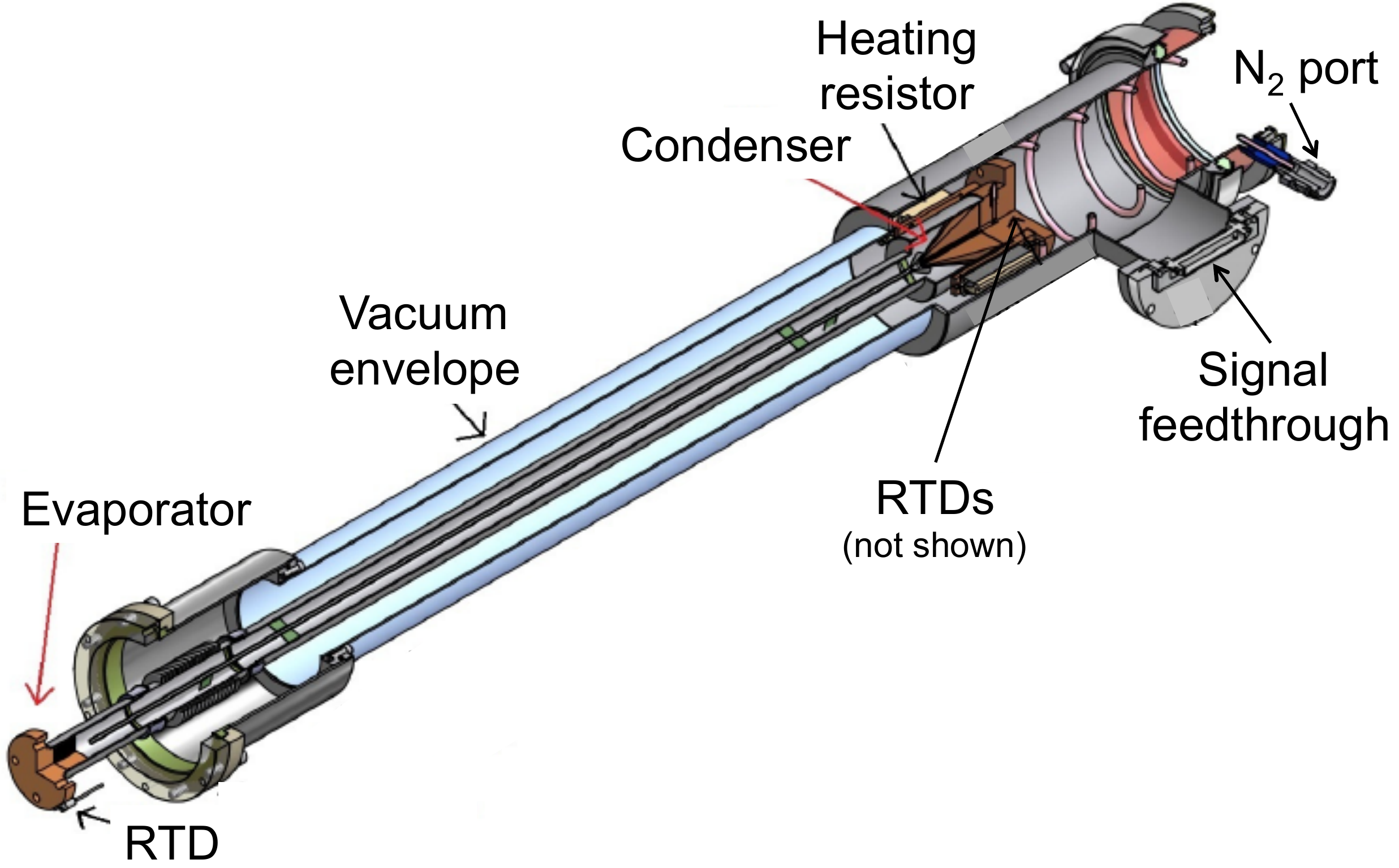}
\end{center}
\caption{\label{fig:thermosiphon}Three-quarter cross-section of thermosiphon. The cold head of the pulse-tube cooler, not shown, mates to the top of the condenser.}
\end{figure}

Figure~\ref{fig:thermosiphon} shows a cutaway view of the thermosiphon. At the cold end, the pulse-tube-cooler cold head bolts to a conical copper condenser. Condensed nitrogen droplets flow to the tip of the condenser and drop to the copper evaporator, 0.78~m below, through a central, 3-mm stainless-steel tube. Nitrogen vapor rises to the condenser in a concentric 25-mm tube, the adiabatic region of the thermosiphon. The evaporator area of 0.0108~m$^2$ was chosen to support a 60-W boiling rate, assuming complete immersion and a heat flux rate of 6200~W/m$^2$, the film boiling minimum in a similar system~\cite{Jin:BoilingHeatTransfer09}. An external nitrogen reservoir, held at room temperature, is connected to the condenser by a spiral tube with low thermal conductance. Platinum resistance temperature detectors (RTDs) sense the temperature at the evaporator and condenser. A pair of resistors, controlled by a proportional-integral-derivative (PID) loop, maintain a condenser temperature above the freezing point of nitrogen. 

The pulse-tube cooler and thermosiphon are held at ground potential and must therefore cool the entire bulk of the post-acceleration electrode in addition to the detector and electronics. As a result, it takes the system two days to equilibrate. During normal operation, this system can maintain a detector flange temperature below 0$^{\circ}$C and electronics temperatures below 20$^{\circ}$C, with drift within about 0.5$^{\circ}$C/day. 

\subsection{Shield and Veto System}
\label{sec:apparatus:veto}

\begin{figure}
\begin{center}
\includegraphics[width=\columnwidth]{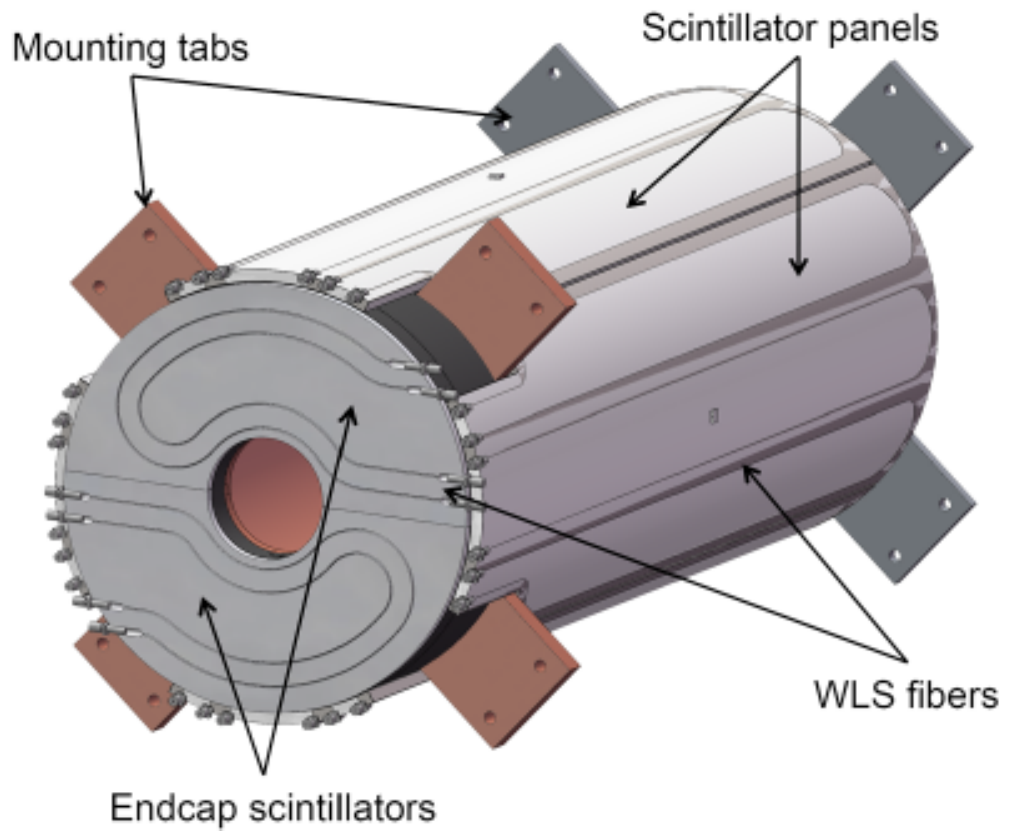}
\end{center}
\caption{\label{veto_panels}Veto scintillator panels, with WLS-fiber layout.}
\end{figure}

The FPD system's radiation shield consists of two nested cylindrical shells: a 3-cm thickness of lead that reduces the $\gamma$ background by an estimated factor of 20, and an inner 1.27-cm thickness of oxygen-free, high-conductivity copper to block lead X-rays. It is surrounded by a veto system (Fig.~\ref{veto_panels}) to tag incoming muons: a four-panel barrel with its downstream end partially covered by two semicircular endcap panels.  Each curved, 1-cm-thick panel is made of St.\ Gobain Bicron-408 plastic scintillator wrapped in Gore Diffuse Reflector Product, with an outer layer of adhesive-backed aluminum foil. The whole assembly is 38~cm in outer diameter and 106.3~cm in length. Embedded single-clad St.\ Gobain wavelength-shifting (WLS) fibers collect scintillation light and transport it to photon detectors.  There are three WLS fibers in each of the four long panels and two in each of the two endcap panels; the fibers are bent into U shapes and read out at both ends. Outside the scintillator panels, each WLS fiber end couples to a clear plastic optical fiber, which has a longer attenuation length.

Individual multi-pixel photon counter (MPPC) units from Hamamatsu (S10362-11-050P), selected for their ability to operate in high magnetic fields, detect light from each of the 32 fiber ends.  The units operate in a dry-nitrogen enclosure at individually optimized bias voltages of 70~V on average.  MPPCs suffer from high dark rates, which decrease exponentially with temperature. At room temperature, dark rates are on the order of 1~Mcps; cooling to $-18^{\circ}$C reduces dark rates to $\sim 4.7$~kcps/channel. The MPPCs are maintained at that temperature by two water-cooled Peltier coolers.

Each of the veto's four front-end circuit boards, developed at the Karlsruhe Institute of Technology, accommodates eight preamplifier channels with individually adjustable bias voltages, along with a bias-voltage regulator, a temperature readout, and power-conditioning circuitry. To reduce rates from thermally induced signals, the preamplifier for each MPPC was equipped with a baseline clip to reduce pulses by at least the single-photoelectron peak height. The board outputs drive standard 50-$\Omega$ coaxial cables with an amplification of 235x. On average, the signal from an incident muon produces 4 or 5 photoelectrons, yielding a post-preamplifier pulse of some 50~ns full-width at half-maximum (FWHM) with an amplitude of $-50$ mV per photoelectron. 

The relative efficiencies of the panels were determined by measuring coincidences with the panels arranged in a vertical stack. In this hodoscope configuration, the efficiency for each long panel was determined from the number of times it fired in coincidence with a particle detection from three other long panels; each endcap panel was compared to two other endcap panels.  These tests demonstrated efficiencies in excess of 91\% for five panels, with the sixth, a long panel, at 84\%.

\subsection{Calibration Systems}
\label{sec:apparatus:calibration}

Two systems, one providing $\gamma$s and one providing photoelectrons with adjustable energies, calibrate the detector's response to incident particles (Sec.~\ref{sec:apparatus:eandgammasource}). Monitoring the photocurrent of the latter source (Sec.~\ref{sec:apparatus:pulcinella}) also allows measurements of the detector efficiency.

\subsubsection{Calibration Sources}
\label{sec:apparatus:eandgammasource}

\begin{figure}[tbp]
\begin{center}
        \includegraphics[width=\columnwidth]{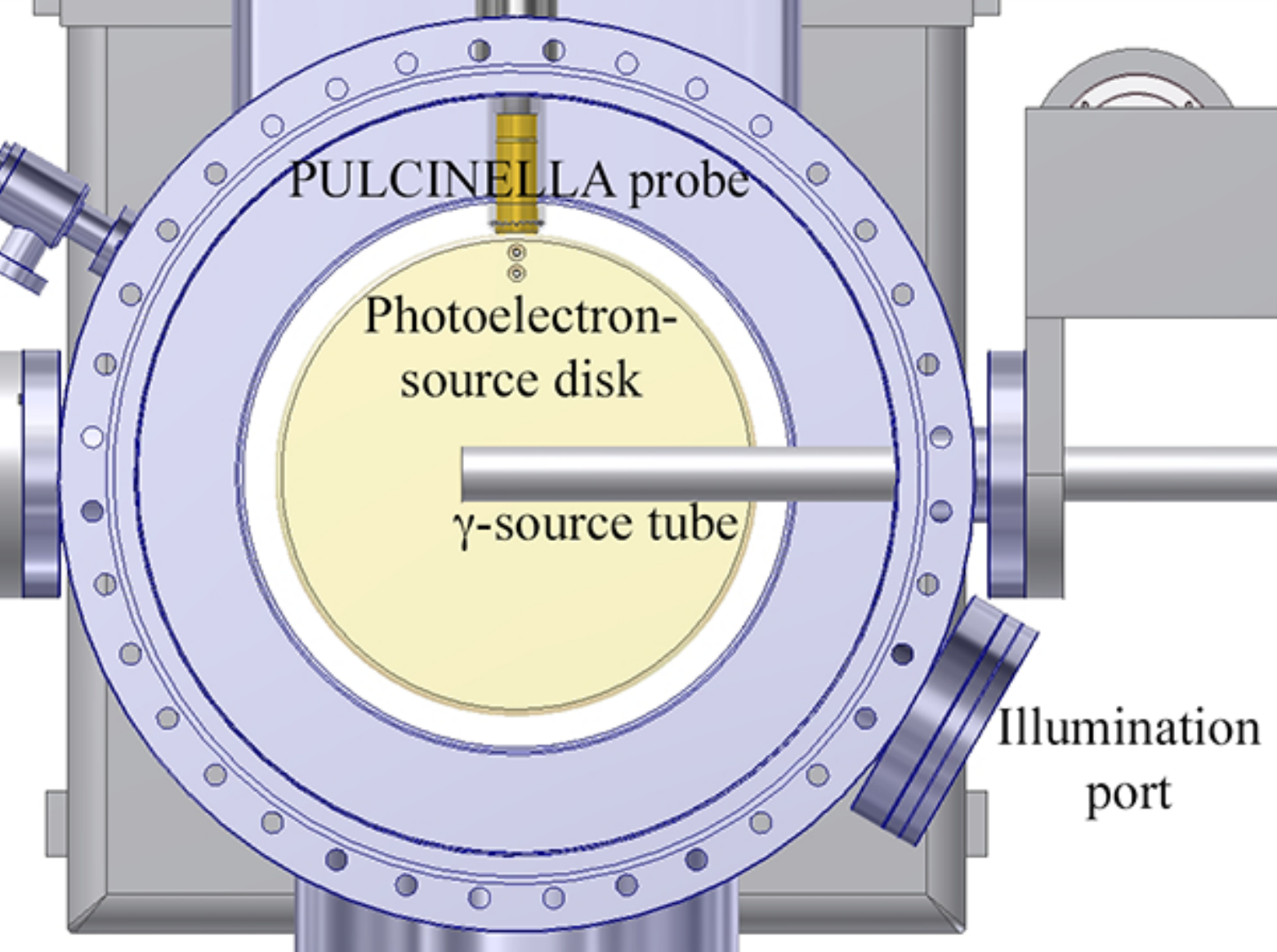}
        \caption{View from the detector looking upstream toward the main spectrometer, showing the inserted $\gamma$ source, the inserted disk of the photoelectron source, and the port through which the photoelectron source is illuminated with UV light.}
        \label{upstream_esourceview}
\end{center}
\end{figure}

Figure~\ref{upstream_esourceview} shows the calibration systems as seen from the detector. KATRIN will use two encapsulated $\gamma$ sources, which provide an absolute energy-scale calibration independent of dead-layer effects.   The first source, which is presently in use, is 0.5~mCi of $^{241}$Am. This source emits $\gamma$s at 59~keV and 26~keV, as well as X-rays from $^{241}$Am daughters.  The planned second source is 0.5~mCi of $^{109}$Cd, which will emit 88-keV $\gamma$s as well as lower-energy X-rays. The experimenter can move the $\gamma$ source into the detector line of sight, without breaking vacuum, via an air motor and bellows with a thin aluminum window.

Gamma radiation liberates electrons from the source holder via the photoelectric effect. When the FPD-system magnets are energized, these high-rate photoelectrons are guided to specific regions of the FPD wafer. The photoelectron continuum obscures the $^{241}$Am calibration lines in the affected pixels, which number between four and six depending on the field strength and source location. A full calibration of the entire wafer under magnetic field thus requires data taken at two distinct source locations, chosen so that the affected regions of the wafer do not overlap.

Stable photoelectron sources allow measurement and monitoring of the detector response to electrons of a chosen energy. In our apparatus, bellows and air motor allow the operator to change the position of a photoelectron source disk made of titanium, without breaking vacuum. When the disk is in place, a high voltage is applied to it, raising the energy of the photoelectrons emitted when the disk is illuminated with an ultraviolet (UV) light-emitting diode (LED). Due to space and vacuum constraints, the 255-nm UV source is located outside the vacuum system. An adjustable illumination assembly projects its light onto the disk. The resulting photoelectrons are adiabatically guided to every pixel of the detector by the detector-magnet field (Sec.~\ref{sec:apparatus:magnet}). The illumination device also includes a red LED whose pulsed light can be used for linearity measurements (Sec.~\ref{sec:performance:linearity}).

\subsubsection{Photocurrent Measurement}
\label{sec:apparatus:pulcinella}

The Precision Ultra-Low Current Integrating Normalization Electrometer for Low-Level Analysis (PULCINELLA) is designed to measure picoamp-scale currents with 3\% accuracy. PULCINELLA permits a determination of the absolute efficiency of the detector via a comparison of the measured photocurrent leaving the photo\-electron source to the recorded event rate. In principle, the ability to measure current on the photo\-electron source also allows the disk to be used as a Faraday cup when commissioning new high-voltage configurations for the experiment. In this mode, the disk shields the detector from currents arising from discharges upstream, while PULCINELLA provides some ability to measure the frequency of such discharges.

The PULCINELLA meter board uses a charge-integrating Texas Instruments DDC-114 analog-to-digital converter (ADC) 
controlled by an on-board field programmable gate array (FPGA) to continuously sample current. The ADC has a full-scale range of 12~pC; the null current is shifted to mid-range via a bias resistor. Jumper pins select an integration time from 10~ms to 2.55~s. The FPGA receives data from the ADC, appends the channel number and an eight-bit clock register for error checking, and then sends the data optically to a receiver board at ground potential. The meter board floats at high voltage but uses a low-voltage power source: a photovoltaic module illuminated by an array of LEDs. The receiver board adds a bit for phase synchronization and transmits the data for readout using standard DAQ software (Sec.~\ref{daq_sw}).

The electronics for PULCINELLA are contained inside a series of nested boxes. The outer box is held at the same potential as the UHV chamber and supports the LED array. The middle and inner boxes connect to the photoelectron-source voltage through two RC filters. The photovoltaic module is housed in the middle box; the innermost box contains the meter board, which connects directly to the electron emitter through a vacuum feedthrough.

This feedthrough connects the meter to the photo\-electron-source disk by a 6.35-mm-diameter rod. Around the rod are a 12.7-mm-diameter tube electrically connected to the inner box, and a 19.05-mm-diameter tube electrically connected to the middle box at one end and to the electron-source disk on the opposite end.   A 25-mm glass tube near the disk end separates the two potentials on this outer tube, which forms the vacuum boundary for the probe. A spring at the top of the feedthrough pushes up on the top of the rod, opposing atmospheric pressure and the weight of the disk and thereby reducing stress on the glass electric break. To prevent photo\-electron emissions in response to UV illumination of the probe, the disk end of the probe is coated in gold up to the electric break. Nearly the whole length of the outer tube is surrounded by a 28-mm-outer-diameter silica sleeve, which extends down to the disk side of the electric break and further prevents electron emissions. 

\subsection{Data Acquisition}
\label{sec:apparatus:daq}

The DAQ system developed for the FPD and veto consists of two closely coupled parts: the DAQ electronics (Sec.~\ref{daq_hw}) and the DAQ  software (Sec.~\ref{daq_sw}). Devices that are not part of the FPD-system signal chain are monitored via the slow-controls system (Sec.~\ref{sec:apparatus:slowcontrols}).

\subsubsection{KATRIN DAQ Hardware}
\label{daq_hw}

\begin{figure}
\begin{center}
\includegraphics[width=\columnwidth]{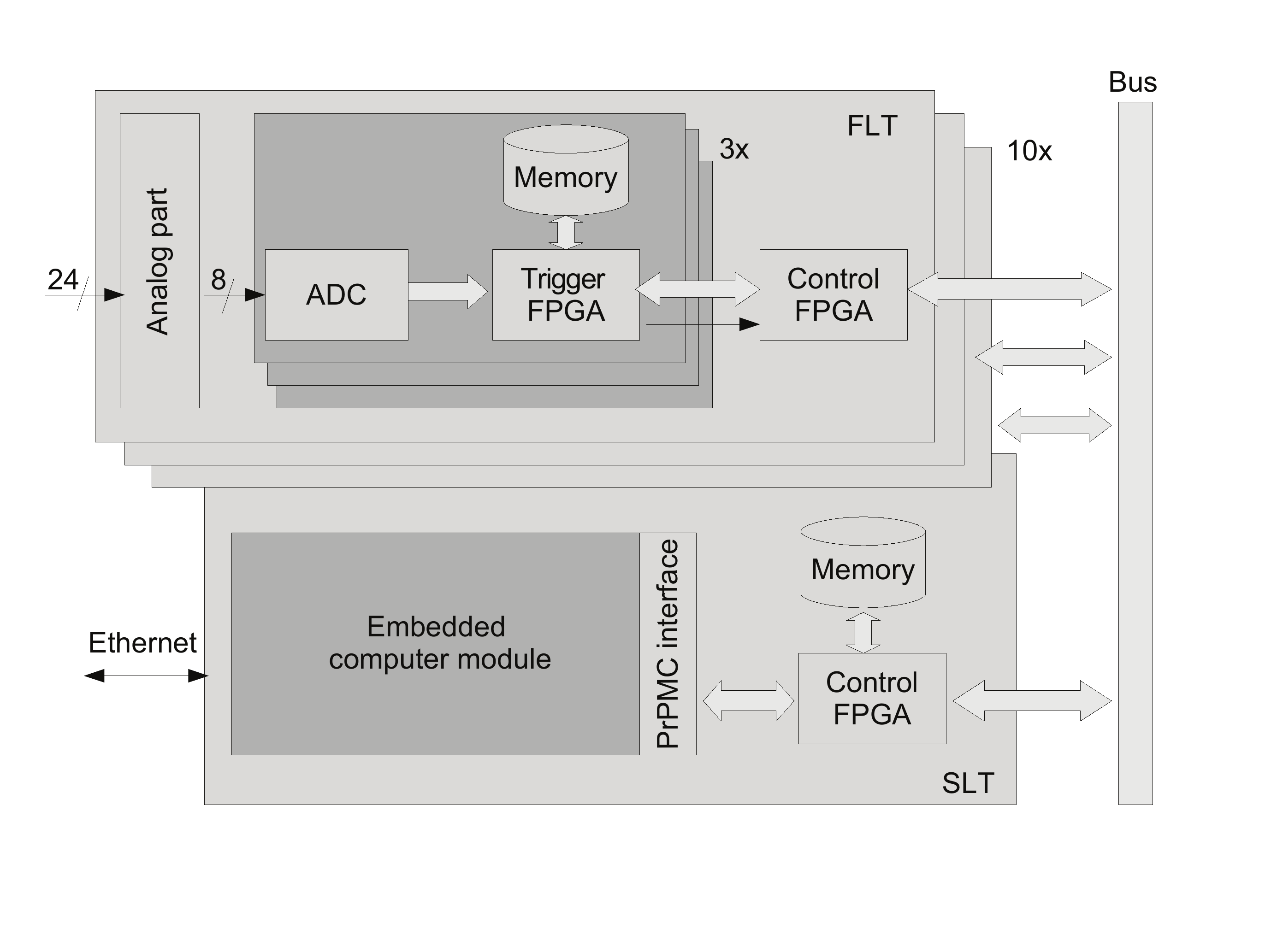}
\end{center}
\caption{\label{DAQ-arch} Simplified architecture of the DAQ electronics. The setup consists of 10 first-level trigger cards and one second-level trigger interface card in a subrack. The GTL control bus is not shown.}
\end{figure}

Our custom DAQ electronics, shown schematically in Fig.~\ref{DAQ-arch}, follow the same concept used for the Pierre Auger Cosmic Ray Observatory~\cite{Gemmeke:AUGER01}. The analog part of the chain is KATRIN-specific, while the digitizers and the digital components are fairly universal. Programmable logic, implemented in FPGAs, allows the digital filtering and trigger system to be changed at any time.

The DAQ electronics consist of first-level trigger (FLT) and second-level trigger (SLT) cards. Each FLT card can handle 24 channels of analog signal conditioning and processing, using analog differential receivers with programmable offsets; programmable amplifiers; bandpass filters; digitizer drivers; serial ADCs with 12-bit precision and a 20-MHz sampling rate; and auxiliary memory for the ADCs. Altera Cyclone II EP2C35 FPGAs control acquisition and preprocessing while an additional central-control FPGA performs time synchronization and readout for each card. Eight FLT cards serve the main detector, while two serve the veto. 

A single SLT card provides a single-board computer (SBC) -- an embedded 1.4-GHz Pentium M processor running Linux -- that initializes and coordinates all ten FLT cards over a multipoint LVDS bus with a maximum measured throughput of $\sim 70$~MB/s~\cite{bergmann:2012}. The SLT communicates with the DAQ computer via a fast Ethernet interface, with synchronization using the network time protocol. An additional synchronization unit delivers a 10-MHz signal as well as a 1-Hz signal that synchronizes the internal counters of the DAQ electronics. This allows searches for coincident events across different DAQ systems connected to the synchronization unit, e.g.\ between FPD events and events in the main-spectrometer muon-detector system.

The FPD DAQ digital electronics include a ring buffer to record ADC traces, a filter system to estimate the energy of the pulses, and a trigger based on the output of the energy filter. While the expected rate during neutrino-mass data-taking is expected to be less than 1~cps, the system must be able to run without deadtime at rates as high as 1~Mcps during calibration runs. To cope with changing rates, the experimenter can tune the amount of information recorded for each FPD event by choosing from three DAQ modes, summarized in Table~\ref{DAQ-modes}. Energy mode is the primary data-taking mode and records energy and timing for each event. Trace mode adds the 2048-point ADC waveform of the event. In histogram mode, designed for very high rates, the FLT hardware fills a 2048-bin energy histogram for each channel~\cite{kop08}. The user may specify a bin size and therefore an energy range, as well as the interval at which histograms are written to disk. Individual events are not recorded in this mode.

\begin{table}
\begin{center}
\begin{tabular}{lccc}
\hline
		DAQ mode	&  Trace & Energy & Histogram \\
\hline \hline
ADC trace 			&  yes			&		no 	&	no	\\
\hline
Event identifier				&  yes			&		yes	&	no 	\\
Time stamp			&  yes			&		yes	&	no	\\
Energy				& yes 			&		yes	&	no	\\
Channel map			&  yes			&		yes	&	yes	\\
\hline
Trigger rate			&  yes			&		yes	&	yes	\\
Energy histogram		&  no			&		no	&	yes	\\
\hline
\hline
Event size			& 4 kB			&	     12 B	&	$-$	\\
Histogram size 		& 		$-$	&		$-$	&	8 kB 	\\
\hline
Max. acq. rate  & & & \\
(deadtime-free) & 8 kcps & 108 kcps & 3.3 Mcps \\
\end{tabular}
\end{center}
\caption{\label{DAQ-modes} Data-record characteristics for each DAQ mode.}
\end{table}

The ADC ring buffer has 64 pages, each with 2048~samples, as well as four bit-status indicators. With every trigger, the recording stops, and subsequent event data are recorded in the next page. This paging allows a deadtime-free operation for rates up to those specified in Table~\ref{DAQ-modes}: during recording in one page, the DAQ computer reads out the results of all other pages. Note that high-rate data will suffer distortion due to pileup even if recording is deadtime-free (Sec.~\ref{sec:high_rates}).

\begin{figure}[tbp]
\begin{center}
\includegraphics[width=\columnwidth]{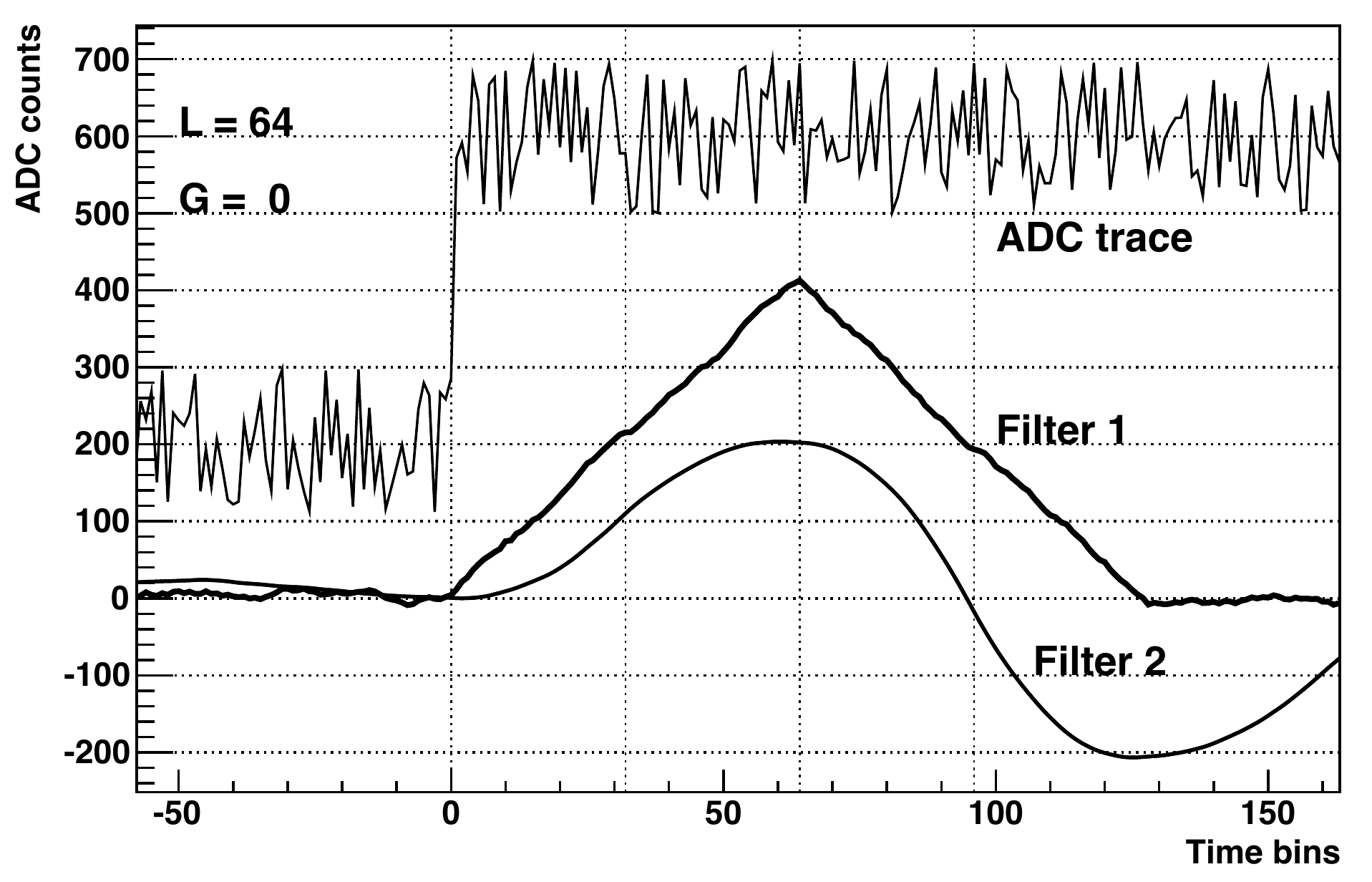}
\end{center}
\caption{\label{DAQ-filter_application} Application of the trapezoidal filter series (with $L=64$ and $G=0$) to a sample ADC trace with noise. The step in the trace at $t=0$ corresponds to the detection of an electron. The filter response is normalized to the ADC counts. Vertical lines indicate the trigger time of zero, and multiples of $L/2$.}
\end{figure}

\paragraph{Trigger} An electron incident on the detector causes a step response in the signal, with step height proportional to the electron energy. The typical rise time is on the order of 200~ns with a fall time on the order of 1~ms. A cascade of two trapezoidal filters~\cite{jor94} detects these steps in the digitized waveform. A trapezoidal filter is defined by its shaping length $L$ and its gap length $G$, both measured in ADC time bins. For each 50-ns time bin $i$, the filtered signal output $S_i$ is computed  from a moving sum of the previous $2L+G$ trace values ($v_i$):

\begin{equation}
  S_i = \sum_{j=0}^{L}v_{i-j} - \sum_{k=L+G}^{2L+G}v_{i-k}.
\end{equation}

\noindent The maximum of the output signal from the first filter occurs $L$ time bins after the step and is equal to the step height. The second filter, applied to the output of the first filter, has a shaping length set to $L/2$. A maximum in the first filter corresponds to a zero crossing in the second filter. The trigger is implemented based on the output of both filters: for each zero-crossing in the second filter, the amplitude from the first filter is compared to a programmable energy threshold. A large value of $L$ optimizes energy resolution while a short $L$ optimizes timing resolution. Figure~\ref{DAQ-filter_application} shows the application of the cascade to a sample signal.

MPPCs in the veto system (Sec.~\ref{sec:apparatus:veto}) generate short pulses rather than step waveforms. Since the pulse length is comparable to the Nyquist limit for the system, the first filter in the cascade was replaced with an adjustable ($2-4$-sample) boxcar filter. This allows the total trace to be captured, eliminating signal-clipping effects induced by aliasing and providing a more accurate energy reading. The second stage, a trapezoidal filter, determines the timestamp. 

In veto mode, each FPGA handles all the fiber-end signals associated with a single scintillator panel. To identify muon events, an FPGA searches for coincidences between a user-specified number of fiber-end signals (typically two), and triggers a readout when such a coincidence occurs at the same time that the sum of all signals in that panel exceeds a threshold. Summing is performed on dedicated analog summing boards, but the FPGAs calculate all coincidences, so the coincidence interval may be set by the experimenter. 

The FPGAs on the FLTs can accommodate the logic for all the data-taking modes of the FPD and veto, so changing modes does not require reprogramming the card. A change to the programmable logic triggers an automated test of the system, including a comparison of the on-board trapezoidal-filter cascade to an offline software version.

\subsubsection{DAQ Software}
\label{daq_sw}

\begin{figure}
\begin{center}
\includegraphics[width=\columnwidth]{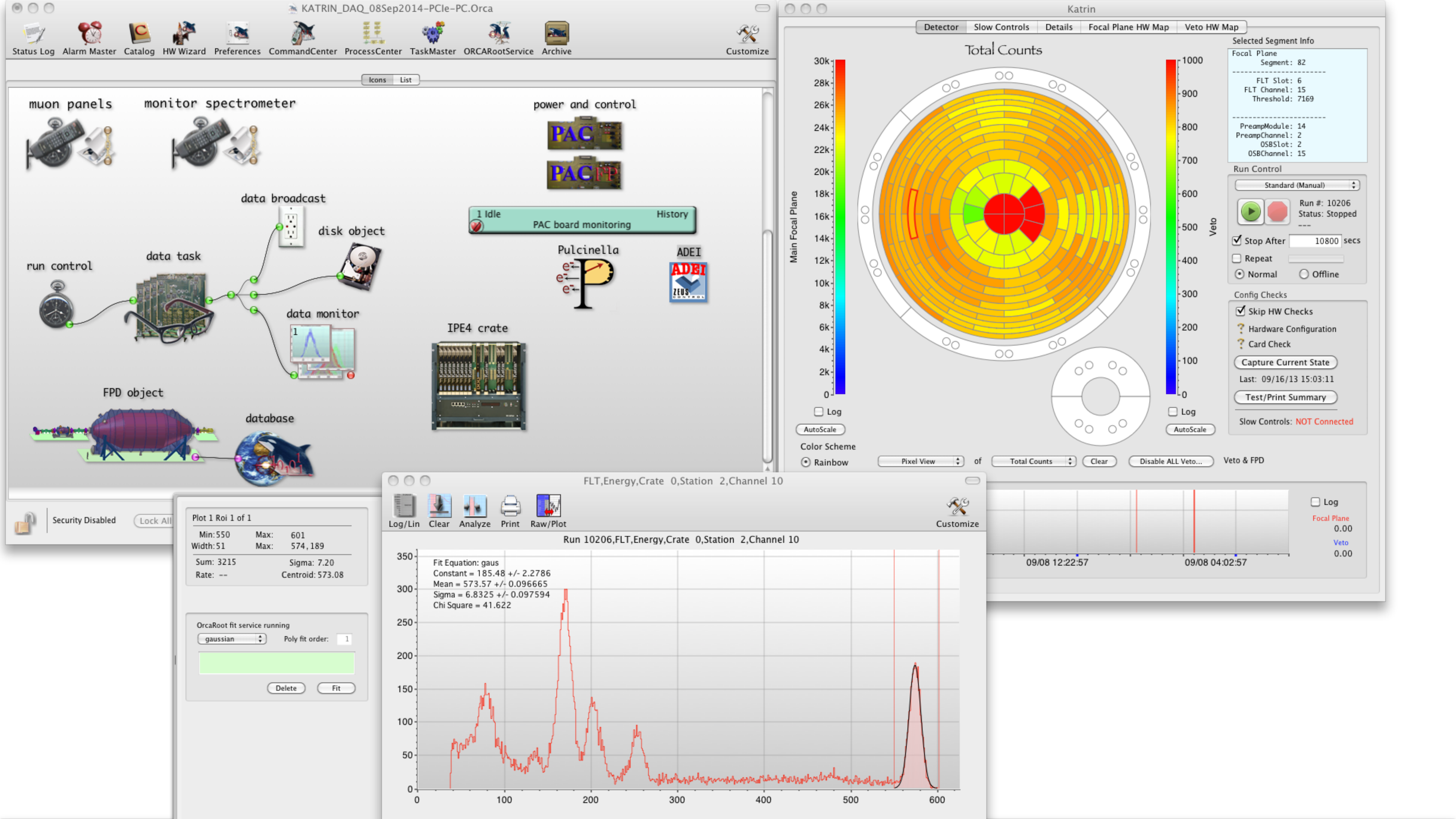}
\end{center}
\caption{\label{ORCA_screenshot} Sample screenshot of ORCA software during a calibration run with a $^{241}$Am source and the magnets energized. Top left: the graphical DAQ configuration constructed by the experimenter with a drag-and-drop interface. Top right: FPD status summary with pixel-by-pixel accumulated count display. Bottom: real-time ADC histogram for a sample channel, showing ability to apply peak fitting in real time.}
\end{figure}

The DAQ software is based on a package called ORCA (Object-oriented Real-time Control and Acquisition)~\cite{how04} that was developed at the University of Washington and the University of North Carolina at Chapel Hill.  It was written in Objective-C for the Mac\-OS X operating system and uses object-oriented programming techniques to encapsulate hardware elements and DAQ concepts into software objects. A graphical user interface (GUI) allows the user to drag icons from an object catalog into a configuration window, where they are grouped or connected together to represent the hardware setup. All configuration can be done at run-time with no compilation needed.  ORCA is fully scriptable and can operate continuously. A separate analysis framework, OrcaROOT, provides decoders that use the ROOT software packages~\cite{Brun:ROOT97} to analyze data generated through ORCA.

ORCA can control low-level detector tests, such as leakage-current and temperature measurements, and read out the resulting data. It also provides access to all hardware and mode parameters of the FLT and SLT cards and directs all hardware initialization of the cards. ORCA supplies the SLT-embedded SBC with code to handle data readout. The resulting data are consolidated into large packets and transferred to ORCA for monitoring and storage~\cite{how08}. An interface to the slow-controls system (Sec.~\ref{sec:apparatus:slowcontrols}) allows access to all hardware readouts, as well as direct hardware control. A special operator dialog provides a single-window status summary, including overall signal rates and a pixel-by-pixel view of the FPD. The experimenter can view ADC histograms, filled in real time, for individual channels or groups of channels during a run. Figure~\ref{ORCA_screenshot} is a screenshot displaying several of these capabilities.
  
The experimenter can control runs manually, or set them to stop or repeat automatically after a specified duration.  Any number of runs or subruns of varying length, with varying hardware settings, can be implemented in scripts. 

ORCA also optionally stores status information in a remotely accessible SQL database. An experimenter can then monitor data rates, energy histograms, alarms, and the run state over the Internet (via ORCAWeb) or from iPads or iPhones (via iORCA).

\subsection{Slow Controls}
\label{sec:apparatus:slowcontrols}

\begin{figure}
\begin{center}
\includegraphics[width=\columnwidth]{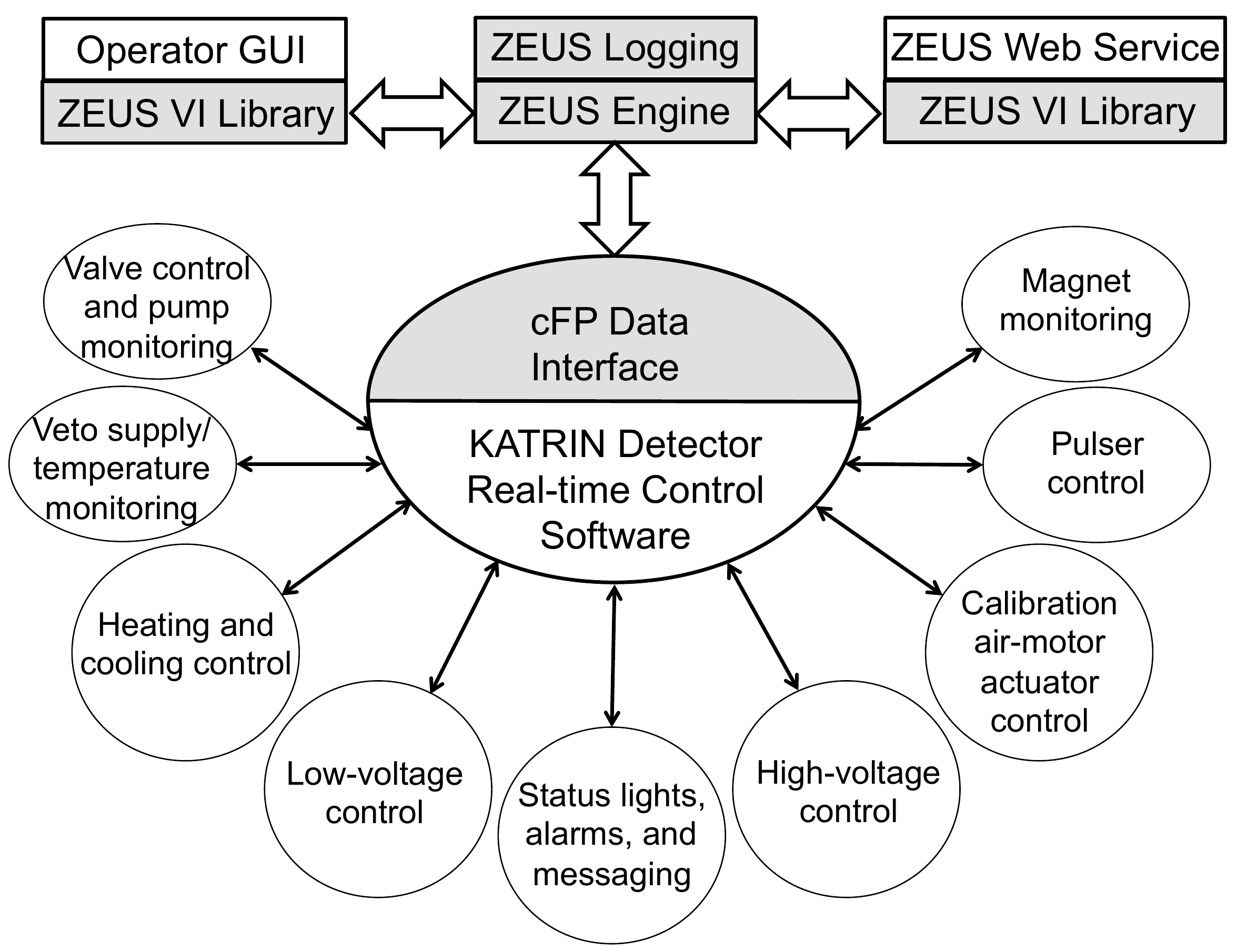}
\end{center}
\caption{\label{SlowControls}Schematic layout of the slow-controls system. The KATRIN-specific implementations are shown in white, and general ZEUS components in gray.}
\end{figure}

The slow-controls system provides a software interface for FPD-system hardware elements outside the FPD and veto signal chains. Devices communicate with the system via two compact Field Point (cFP) devices from National Instruments. The ZEntrale datenerfassung Und Steuerung (Central DAQ and Control System, or ZEUS)~\cite{LefhalmZEUS05} software, developed at the Karlsruhe Institute of Technology, provides overall management. For each cFP, three ZEUS components run on a host computer: the ZEUS engine, which communicates with the cFP to obtain sensor data and change set points; the ZEUS logger, which records data and set points; and the ZEUS VI library, which extends the system's features into LabVIEW~\cite{LabviewPage} applications and a GUI. Figure~\ref{SlowControls} summarizes this architecture schematically. A GUI for each subsystem allows the operator to make quick status checks and setting changes. Through ZEUS web services, the ORCA DAQ software (Sec.~\ref{daq_sw}) may also issue device instructions.

Each cFP runs VxWorks and incorporates a network controller, an eight-slot backplane, and several input/output modules. In order to isolate safety-related functions from the computer network, the local cFP controller program monitors interlocks and generates alarms when interlocks are activated or hardware readbacks go out of tolerance. When an alarm is triggered, the system automatically sends text messages or emails to subsystem experts and takes pre-programmed safety measures, e.g. ramping down all high voltages.

As Fig.~\ref{SlowControls} shows, this allows control of a wide range of subsystems, from the pumps and valves of the vacuum system and pumping stations to the positions of the calibration sources. Cryo\-pump regeneration and vacuum-system bakeout are achieved by temperature-control channels based on PID loops. Low-voltage power supplies for the front-end readout and ambient-air electronics, and high-voltage power supplies for the photoelectron source and the post-acceleration electrode, may be ramped at adjustable rates with adjustable over-current protection. 

\subsection{Data Management}
\label{sec:apparatus:datamanagement}

The Advanced Data Extraction Infrastructure (ADEI)~\cite{Chilingaryan10} is a highly modular data-management tool that has been extended to manage FPD-system and other KATRIN data. ADEI integrates heterogenous subsystems and data sources to archive all relevant data centrally. It provides worldwide, interactive access to the data, with navigation methods optimized for a fast response time. ADEI also includes a layer of abstraction so as to ensure a stable application programming interface over the course of the experiment and its analysis, regardless of changes to the data format, to the number of data streams, or to the underlying database engine.

\begin{figure}
\begin{center}
\includegraphics[width=\columnwidth]{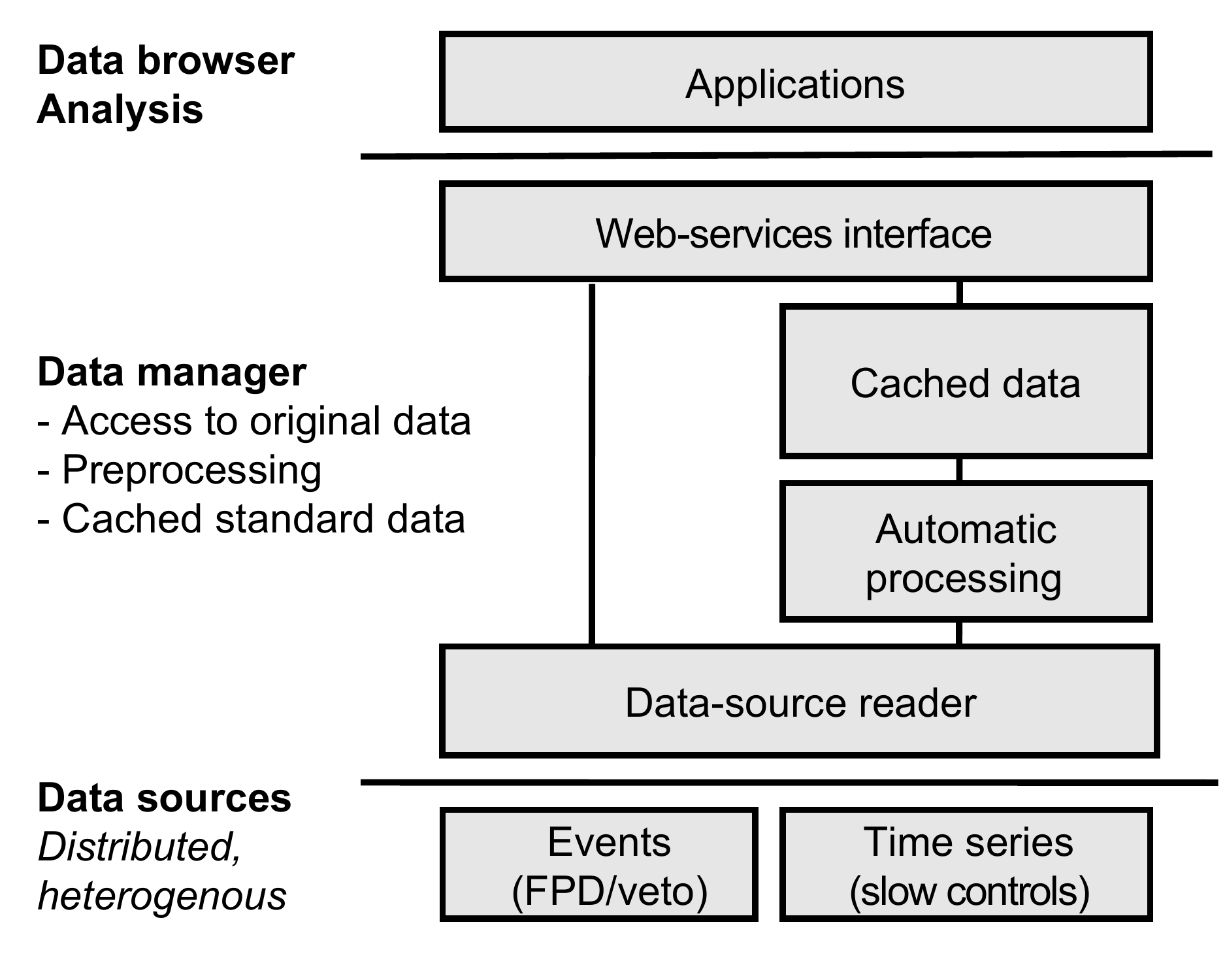}
\end{center}
\caption{\label{ADEI-principles} Data-management architecture for KATRIN. ADEI provides an interface between applications, such as analysis tools and data browsers, and data sources, such as event data from the detectors and time-series data from slow controls. As requested by the application, ADEI provides access either to the raw data or to cached data after aggregation or pre-processing.}
\end{figure}

The FPD system collects data of two types: time-series data from slow-controls readouts and event-based detector data from the FPD and veto. Each subsystem records data in a local database; the data are then copied to the main KATRIN database and permanently archived. The ADEI library defines an abstract reader interface that specifies formats for data requests and retrieval. This interface provides raw data either directly to the requesting application or to the ADEI caching daemon, running on a backend server. The daemon pipes the extracted data through a series of filters that may check data quality, perform low-level analysis, or drop invalid data. It continuously aggregates time-series data over several predefined intervals (minutes, hours, weeks, etc.), calculating and storing statistical properties. The system can be configured so that the creation of a new detector run file automatically starts the default KATRIN detector analysis chain, including information retrieval from the slow-controls system and data conversion to ROOT format. The internal caching database stores characteristic values for each run.

The standard ADEI web-service interface disseminates data upon request while encapsulating the details of data formats, thus assuring platform-independent data retrieval for the lifetime of the experiment. ADEI web services also provide communication between ORCA and the slow-controls system. The KATRIN analysis framework connects to ADEI via KaLi~\cite{kleesiek:2014}, a library of C++ functions developed by the KATRIN collaboration.  

\begin{figure}
\begin{center}
\includegraphics[width=\columnwidth]{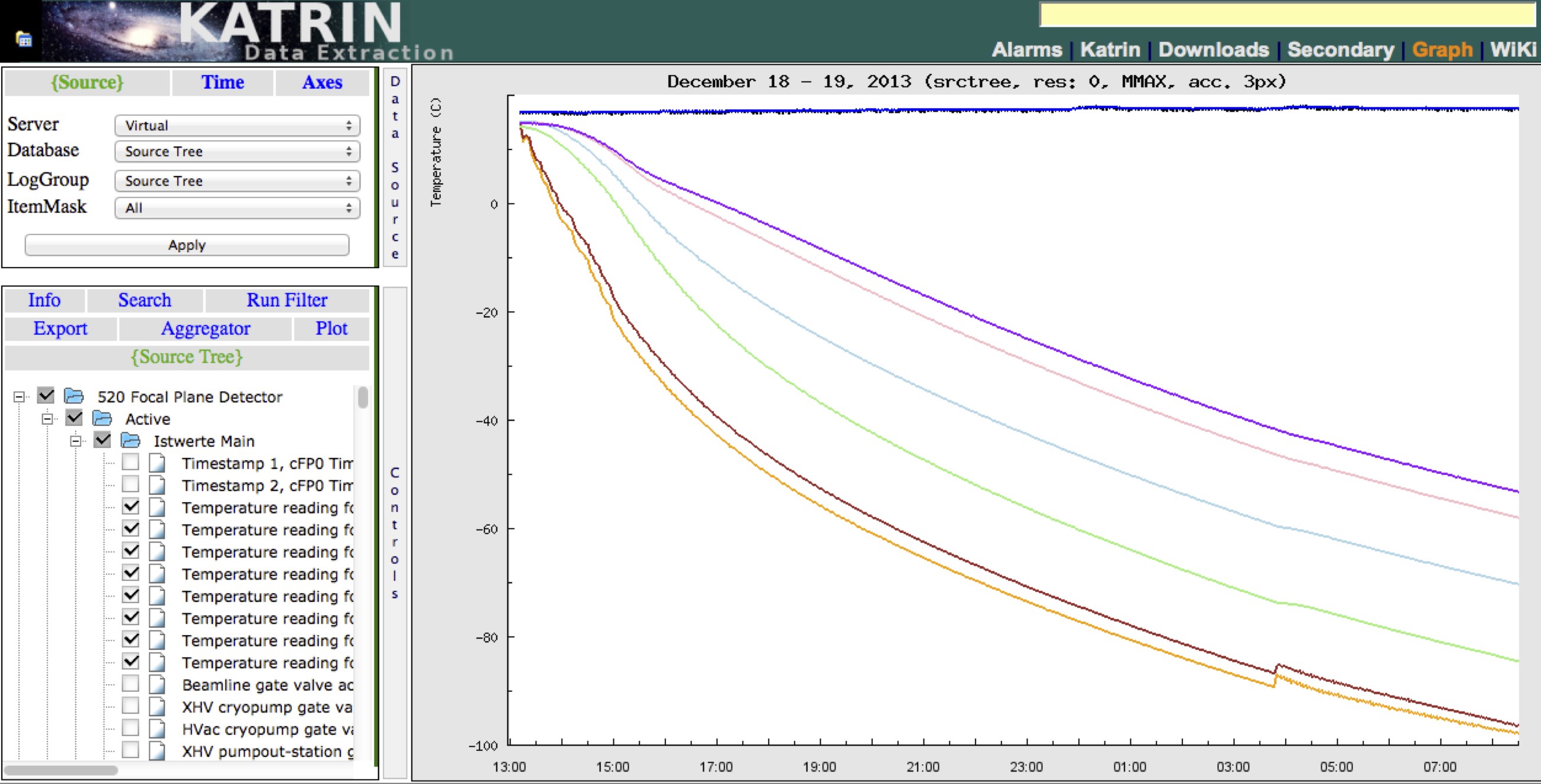}
\end{center}
\caption{\label{ADEI-screenshot} Sample screenshot of the KATRIN data portal, showing the time evolution of  temperature readouts from several parts of the system during cooldown with the thermosiphon (Sec.~\ref{sec:apparatus:thermosiphon}).}
\end{figure}

The most prominent implementation of the ADEI web services is the interactive KATRIN data portal, which provides graphical access to data from slow controls, the FPD, and the veto. Figure~\ref{ADEI-screenshot} shows a screenshot of this portal and its hierarchical list of the available data sources; the user may also select a time interval and value range.  To improve performance, we use intelligent aggregation techniques to distill a few thousand data points from the millions that may span the time interval; caching techniques to speed data access; and asynchronous communication between server and user. The complete plot-generation time does not usually exceed 500~ms. Every generated view of the data portal can be retrieved with a unique, persistent reference that can be included in documentation. An adaptable ``front page'' gives an overview of the current hardware configuration and status, using markup extensions to common wiki functions. 

\begin{figure}
\begin{center}
\includegraphics[width=\columnwidth]{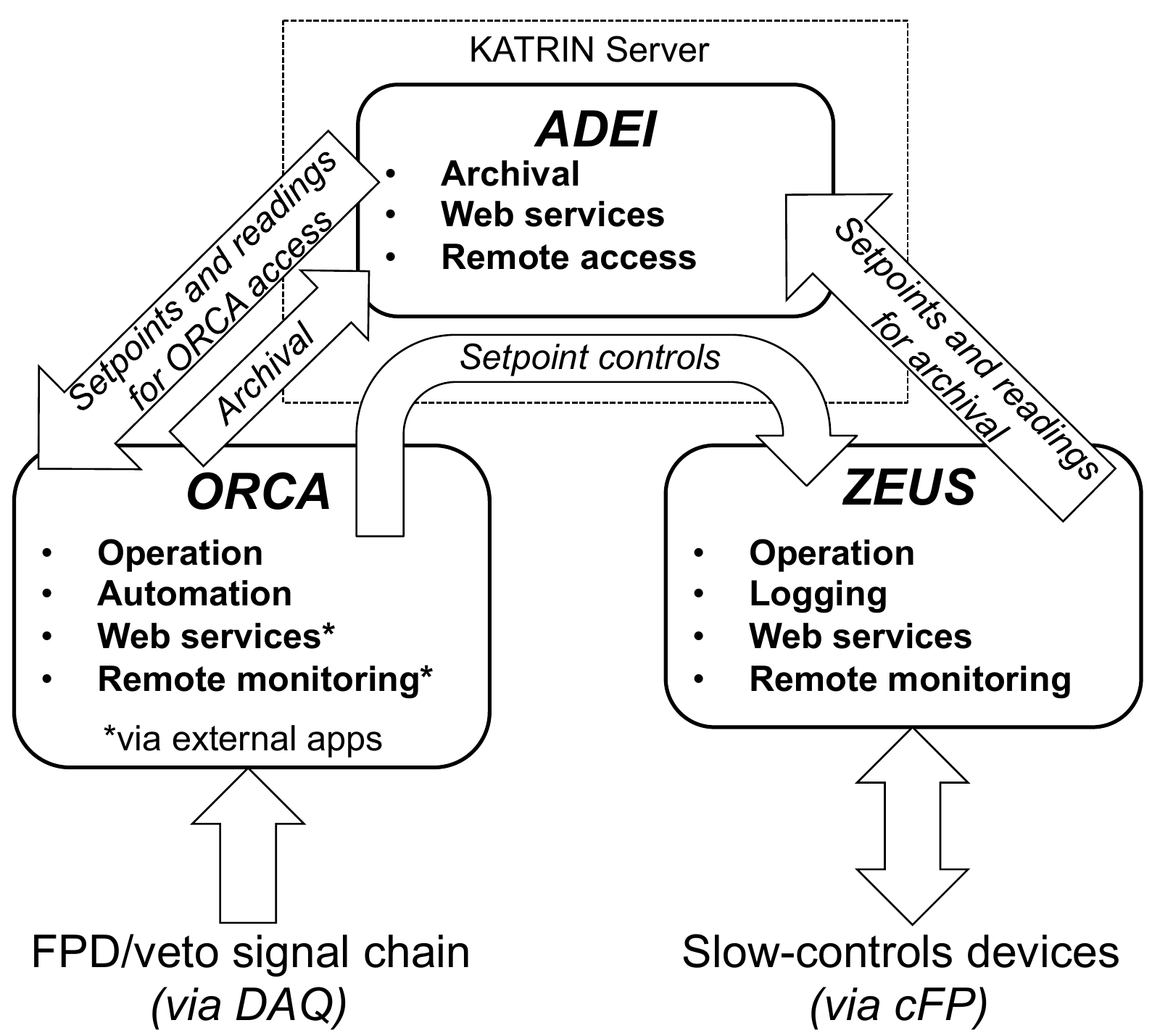}
\end{center}
\caption{\label{fig:OverallSoftwareArchitecture} Relationships between the three primary software systems for the KATRIN FPD system: ORCA for the DAQ (Section~\ref{daq_sw}), ZEUS for slow controls (Section~\ref{sec:apparatus:slowcontrols}), and ADEI for data management.}
\end{figure}

Figure~\ref{fig:OverallSoftwareArchitecture} shows the interplay between the software packages for DAQ, slow controls, and data management.

\section{Detector Performance}
\label{sec:performance}

In this section, we discuss the performance of the FPD system. Data presented were taken in Karlsruhe, unless otherwise specified. We begin with determinations of the detection efficiency (Sec.~\ref{sec:performance:efficiency}) and energy resolution (Sec.~\ref{sec:performance:resolution}). Section~\ref{sec:performance:linearity} treats the linearity of the system. Timing resolution is discussed in Sec.~\ref{sec:performance:timingresolution}, the time intervals between events are analyzed in Sec.~\ref{sec:time_interval}, and the system performance at high rates is treated in Sec.~\ref{sec:high_rates}. Section~\ref{sec:energy_stability} discusses the stability of the FPD energy calibration.

\subsection{Detection Efficiency}
\label{sec:performance:efficiency}

During commissioning in Karlsruhe, two adjacent pixels had to be removed from the data stream due to noise. Microscopic inspection of the detector wafer revealed that these pixels were shorted together. The replacement wafer from the same production batch was found to have the same problem in a different location. The cause has been traced to a production problem that is being addressed by the manufacturer. The remaining 146 pixels were fully functional.

A measurement of the detection efficiency requires photocurrent data from PULCINELLA (Sec.~\ref{sec:apparatus:pulcinella}). While current measurements accurate to 1~fA can be obtained in less than a second by the meter board itself, use at high voltage with the FPD vacuum system is more difficult. The photoelectron-source probe has a capacitance of 17~pF to ground and the meter picks up the charging current as the probe vibrates and its capacitance changes, introducing a noise proportional to the photoelectron-source bias. With modulation by a square wave with a 60-s period, and the introduction of a status signal from the UV LED drive pulser, noise was reduced from 5~pA/$\sqrt{\mathrm{Hz}}$ to 400~fA/$\sqrt{\mathrm{Hz}}$ with the photoelectron source and the post-acceleration electrode carrying potentials of 7.6~kV and 11.0~kV respectively. In order to ensure all electrons from the front of the photoelectron-source disk struck the FPD wafer, the detector magnetic field was raised to 4.68~T with the pinch magnet at 6~T. Tests with different field configurations indicate that $3.8 \pm 0.4\%$ of the photoelectron flux is from the back of the disk, and is therefore not detected in the efficiency measurement; the flux to the detector must be corrected accordingly. 

With this implementation, we measured the electron-source current and FPD hit rate for forty hours in order to extract the detector efficiency. We corrected for noisy pixels by estimating their expected rates from the rates of neighboring good pixels, resulting in a raw detection efficiency of 92.9\% with all pixels working. After all corrections, including a reduction of the measured efficiency by 1.5\% to account for crosstalk, we obtained a measured detection efficiency of $95.0\% \pm 1.8\%_{\mathrm{stat}} \pm 2.2\%_{\mathrm{syst}}$. The efficiency is less than 100\% mainly because of events falling below threshold due to dead-layer losses.

\subsection{Energy Resolution}
\label{sec:performance:resolution}
 
\begin{figure}[tbp]
\begin{center}
\includegraphics[width=\columnwidth]{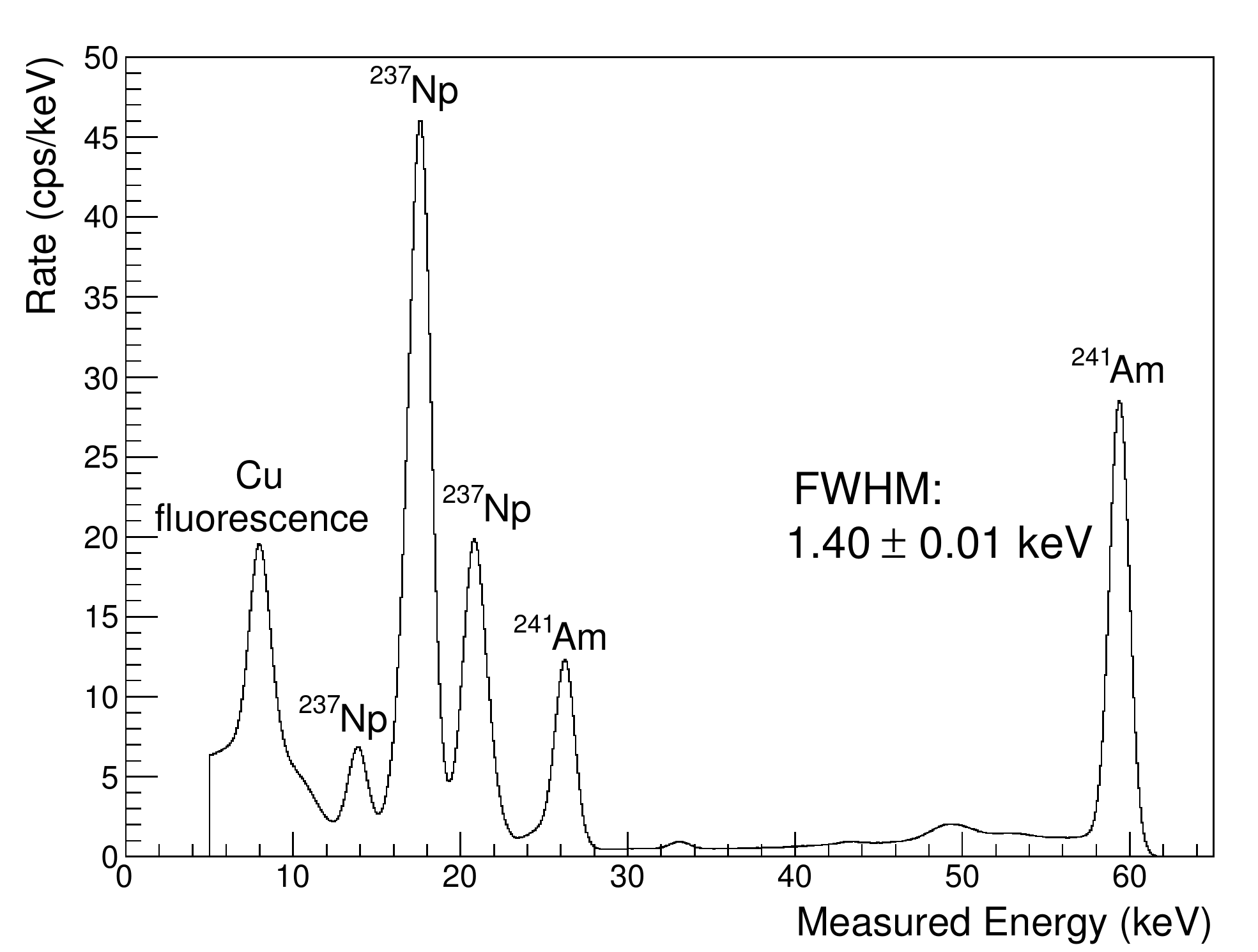}
\end{center}
\caption{\label{241Am}Global $^{241}$Am spectrum for 146 detector channels. No magnetic field was applied. Individual $^{237}$Np lines, within the three visible peaks, cannot be resolved. Below the 5-keV threshold, the spectrum is dominated by noise.}
\end{figure}

 \begin{figure}[tbp]
\begin{center}
\includegraphics[width=\columnwidth]{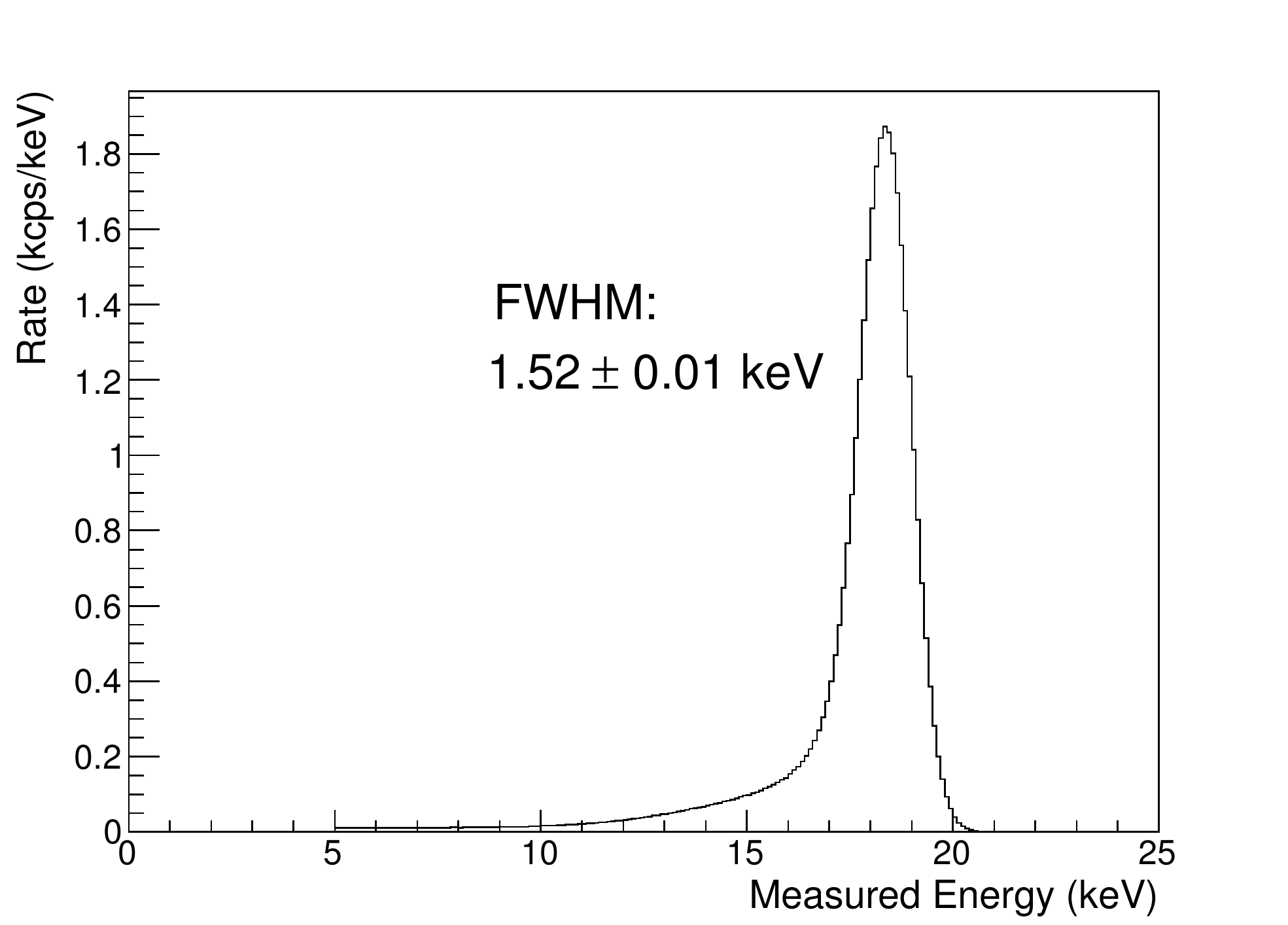}
\end{center}
\caption{\label{186_spectrum} Global spectrum for 18.6-keV electrons from the photoelectron source, for 146 detector channels. The pinch magnet was set to 5~T and the detector magnet to 2~T.}
\end{figure}

 \begin{figure}[tbp]
\begin{center}
\includegraphics[width=\columnwidth]{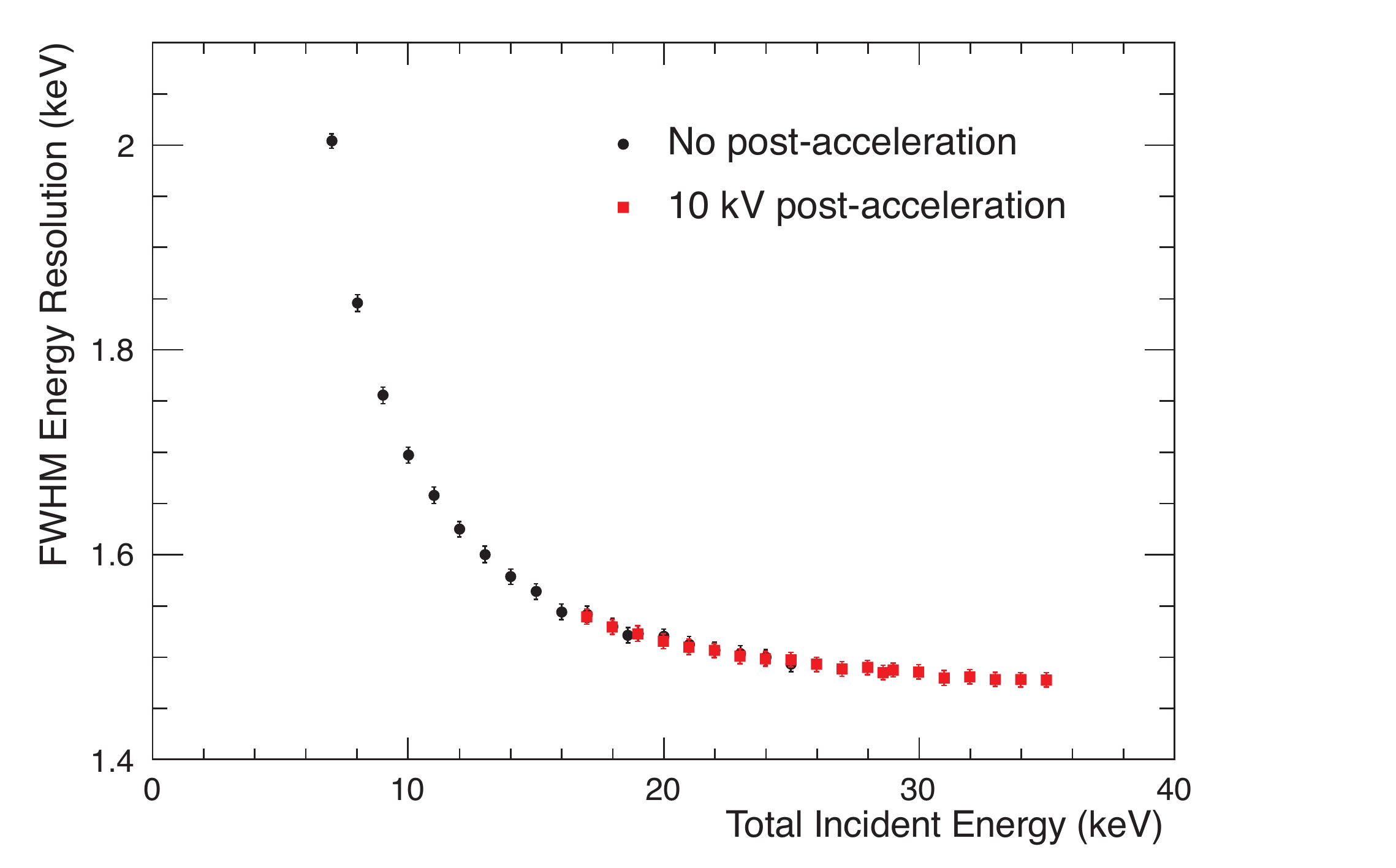}
\end{center}
\caption{\label{e_spectra} Mean FWHM energy resolution over 146~channels as a function of electron energy at two post-acceleration voltage settings. The pinch magnet was set to 5~T and the detector magnet to 2~T.}
\end{figure}

We combined data from 146 working channels, taken with a shaping length $L=6.4$~$\upmu$s, to obtain global spectra from both calibration sources (Sec.~\ref{sec:apparatus:eandgammasource}). In each case the energy resolution was determined as the FWHM of a Gaussian distribution fit to the portion of the spectrum within 50\% of the peak amplitude, where Gaussian noise sources dominate the energy smearing. The results of these simple fits are consistent with more sophisticated approaches that fit the low-energy tail as well as the peak. Figure~\ref{241Am} shows the $^{241}$Am spectrum, with a resolution of $1.40 \pm 0.01$~keV (FWHM) for the 59.5-keV peak. Measurements with the electron source were taken with the pinch magnet at 5~T and the detector magnet at 2~T, so that the source illuminated all pixels of the detector. Figure~\ref{186_spectrum} shows the resulting spectrum for 18.6-keV photoelectrons. The reported global energy resolution of $1.52 \pm 0.01$~keV (FWHM) is the mean of the energy resolutions measured for each channel. For both sources, the RMS deviation of the channel-by-channel energy-resolution distribution is less than 0.1~keV.

Figure~\ref{e_spectra} shows that the global energy resolution for electrons worsens at lower energies, as expected due to dead-layer effects.

\subsection{Linearity}
\label{sec:performance:linearity}

\begin{figure}[tbp]
\begin{center}
\includegraphics[width=\columnwidth]{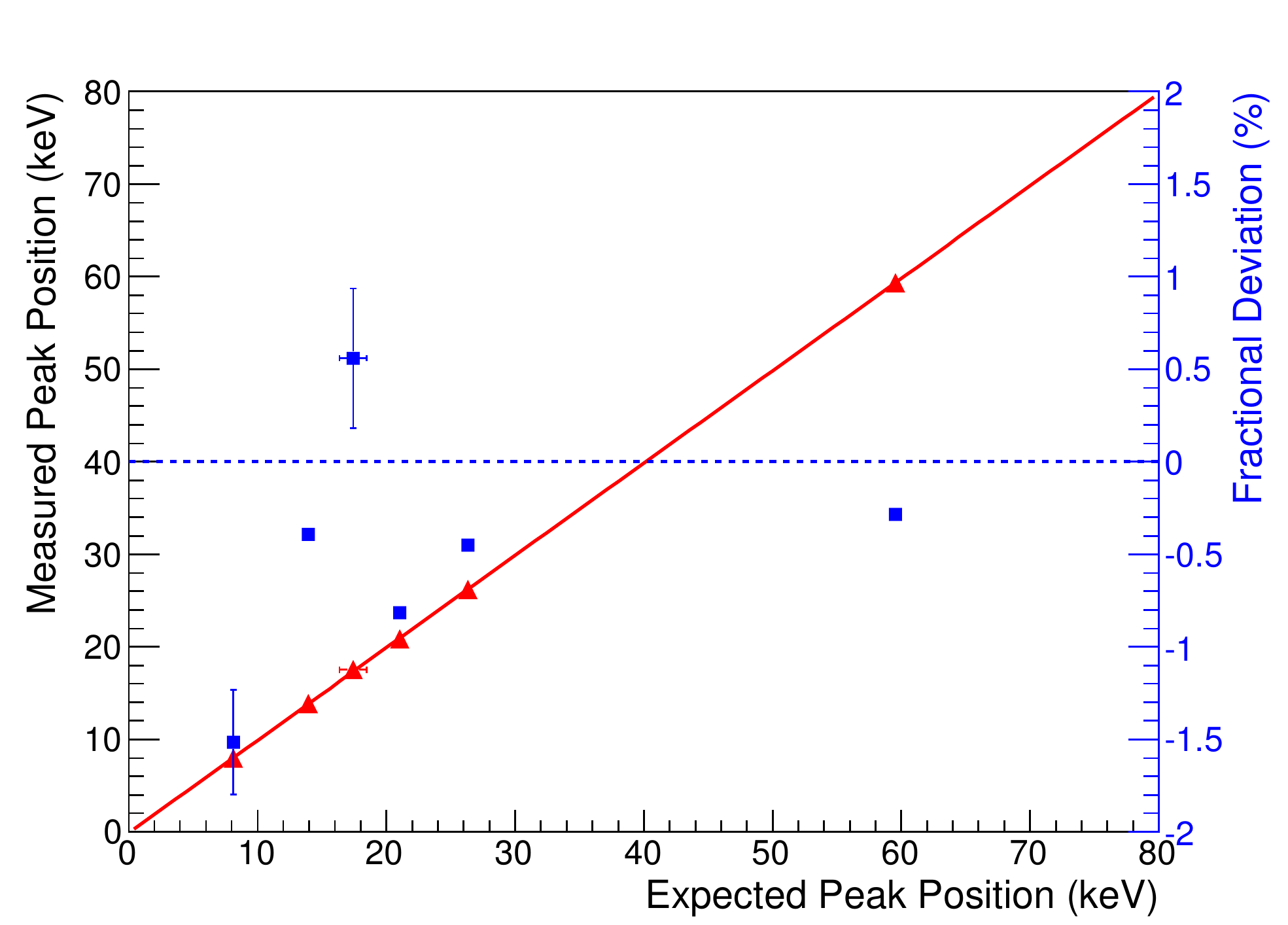}
\end{center}
\caption{\label{fig:Am241_linearity} The triangles (left axis) show the observed X-ray and fluorescence peak positions in the $^{241}$Am spectrum, measured over 146 detector channels, plotted against the expected energies and fit to a line. The squares (right axis) show the fractional deviation of each measured point from the expected value~\cite{schwarz:2014}.}
\end{figure}

Since signals must be transmitted across the post-acceleration potential via fiber optics, electronic pulsers are not suitable for studying the energy response of the system: the optical transmitter distorts the pulser signal.   Instead, we must use one of two alternate approaches. The first method compares the six visible peaks of the $^{241}$Am spectrum (Fig.~\ref{241Am}) to the known X-ray and fluorescence lines between 5~and 60~keV. We calculated the expected location of each peak based on line energies obtained from reference tables for radioactivity~\cite{TORI} and fluorescence~\cite{xraydatabooklet:2009}, applying known attenuation coefficients for silicon~\cite{xcom:2010} to compute the expected deposited energy. Where multiple unresolved lines contribute to a measured peak, the known line intensities must also be considered in calculating the expected peak location, and a significant uncertainty arises. The measured energy was determined from a calibration assuming a straight line between 0~keV and the 26-keV and 59-keV $^{241}$Am lines. A comparison of the expected and measured energy values, shown in Fig.~\ref{fig:Am241_linearity}, tested the linearity of response within the $^{241}$Am calibration range. The weighted linear fit found a root-mean-squared non-linearity below 0.2\%. The maximum energy offset seen by any channel was 180~eV.

\begin{figure}[tbp]
\begin{center}
\includegraphics[width=\columnwidth]{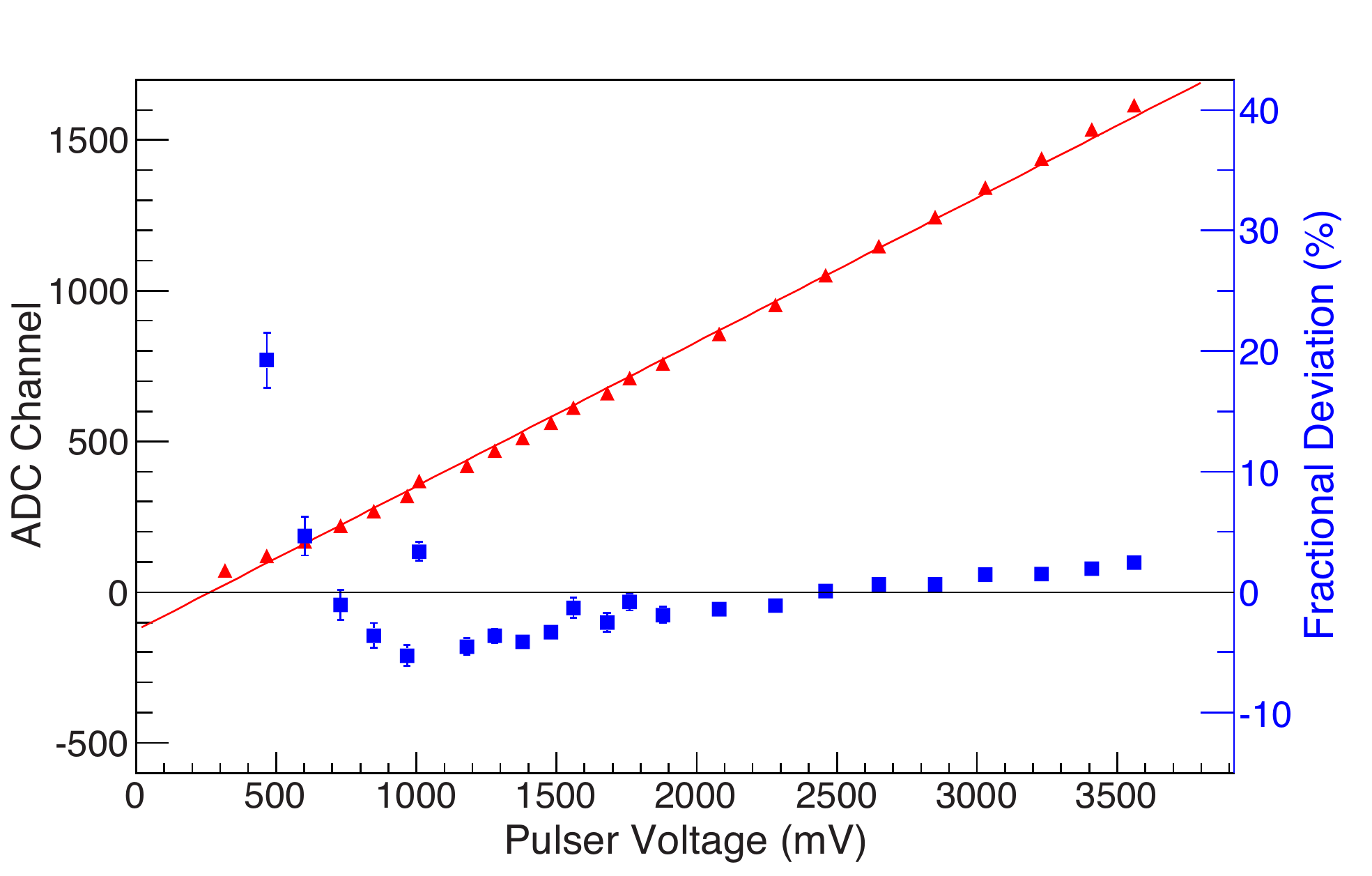}
\end{center}
\caption{\label{fig:LEDlinearity_ch44} Measured ADC values (triangles, left axis) for red-LED pulses, plotted as a function of pulse amplitude for a representative FPD channel. The highest ADC value corresponds to a reconstructed energy of about 160~keV. The squares show the fractional deviation (right axis) from the linear fit.}
\end{figure}

The second approach uses direct illumination from a pulsed red LED, applied through the same window used for the UV illumination of the photoelectron source. During each pulse, red light reflects from interior surfaces, flooding FPD pixels with energies up to the equivalent of about 200~keV. The driving pulse, supplied by an Agilent 33220A pulser, has a fixed width of 5~$\upmu$s, but the amplitude may be varied from run to run, changing the current through the LED and thus the light intensity. The long rise time ($\sim 10$~$\upmu$s) of the observed FPD pulse, as well as the instability of the pulse shape, precludes the use of these data for linearity measurements. However, the data do permit us to conclude that the measured detector energy increases monotonically with the injected charge: as Fig.~\ref{fig:LEDlinearity_ch44} shows for a typical FPD channel, we observe no saturation in the detector or in the front-end electronics out to reconstructed energies in excess of 160~keV. The saturation limit of the 12-bit ADCs (Sec.~\ref{daq_hw}) may be moved arbitrarily by modifying the gain settings.

\subsection{Timing Resolution}
\label{sec:performance:timingresolution}

The DAQ system determines the timing of an input signal from the zero-crossing of the second trapezoidal-filter output (Sec.~\ref{daq_hw}). For an ideal step-shape input, this time has a constant latency of $3L/2+G/2$. For real signals, this relation is not exactly true, but the latencies should be independent of signal amplitude since the trapezoidal filter logic is linear. The charge-collection time of the p-i-n diode and response time of the FET are relatively fast compared to the 50-ns sampling interval of the DAQ system, so the overall timing resolution is expected to be dominated by the sampling interval and by the influence of noise on the trapezoidal filter stages. In contrast to the energy resolution, a longer shaping length results in a worse timing resolution due to the inclusion of sample points further from the signal step. Larger input signals give better timing resolution due to their higher signal-to-noise ratio, so applying a post-acceleration potential (Sec.~\ref{sec:apparatus:vacuum}) improves timing resolution without sacrificing energy resolution.

\begin{figure}[tbp]
\begin{center}
\includegraphics[width=\columnwidth]{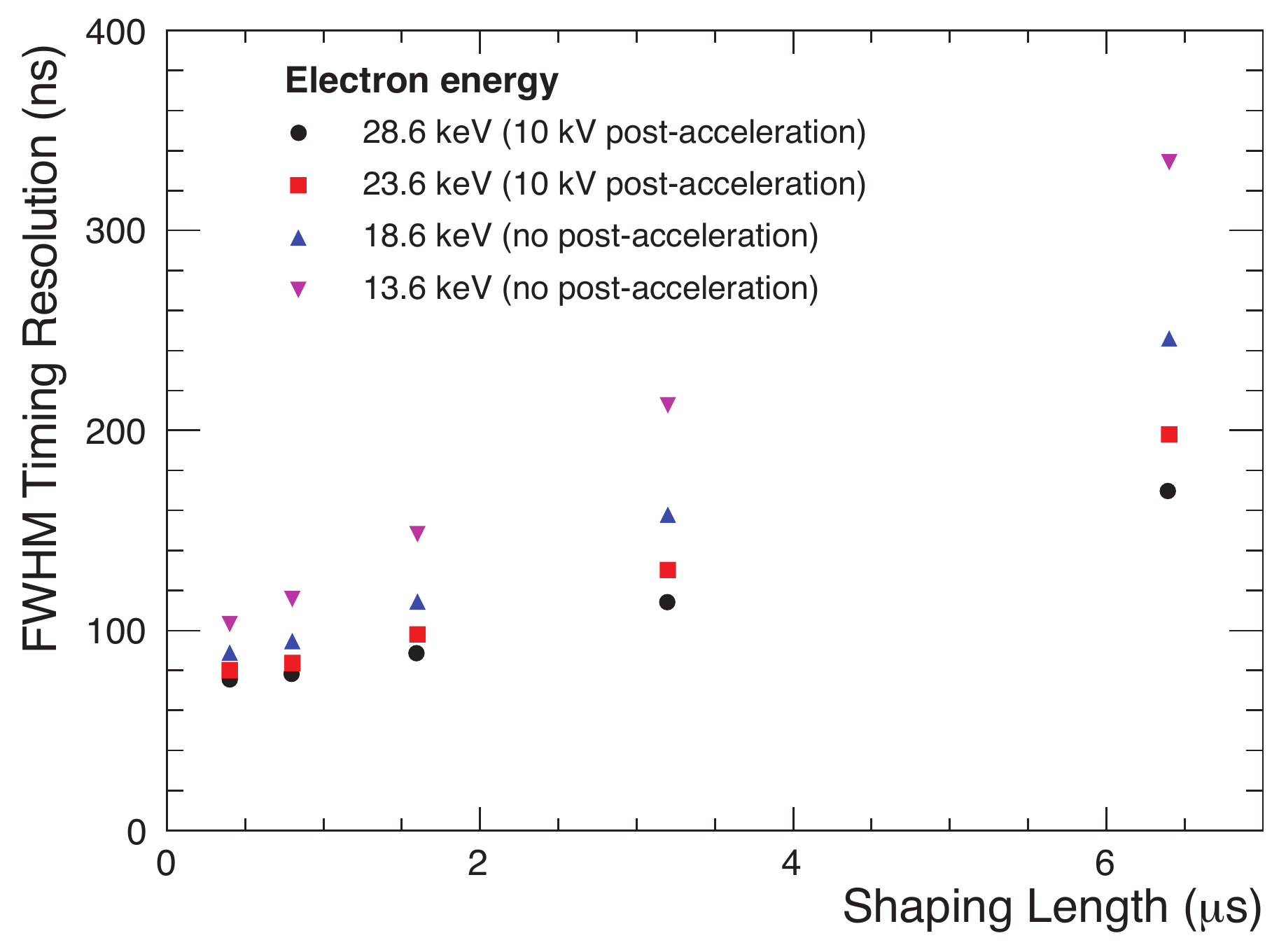}
\end{center}
\caption{\label{fig:TimingResolution} Mean channel-by-channel FWHM timing resolution for various electron energies and shaping lengths. The pinch magnet was set to 5~T and the detector magnet to 2~T. Statistical error bars are too small to be visible.}
\end{figure}

We determined a bound for the FPD-system timing resolution by applying 20-ns pulses, with rise time 5~ns, to the photoelectron-source UV LED (Sec.~\ref{sec:apparatus:eandgammasource}) and measuring the time jitter between the pulse timing and the detector-signal timing. This method gives an upper limit because the measured jitter is also affected by the jitter and duration of the source pulse.

Figure~\ref{fig:TimingResolution} shows the measured timing jitters for electron energies from 13.6 to 28.6~keV and filter shaping lengths from 0.4 to 6.4~$\upmu$s. Each reported timing resolution is the mean of the measurements for each of the 146 working channels. The FWHM timing resolution for 18.6-keV electrons at the 6.4-$\upmu$s shaping length is 246~ns, which meets specifications for normal KATRIN operation. With a 0.8-$\upmu$s shaping length, we achieved an FWHM timing resolution of 95~ns at the same electron energy, meeting the 100-ns requirement for KATRIN time-of-flight measurements. As expected, the timing resolution is improved by applying post-acceleration, even with the shaping length held constant. 

\subsection{Analysis of Time Intervals Between Events}
\label{sec:time_interval}

The distribution of time intervals between detected events provides rich information for detector diagnosis: it is sensitive to deadtime, cross-talk, external noise events, false triggers from ringing and oscillations, timestamp errors and data corruption. This distribution is also sensitive to a variety of physical processes beyond the detection of individual electrons, including correlated events caused by trapped particles, cosmogenic and radiogenic multi-particle events, multi-hit events from back-scattered electrons, and multi-pixel events from charge splitting. After accounting for these possibilities, the interval distribution must be consistent with the total average count rate, providing another comprehensive consistency check.

With the Seattle prototype, we investigated event-interval distributions both for low-rate background runs and for high-rate calibration runs, using interval-distribution histograms for time scales ranging from 10~$\upmu$s to 1000~ms. Based on these distributions, we immediately identified eight channels with cross talk, two channels with oscillations, one malfunctioning channel and one subtle logic problem in FPGA timestamping. After excluding the bad channels, we found an additional channel with abnormally high hit-rate fluctuations over long time scales, and removed it from this interval analysis as well.

\begin{figure}[tbp]
\begin{center}
\includegraphics[width=\columnwidth]{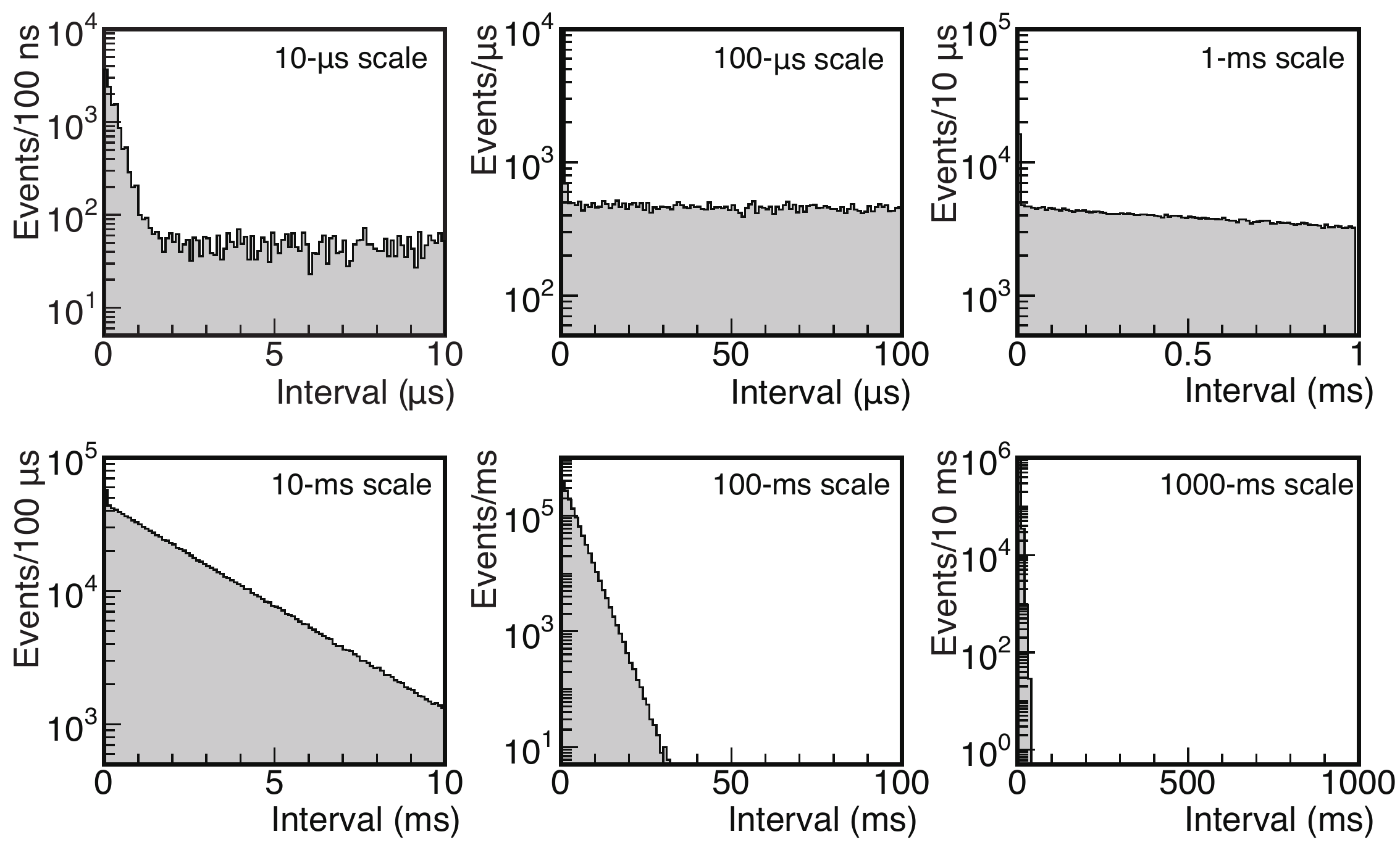}
\end{center}
\caption{\label{fig:EventInterval} Measured distribution of time intervals between events during a photoelectron-source calibration run in Seattle. The interval time constants are $354.9 \pm 6.6$~s$^{-1}$ (1-ms scale), $357.6 \pm 0.5$~s$^{-1}$  (10-ms scale) and $356.5 \pm 1.6$~s$^{-1}$ (100-ms scale), all of which are consistent with the average count rate of this run, $358.1 \pm 0.3$~cps.}
\end{figure}

Figure~\ref{fig:EventInterval} shows the time-interval distributions of a Seattle calibration run after the exclusion of 24 bad channels. We interpret the peak centered at an interval of zero as multi-pixel events; the width of the peak is consistent with the timing resolution. For all time scales greater than 10~$\upmu$s, the distributions are exponential with time constants that are consistent with the average count rate, after applying a correction for the multi-pixel events.

After subtracting accidentals, the multi-pixel events result either from multi-particle events or from charge splitting across several pixels. For photo\-electron-source measurements, charge-splitting events can be identified by two criteria: all involved pixels are adjacent, and the summed charge is the same as that of the single-pixel events. For 18-keV electrons, we used this method to estimate the probability of charge splitting as approximately 1\%, with a higher probability for outer pixels. If this probability is translated to an isotropic charge-dispersion cloud size in the p-i-n-diode array, it corresponds to a scale of several tens of $\upmu$m. This scale is implausible for our detector, suggesting that external processes contribute to this probability. The rate of the remaining multi-pixel events is on the order of 1~mcps, which is consistent with our estimation of multi-particle background events (Sec.~\ref{sec:sim_bg}).

\subsection{Performance at High Rates}
\label{sec:high_rates}

\begin{figure}[tbp]
\begin{center}
\includegraphics[width=\columnwidth]{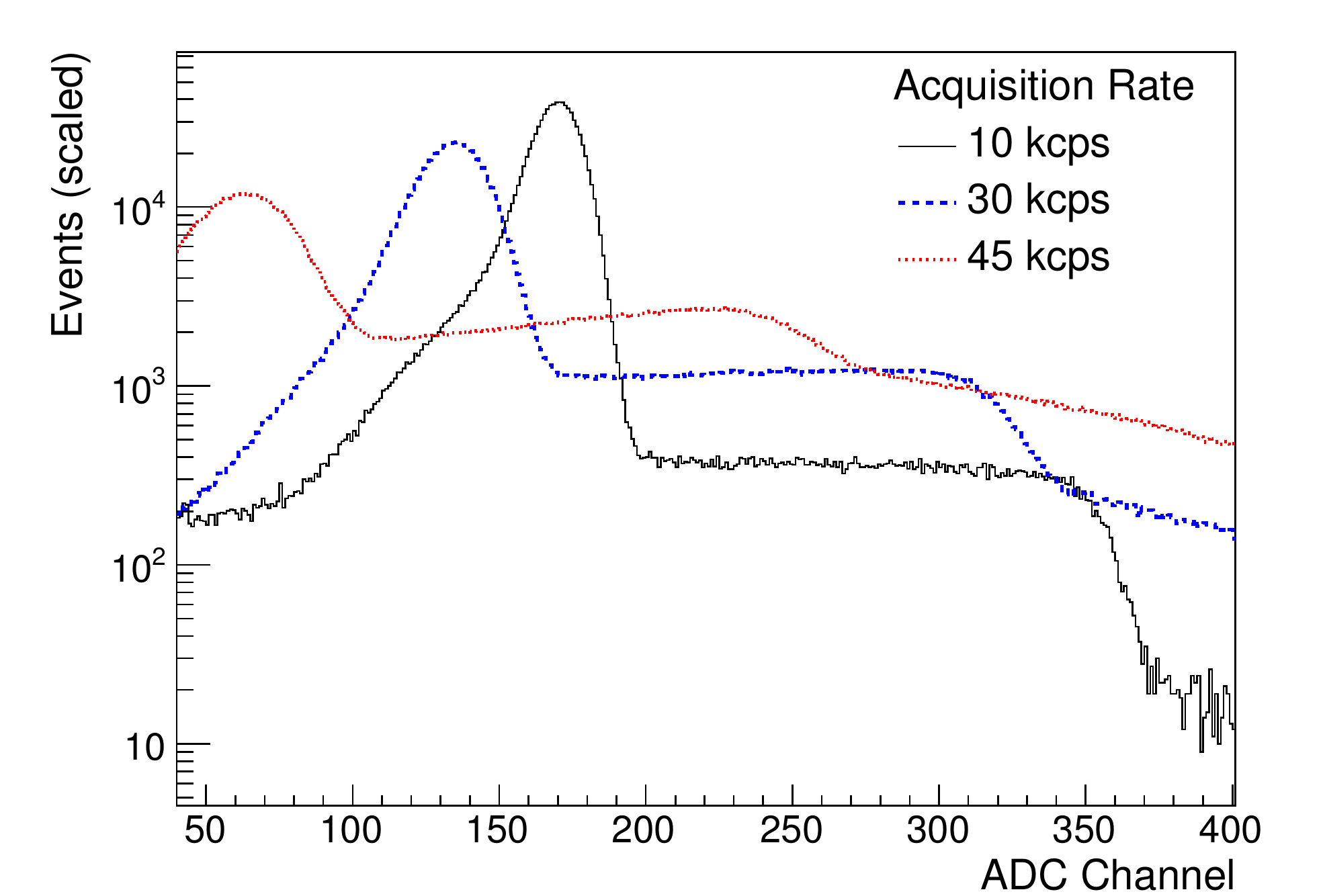}
\end{center}
\caption{\label{fig:RateDependence} ADC spectra of 18.6-keV photoelectrons at three different rates, achieved by varying the voltage on the UV LED illuminating the source disk. The spectra have been normalized to equal area; the trigger threshold was set at 40~ADC channels.}
\end{figure}

At data rates in the tens of kcps and above, distortions appear in measured photoelectron spectra due to pile-up effects. Peak pile-up, in which the interval between electrons is not long enough for separate triggers, suppresses the measured event rate; the energy recorded for such a multiple-occupancy event is sensitive to the exact timing difference. Tail pile-up, in which the pulse from one trigger bleeds into the next, reduces the recorded energy for the second event. Both effects are evident in the photoelectron spectra shown in Fig.~\ref{fig:RateDependence}; note that, due to peak pile-up, the actual electron rate cannot easily be determined from the measured acquisition rate. At these high acquisition rates, waveform data cannot be recorded, so offline correction is not possible. 

This effect would have a negligible impact on the KATRIN neutrino-mass measurement, during which a rate of less than 1~cps is anticipated. However, higher rates are required during commissioning, particularly while measuring the transmission function of electrons through the main spectrometer. Guided by a simulation of the FPD-signal readout chain, we are developing a modification to the FPGA filter logic to address this problem. 

\subsection{Stability of Energy Calibration}
\label{sec:energy_stability}

Occasional shifts in the $^{241}$Am calibration peak locations, of up to 40~ADC channels or $\sim 4$~keV, were observed during running. Each shift was sudden and affected only a single channel, but changed only the peak location, not its width. This phenomenon was correlated with mechanical disturbances to the optical-fiber connections to the optical receiver boards (Sec.~\ref{sec:apparatus:fpd_electronics}), which occur primarily during work on the DAQ system. We are investigating strain-relief options for these connections.

\begin{figure}[tbp]
\begin{center}
\includegraphics[width=\columnwidth]{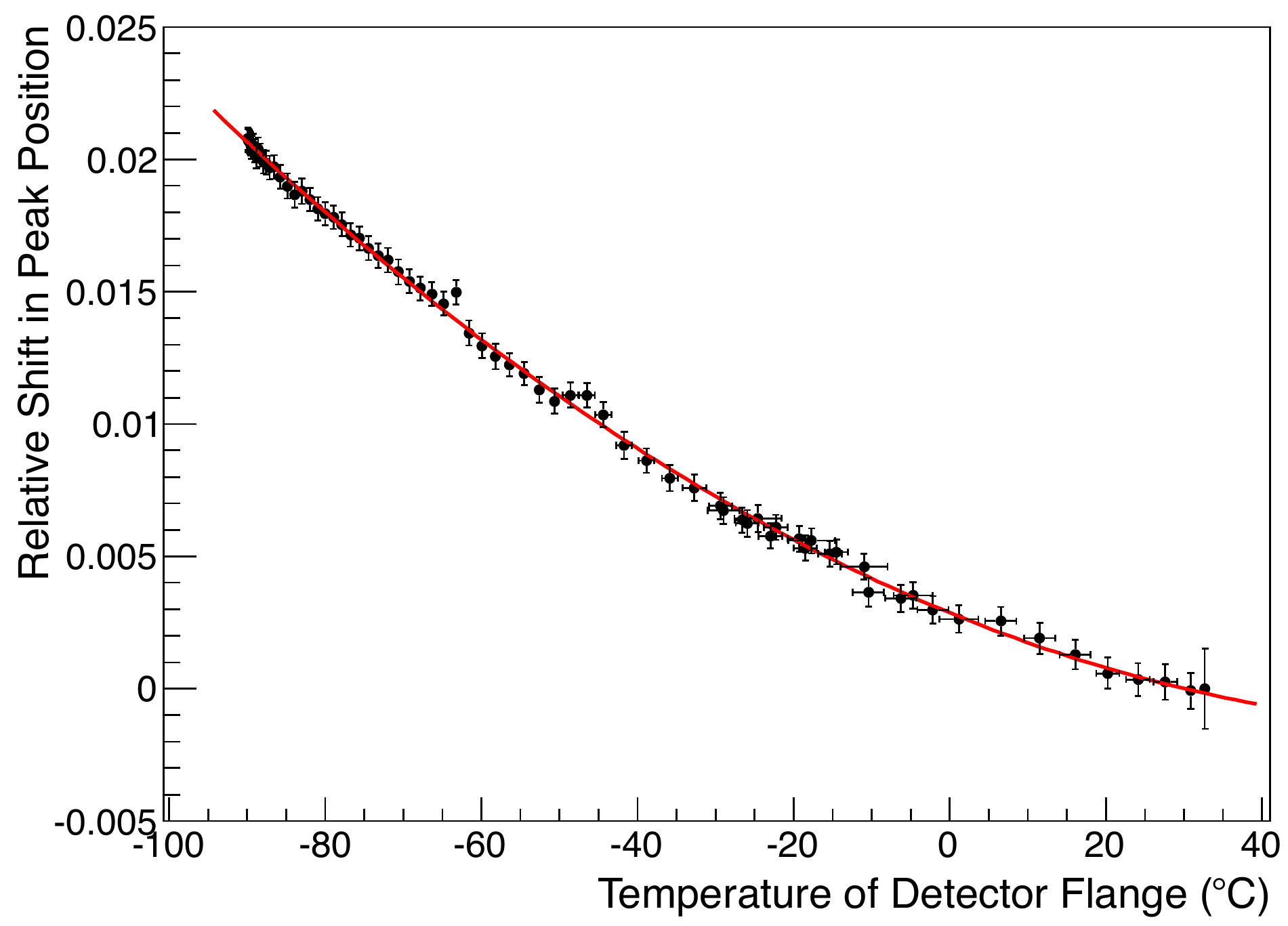}
\end{center}
\caption{\label{fig:TempDependenceAmPeak} Variation of the 59-keV $^{241}$Am calibration peak location as a function of temperature, measured during a cooldown and subsequent warmup of the system. The plot shows the average, over 146 working channels, of the relative deviation from the last measurement of the peak location in each channel. The line shows a quadratic fit to the data with $\chi^2/ndf = 19.28/74$.}
\end{figure}

The temperature of the FPD wafer and readout electronics affects the absolute energy calibration of the system in addition to the energy resolution. To quantify this effect, continuous one-hour $^{241}$Am calibration runs were taken over a period of 90~hours that included the final stages of a system cooldown, some 40~hours of operation at a stable temperature, and the early stages of system warmup. Figure~\ref{fig:TempDependenceAmPeak} shows the relative shift in the location of the 59-keV $^{241}$Am calibration line, averaged over 146~working channels. 

\begin{figure}[tbp]
\begin{center}
\includegraphics[width=\columnwidth]{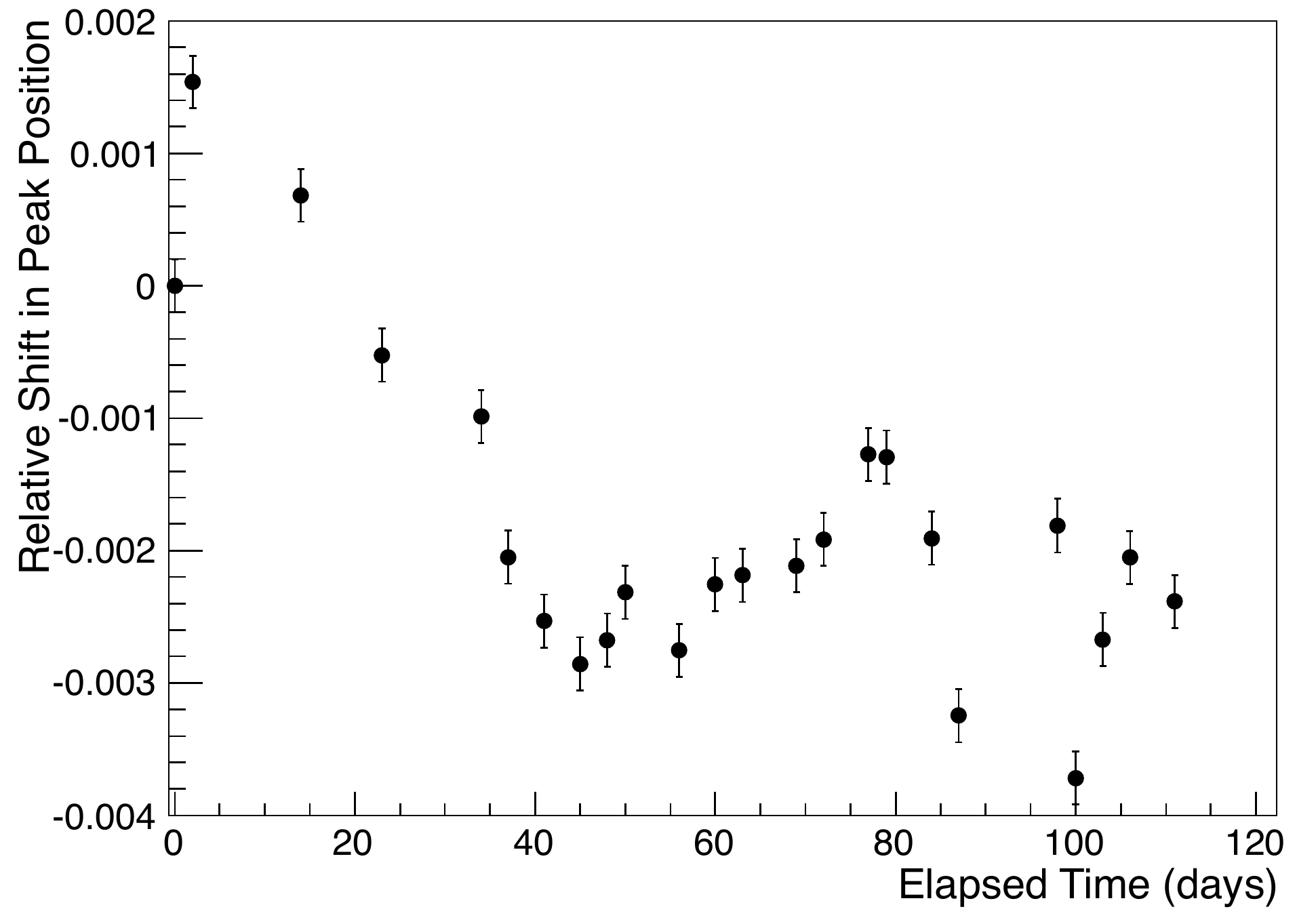}
\end{center}
\caption{\label{fig:SDS_AmPeakStability} Stability of the 59-keV $^{241}$Am calibration peak location during the 118-day main-spectrometer commissioning in summer 2013. The plot shows the average, over 146 working channels, of the relative deviation from the first measurement of the peak location in each channel~\cite{schwarz:2014}.}
\end{figure}

In summer 2013, the FPD system operated for 118~days during the first commissioning phase of the KATRIN main spectrometer. Figure~\ref{fig:SDS_AmPeakStability} shows that the position of the 59-keV $^{241}$Am calibration was, on average, within 0.4\% of its initial value during that time. With the exception of a few channels that experienced large, isolated calibration shifts, the individual channel calibrations evolved in similar ways. The drift seen during the first part of commissioning is temperature-related.

\section{Backgrounds and Figure of Demerit}
\label{sec:det_sys_environment}

The KATRIN neutrino-mass experiment probes a low-statistics region of the tritium beta-decay spectrum and is thus sensitive to background. The primary source of background from the main spectrometer is slow electrons near the analyzing plane, which produce a sharp peak in the measured energy spectrum. By contrast, we expect the intrinsic FPD-system background, arising from cosmic rays and radioactivity, to form a mostly featureless continuum with a gentle energy dependence.  The background spectrum may have isolated features due to X-rays or low-energy gammas, and there is a step in the continuum at about 100~keV, corresponding to the smallest possible energy deposition of minimum-ionizing through-going particles.   

To achieve a given uncertainty in the neutrino mass with a given detector efficiency, poorer energy resolution requires a larger acceptance region, which increases the background from the FPD system. We have quantified this relationship in a Figure of Demerit, which must be minimized to optimize system performance. Backgrounds can be specified as a rate in a particular region of interest in the energy spectrum, or as an average rate per keV in that region. In this work, the latter formulation is used, since it allows direct comparison between differently sized regions of interest.

Section~\ref{sec:sim_bg} discusses the simulations that translate radioassay data, cosmic-ray fluxes, and measured environmental radiation into background rates in the FPD system.  In Sec.~\ref{sec:performance:backgrounds}, we describe how backgrounds in the detector system are measured, and compare the results to our simulations. Finally, the Figure of Demerit is defined and presented in Sec.~\ref{sec:performance:fom}.

\subsection{Simulated Backgrounds}
\label{sec:sim_bg}

We used Geant4~\cite{al.:2003uq,1610988} to simulate the FPD system prior to its construction. This simulation aided in design decisions, particularly in regard to the radiation shield and veto, and in the selection of construction materials based on radio-purity and proximity to the detector. In addition to the FPD, we simulated five scintillator panels for the veto (Sec.~\ref{sec:apparatus:veto}), with the endcap taken as a single piece. The magnetic field was approximated as homogeneous and constant.  The post-acceleration electric field was not included but should not affect the background estimates, since all components near the detector are held at the post-acceleration potential. 

The background sources range in energy from X-rays of a few keV to cosmic-ray muons of 100~GeV.  Particle interactions in the detector and other materials are defined by the low-energy electromagnetic, high-precision neutron, and low- and high-energy  models released with Geant4 version 9.2 (patch 01). The electromagnetic processes are most important for this simulation and have been extensively validated~\cite{Amako:Geant4Validation2005}.  

Radioactive decays occurring in close proximity to the detector are fully simulated with the Geant4 Radioactive Decay Module, producing alphas, betas, and de-excitation photons and electrons from $^{40}$K, the $^{238}$U chain, and the $^{232}$Th chain.  The default Radioactive Decay Module of version~9.2 produced many of the de-excitation photons under 100~keV with the wrong intensity or energy~\cite{RadDecayModule_BugReports}. We corrected the module by generating photons with energy and intensity matching the Table of Radioactive Isotopes~\cite{TORI}. 

Early background studies~\cite{schwamm:2004} indicated the importance of material selection for reducing the FPD-system background to an acceptable level. Guided by initial simulations of the final system design that used activities estimated from the literature, we identified FPD-system components with critically important radiopurities. These components were radioassayed at the Low Background Facility of the Lawrence Berkeley National Laboratory~\cite{lbl_lowbackground_talk}. These results, given in Table~\ref{assay}, were then included in the simulation program to improve estimates of the detector backgrounds~\cite{Leber_thesis}. A custom generator code, tuned with background measurements in a Ge detector (Sec.~\ref{sec:performance:backgrounds}), simulates penetrating photons produced in radionuclide decays outside the FPD system. 

Cosmic rays at sea level contain components of muons~\cite{miyakeCR}, photons~\cite{Daniels:1974kx}, nucleons~\cite{PDG12}, and low-energy neutrons~\cite{Yamashita:1969qf}. Each component was simulated separately. The simulated veto, with a threshold of 1.2~MeV, is effective at tagging cosmic-ray-muon-related background events, but imperfect efficiency (simulated at 90\%) and incomplete coverage conspire to preserve cosmic-ray muons as a non-negligible contribution to the background.

\begin{table*}
	\begin{center}
	\begin{tabular}{lccc}
	\hline
	Component & $^{238}$U (Bq kg$^{-1}$) & $^{232}$Th (Bq kg$^{-1}$) & $^{40}$K (Bq kg$^{-1}$) \\ \hline \hline
	Pogo pins & $0.86\pm 0.12$ (early) & $<0.041$ & $<0.125$ \\ 
	 & $<0.04$ (late) & & \\ \hline
	Glass feedthrough & $12.3 \pm 1.2$ & $3.66 \pm 0.41$ & $1125 \pm 31$ \\ \hline
	Conflat flange  & $<0.004$ & $<0.004$ & $<0.013$ \\  \hline
	Preamp boards  & $3.85 \pm 1.31$ & $0.7 \pm 0.3 $ & $< 2.2$  \\  \hline
	Magnet coil   & $0.25 \pm 0.20$ & $0.20 \pm 0.041 $ & $0.56 \pm 0.28$  \\ \hline
	Magnet coil  & $10 \pm 1.2$ (early) & $1.0 \pm 0.3$& $<0.6$  \\ 
	banding  & $< 0.25$ (late) & &  \\ \hline	
	\end{tabular}
	\end{center}
	\caption{Measured activities for critical detector-system components, grouped according to decay series. Where possible, the uranium series is further divided into two: early elements are members of the series from $^{238}$U through $^{230}$Th, while late elements are further down in the series, beginning with $^{226}$Ra. Radioassay is by direct gamma counting at the Lawrence Berkeley National Laboratory facilities in Berkeley and Oroville, California~\cite{Smith:privcom}.}
	\label{assay}
\end{table*}

Certain experimental parameters influence the effective background. For example, increasing the detector magnetic field in the range 3.3 to 5~T decreases the flux-tube size and effective area of the detector, thus decreasing the background. Raising the post-acceleration potential shifts signal and spectrometer-side background electrons to a different energy range, possibly separating them from other backgrounds. Table~\ref{tab:sim} summarizes the simulated backgrounds for selected combinations of the free experimental parameters, together with measured background rates. Two offline background-reduction cuts are applied to both measured and simulated background data. The \textit{veto cut} rejects any FPD event occurring in coincidence with a veto trigger (measurement) or with a recorded hit above the single-photoelectron threshold in any veto scintillator panel (simulation). The \textit{multi-pixel cut} rejects any FPD event in which hits are recorded in multiple FPD channels within a certain time window. The coincidence windows for both cuts were set to 1~$\mu$s for data taken with the Seattle prototype; after hardware and firmware modifications in Karlsruhe, they were enlarged to 1.5~$\mu$s.  

Environmental radiation, dominated by electrons from the beta decay of $^{40}$K in the glass feedthrough insulators, is the largest simulated background contribution. This result motivated the installation of shielding for the feedthroughs after the apparatus was moved to Karlsruhe (Sec.~\ref{sec:apparatus:detector}).

\begin{table*}[tbp]
\medskip 
\begin{center}
\begin{tabular}{lcccc}
\hline
 Source & Multi-Pixel, & Post-Accel. & Field & Rate \\
& Veto Cuts & kV & T & mcps/keV  \\
 \hline\hline
 Sim. Cosmic& off & 0 & 3 & 2.22 $\pm$ 0.05  \\
Rays  & off & 10 & 3 & 2.02 $\pm$ 0.06  \\
& on & 0 & 3 & 0.28 $\pm$ 0.01  \\
& on & 10 & 3 & 0.19 $\pm$ 0.01   \\
 \hline
Sim. Env. Bkg. & on & 0 & 3 & 0.49 $\pm$ 0.03  \\
(Seattle) & on & 10 & 3 & 0.38 $\pm$ 0.03  \\ 
\hline
Sim. Env. Bkg. & on & 0 & 3 & 0.22 $\pm$ 0.02  \\
(Karlsruhe) & on & 10 & 3 & 0.17 $\pm$ 0.03  \\
\hline
Sim. & on & 0 & 3 & 0.44 $\pm$ 0.03 \\
Radioactivity  & on & 10 & 3 & 0.27 $\pm$ 0.03  \\
\hline
\hline
Sim. Total & off & 0 & 3 & 3.14 $\pm$ 0.06   \\
(Seattle) & on & 0 & 3 & 1.21 $\pm$ 0.04   \\
 & on & 10 & 3 & 0.84 $\pm$ 0.05   \\
\hline
Measured & off & 0 & 3.3 & $2.87 \pm 0.17$   \\
(Seattle) & on & 0 & 3.3 & $0.95\pm0.08$   \\
 & on & 10 & 3.3 & $1.63\pm0.11$ \\
 \hline
 \hline
Sim. Total & off & 0 & 3 & 2.86 $\pm$ 0.06   \\
(Karlsruhe) & on & 0 & 3 & 0.94 $\pm$ 0.04  \\
 & on & 10 & 3 & 0.65 $\pm$ 0.04    \\
\hline
Measured & off & 0 & 3.3 & $2.44 \pm 0.04$   \\
(Karlsruhe) & on & 0 & 3.3 & $1.05 \pm 0.02$   \\
 & on & 11 & 3.3 & $1.21 \pm 0.04$ \\
\hline
\end{tabular}
\end{center}
\caption{Simulated and measured backgrounds in optimized regions of interest determined from measured data (Tables~\ref{tab:FDR_noPAE} and~\ref{tab:FDR_PAE1}), with statistical uncertainties. The magnetic field values quoted are at the detector, and are thus somewhat smaller than the field in the center of the detector magnet. Neither simulation includes shielding for the glass feedthroughs. 
}
\label{tab:sim}
\end{table*}

\subsection{Measured Backgrounds}
\label{sec:performance:backgrounds}

\begin{figure}[tb]
\begin{center}
\includegraphics[width=\columnwidth]{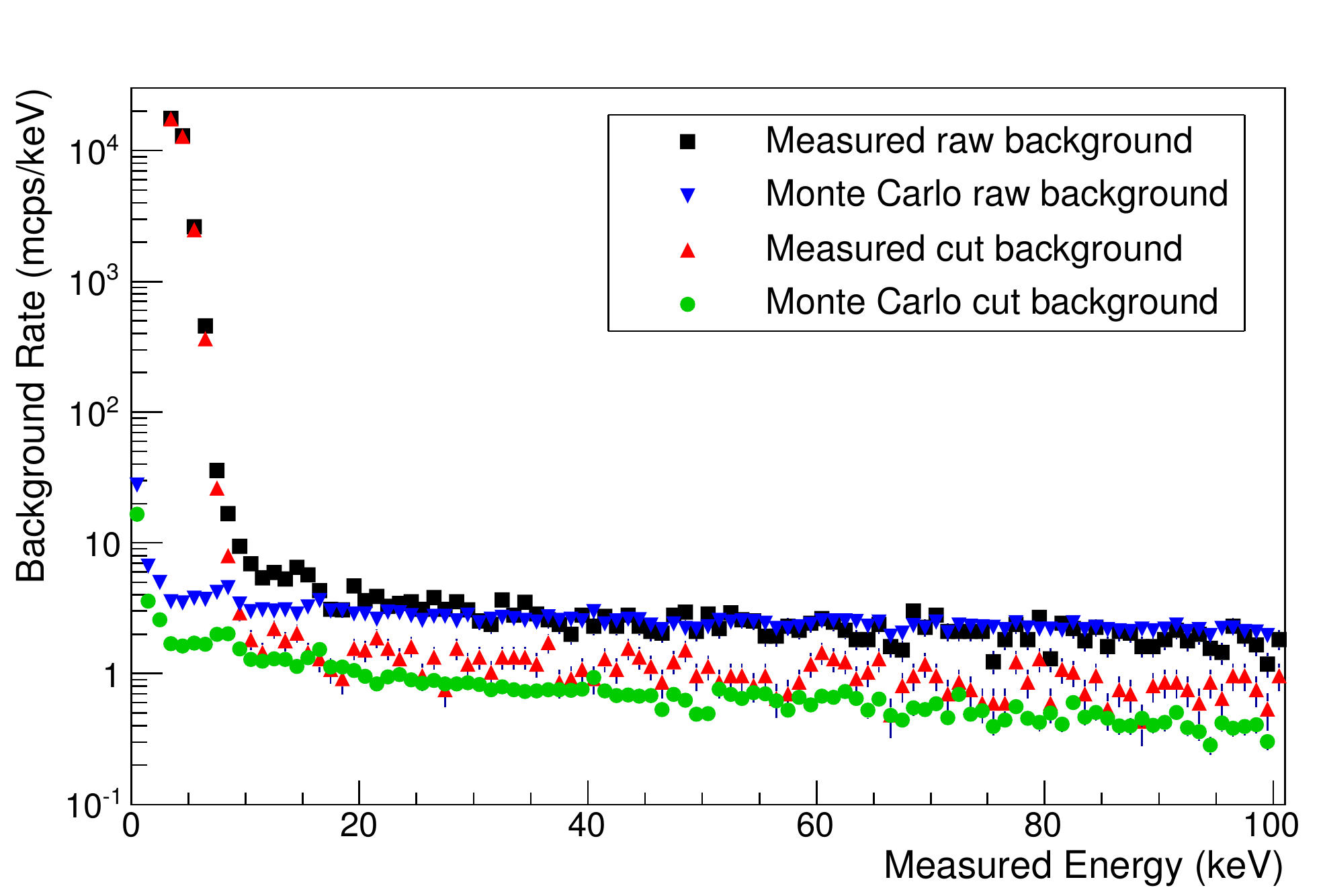}
\end{center}
\caption{\label{fig:UW_backgnd_overlay} Measured and simulated background spectra for the FPD-system prototype in Seattle. The measurement was performed over 127 channels and scaled by $148/127$ to give the effective spectrum over the entire active area of the detector. The spectra are not scaled to one another. }
\end{figure}

\begin{figure}[tb]
\begin{center}
\includegraphics[width=\columnwidth]{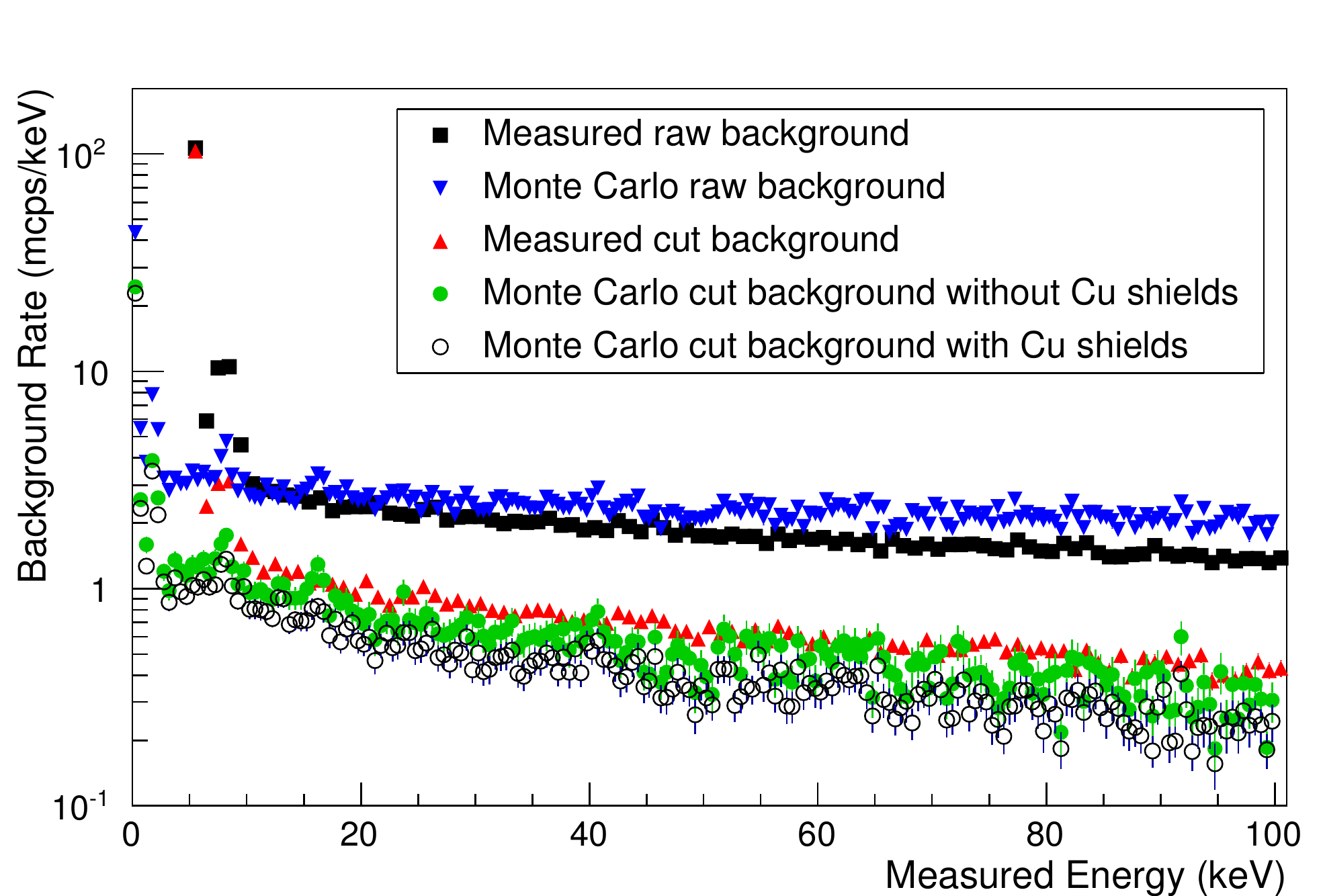}
\end{center}
\caption{\label{fig:KIT_backgnd_overlay} Measured and simulated background spectra for the FPD-system installation in Karlsruhe. The measurement was performed over 146 channels and scaled by $148/146$ to give the effective spectrum over the entire active area of the detector. The spectra are not scaled to one another. }
\end{figure}

The intensity of the environmental radiation, or photons from ambient natural radioactivity, differs between the commissioning laboratory in Seattle and the experimental hall in Karlsruhe. Measurements with Ge detectors quantified these differences and allowed us to tune the simulation that predicts background rates in the FPD. In Seattle, three measurements were made: one with an unshielded Ge detector, one with the detector placed inside the actual KATRIN FPD-system shield, and one with the detector shielded and the upstream shield opening blocked by a 2.5-cm-thick steel plate. Three similar measurements were also made in Karlsruhe: one with the Ge detector unshielded, one with the detector placed inside a 15-cm-thick Pb enclosure with a $25^{\circ}$ funnel-shaped opening, and one inside the enclosure with that opening blocked by a 15-cm-thick Pb wall. Calibration of both detectors with a $^{60}$Co source gave 5\% agreement between the simulations and the measured peak-to-Compton ratio and relative efficiency. The background measurements gave 5\% agreement with simulations in the integral number of counts, and 10\% agreement in the peak amplitudes relative to the continuum.

Background measurements were taken with the FPD-system prototype in Seattle (Fig.~\ref{fig:UW_backgnd_overlay}) and with the FPD system in Karlsruhe (Fig.~\ref{fig:KIT_backgnd_overlay}) and compared with simulations for the two locations. Both the veto and multi-pixel cuts have been applied to the ``cut'' background spectra. The actual efficiency of the veto system is not fully known, but was taken to be 90\% in the simulations. 

In both implementations, the pinch magnet ran with a central field of 6~T and the detector magnet with a field of 3.6~T (3.3~T at the detector wafer); no post-acceleration was applied. The vacuum tube was closed on its upstream side with a blank flange. Below 8~keV, actual channel thresholds vary because they are set automatically by ORCA to produce the same, user-specified, total rate in each channel.  

Figures~\ref{fig:UW_backgnd_overlay} and~\ref{fig:KIT_backgnd_overlay} demonstrate satisfactory agreement between measurement and simulation above 8~keV in both commissioning locations. Below 8~keV, the shape of the total measured spectrum is a composite of the independent noise thresholds of every included channel; this feature is not included in the simulation, accounting for the low-energy discrepancy. In both locations, the simulation slightly overestimates the raw background, especially at higher energies, and slightly underestimates the background after offline cuts. It must be noted that the measured spectra shown in Fig.~\ref{fig:KIT_backgnd_overlay} were acquired after the installation of copper shielding for the glass feedthroughs in the detector flange (Sec.~\ref{sec:apparatus:detector}); simulations that incorporate the shielding significantly underestimate the cut background, indicating the presence of an unidentified low-rate background source. 

Table~\ref{tab:sim} lists the background rates measured in Seattle and in Karlsruhe.  At a post-acceleration voltage of 0~kV, there is good agreement with the simulations, which were completed two years before Seattle data-taking and four years before Karlsruhe data-taking. However, in both locations the measured background with post-acceleration applied was  {\em higher} than without, a feature not seen in the simulations.  This is believed to be due to electrons and delta rays produced by cosmic rays in the blank flange across the upstream beam aperture; these electrons gain energy in the post-acceleration potential.  In the final experiment, this flange will be absent, replacing this source of background with the main spectrometer.
 
\subsection{Detector Figure of Demerit}
\label{sec:performance:fom}

Setting a limit on the measured background rate is an unsatisfactory method of characterizing the FPD-system performance, since a more efficient detector will record a higher background rate. A better metric balances the benefits of a better detector response with the drawbacks of higher background. We have developed the detector figure of demerit, $F$, to characterize the effect of the detector on the overall experiment; for an optimal measurement, this figure must be minimized. To determine a functional form for $F$, we begin with the statistical uncertainty for the square of the neutrino mass ($m_{\overline{\nu}}^2$) for a region of interest (ROI) with lower and upper bounds given by $E_L$ and $E_U$. This may be represented as~\cite{otten:1994} 
 
\begin{equation}
\sigma(m_{\overline{\nu}}^2)=\frac{k b(E_L, E_U)^{1/6}}{r^{2/3}t^{1/2}},
\label{eq:MnuUncertainty}
\end{equation}

\noindent where $k= \left(16/27\right)^{1/6}$ is a constant, $t$ is the measurement time in seconds, $r$ is the normalized count rate in cps/eV$^3$ of tritium beta-decay electrons that have passed through the main spectrometer, and $b(E_L, E_U)$ is the total background rate in cps, integrated over the ROI. This total rate is the sum of the integrated FPD-system background $b_{det}(E_L, E_U)$ and the main-spectrometer background $b_{ms}$, which is taken to be constant in energy at the design goal of 10~mcps~\cite{katrin_dr2004}. This formula assumes that the background and the endpoint energy are known independently (but not absolutely) and with perfect precision. 

The detected event rate is modified by the fraction $f(E_L, E_U)$ of the total spectral intensity that is detected within the ROI. This fraction is a measure of the detector response, and depends on the detector dead layer, kinematics of the incident electron, and energy resolution of the system in addition to the choice of ROI. If we apply this correction to Eq.~\ref{eq:MnuUncertainty} and separate the sources of background, we obtain an uncertainty of 
 
 \begin{equation}
 \sigma(m_{\overline{\nu}}^2)=\frac{k b_{ms}^{1/6}}{r^{2/3}t^{1/2}}\frac{\left( f(E_L,E_U)+\frac{b_{det}(E_L,E_U)}{b_{ms}} \right) ^{1/6}}{f(E_L,E_U)^{2/3}}.
 \end{equation}
 
\noindent From this idealized expression, the figure of demerit $F$ is defined as

\begin{equation}
 F(E_L,E_U)=\frac{\left( f(E_L,E_U)+\frac{b_{det}(E_L,E_U}{b_{ms}}\right) ^{1/6}}{f(E_L,E_U)^{2/3}} \geq 1.
 \label{eq:FigureOfDemerit}
\end{equation}

 \begin{figure*}[tbp]
  \centering
    \subfigure[]{
    \includegraphics[scale=0.24]{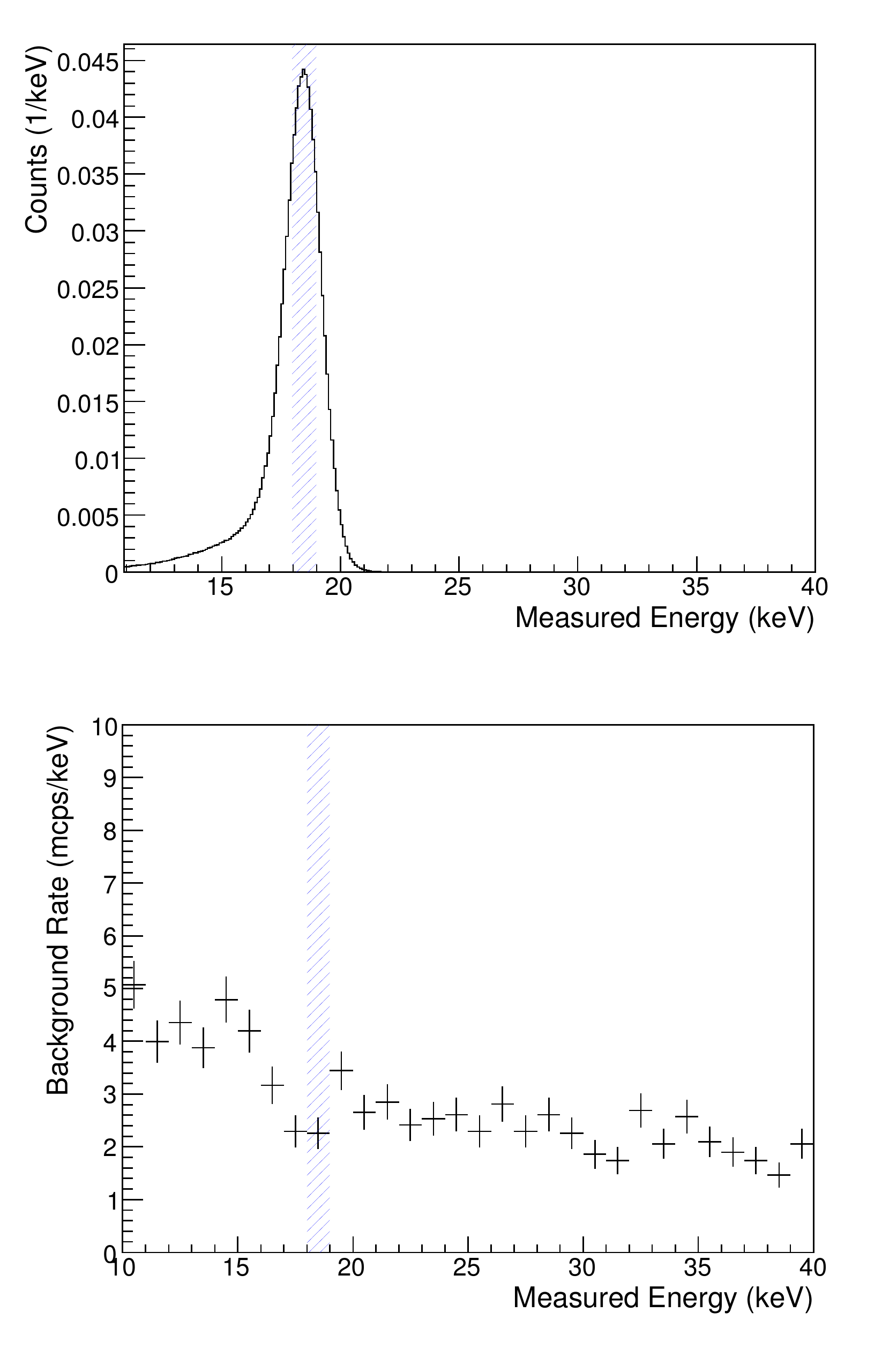}
    \label{fig:optimizeF_toosmall}
  }
    \subfigure[]{
    \includegraphics[scale=0.24]{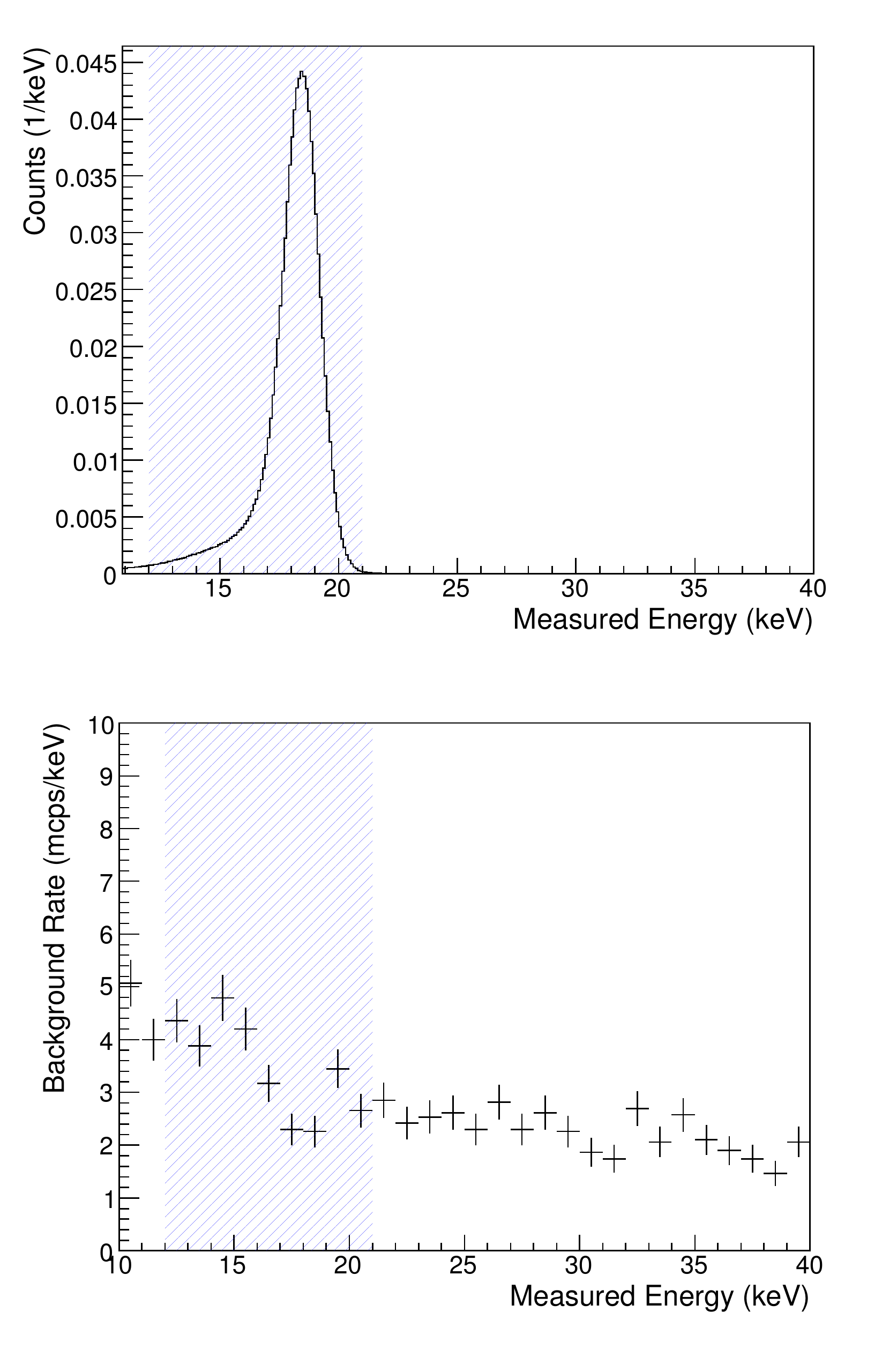}
    \label{fig:optimizeF_toobig}
  }
   \subfigure[]{
    \includegraphics[scale=0.24]{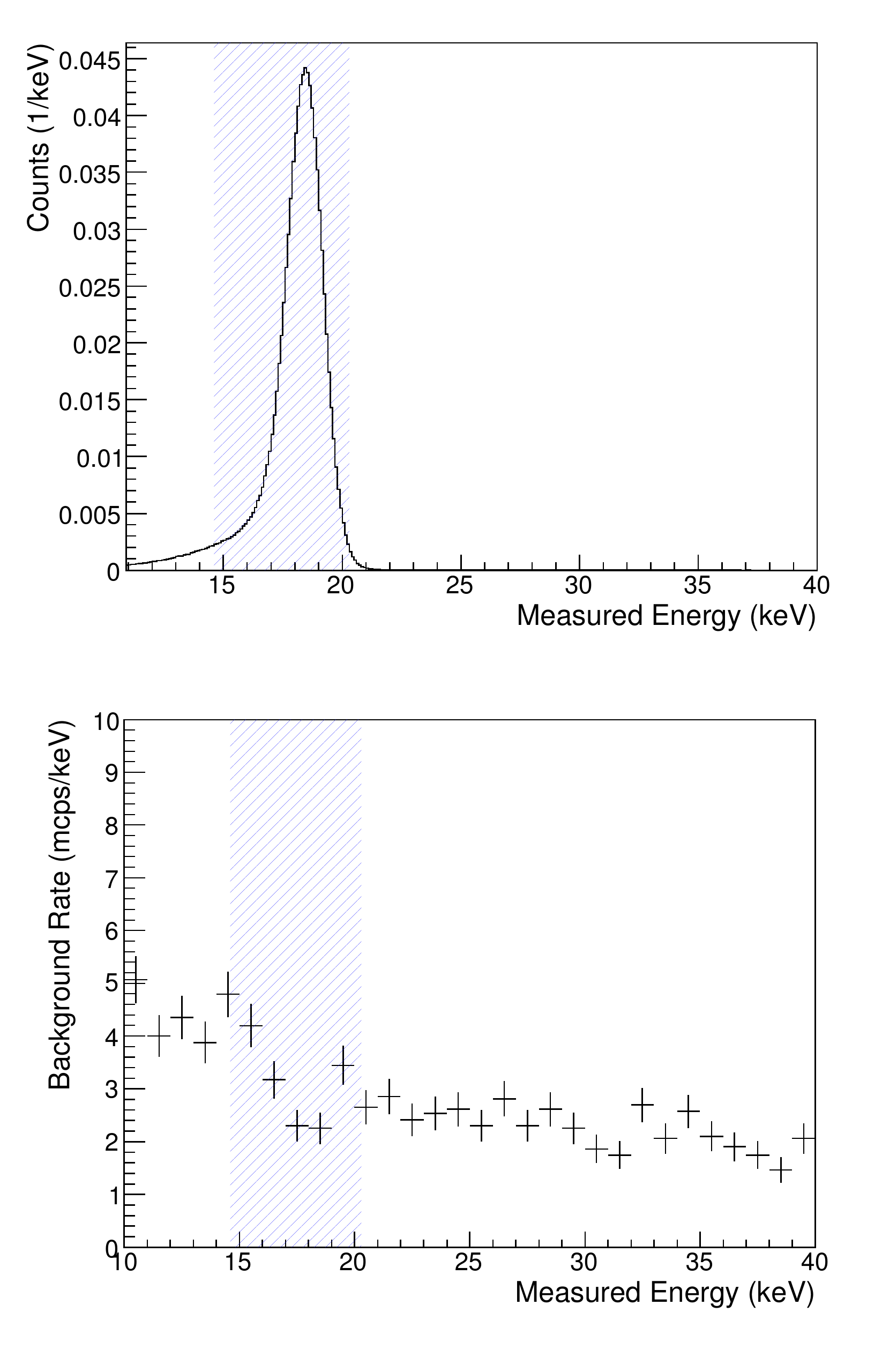}
    \label{fig:optimizeF_justright}
	}
    \caption[]
    {Normalized 18.6-keV electron energy spectra (top) and FPD-system background spectra (bottom) from the Seattle prototype. Each shaded region represents a possible region of interest $(E_L, E_U)$.~\subref{fig:optimizeF_toosmall} For a small ROI, the fraction of signal electrons in the window is small, resulting in a large $F$.~\subref{fig:optimizeF_toobig} A large ROI accepts a large number of background electrons, resulting in a large $F$.~\subref{fig:optimizeF_justright} An optimized ROI balances the signal and background. }
	\label{fig:OptimizingDelta}
\end{figure*}

We can minimize $F$, thus minimizing the effect of the non-ideal response of the FPD system, by varying the ROI.  Figure~\ref{fig:OptimizingDelta} illustrates this iterative process. As the width $E_U-E_L$ is increased, so are the contributions from background, while decreasing $E_U-E_L$ limits the number of accepted signal electrons and thus decreases the detector efficiency.  For a given width, there is an optimum window that balances the accepted background electrons and the rejected signal electrons so as to minimize $F$. Of course, this optimization procedure treats the FPD system by itself; in the final KATRIN experiment, other considerations, including stability, will affect the choice of ROI.

\begin{figure}[tbp]
\begin{center}
\includegraphics[width=\columnwidth]{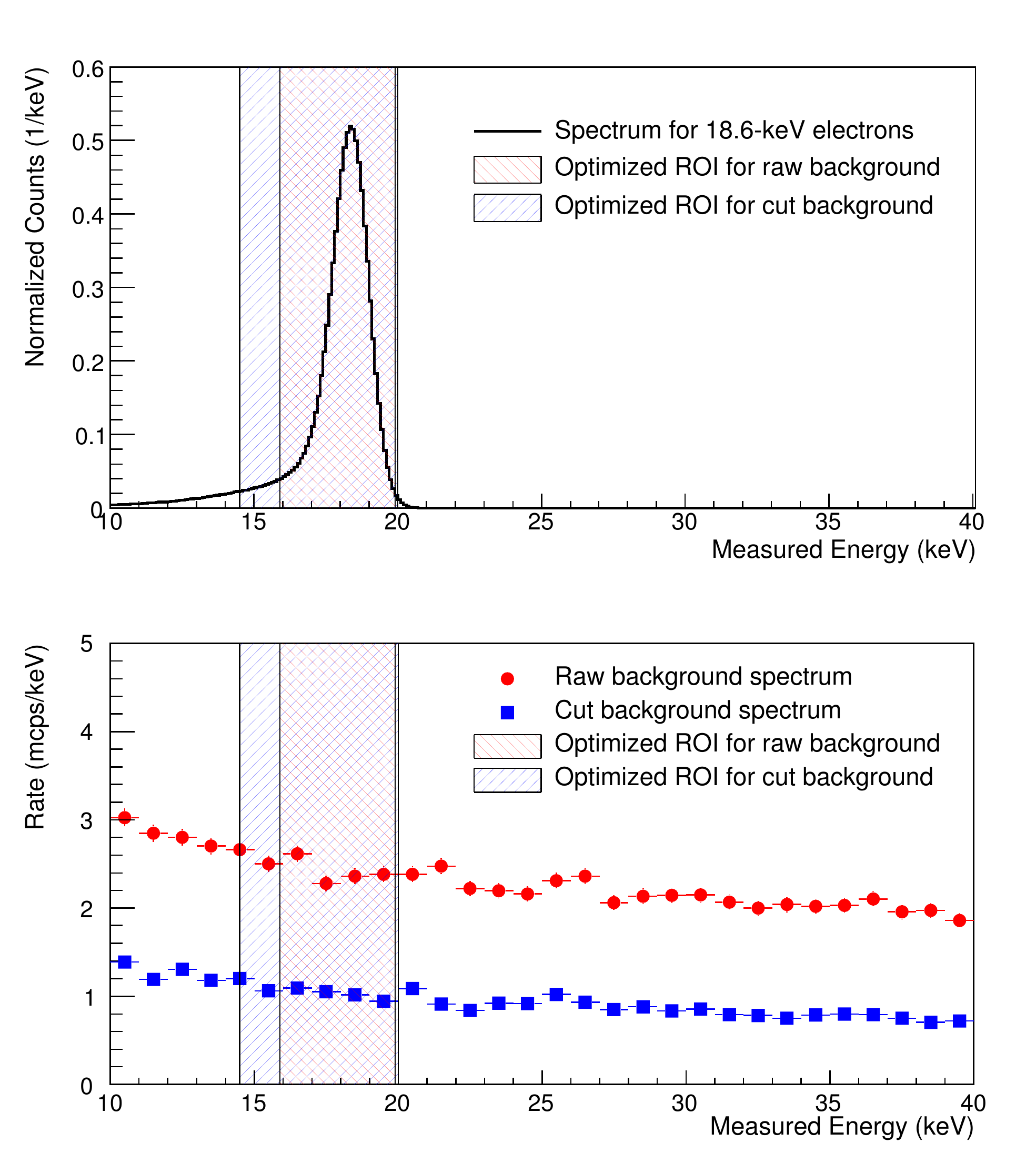}
\caption{18.6-keV electron and background spectra with shaded regions of interest for Karlsruhe data, measured over 146 pixels with no post-acceleration. The background spectra were scaled by 148/146.}
\label{fig:Fspectra}
\end{center}
\end{figure}

\begin{figure}[tbp]
\begin{center}
\includegraphics[width=\columnwidth]{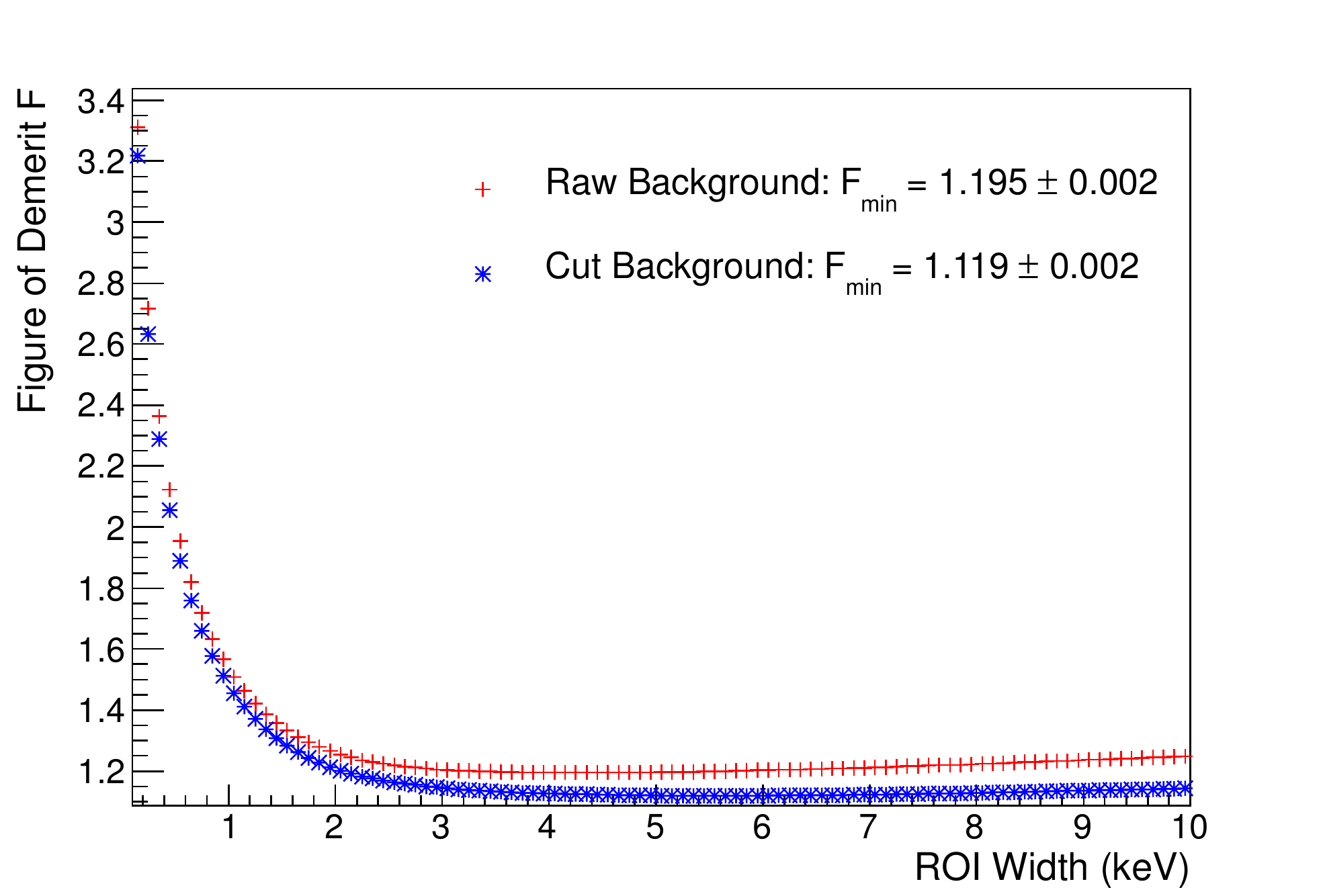}
\caption{Figure of demerit minimization curves for Karlsruhe data, measured over 146 pixels with no post-acceleration. The background spectrum was scaled by 148/146.}
\label{fig:Fminimization}
\end{center}
\end{figure}   

For the determination of $F$ with no post-acceleration, we measured the detector response in calibration runs with the photoelectron source set at the tritium beta-decay endpoint energy of 18.6~keV. For the Seattle data, we averaged FPD-system background spectra over 8~hours of runtime; the Karlsruhe background data are averaged over 81~hours of runtime. Table~\ref{tab:FDR_noPAE} shows the resulting minimum values of the figure of demerit, both before and after applying veto and multi-pixel cuts to the background spectra (Sec.~\ref{sec:sim_bg}); the lower background rate after cuts permits an enlargement of the optimal ROI. Figures~\ref{fig:Fspectra} and~\ref{fig:Fminimization} show the corresponding spectra and minimization curves for Karlsruhe data. All of the data were taken at nominal magnetic field settings: 3.6~T in the center of the detector magnet and 6~T in the center of the pinch magnet. In both locations, the cut background spectrum meets our commissioning goal of $F<1.2$ and is close to our operational goal of $F<1.1$. While the figure of demerit is useful as a performance metric, we note that the analysis of tritium data will be by a maximum-likelihood technique with appropriate probability density functions, or by a similar method.
 
 \begin{table*}[tbp]
    \centering
    \begin{tabular}{cccccc} 
             \hline

       Location & Background    & ROI  & $E_U-E_L$ & $F_{min}$ & $b_{det}$ \\
       & Type & (keV) & (keV) & & (mcps/keV) \\
       \hline \hline
             Seattle & Raw       & $16.1 \pm 0.1 -20.1 \pm 0.1 $  & 4.0 & $1.224 \pm 0.007$ & $2.87 \pm 0.17$ \\
       	    &	Cut   &  $14.6 \pm 0.1 -20.3 \pm 0.1 $ & 5.7 & $1.119 \pm 0.006$ & $0.95 \pm 0.08$ \\
              \hline
              Karlsruhe & Raw & $15.9 \pm 0.1 -19.9 \pm 0.1 $ & 4.0 & $1.195 \pm 0.002$ & $2.47 \pm 0.04$ \\
              & Cut & $14.5 \pm 0.1 -20.0 \pm 0.1 $ & 5.5 & $1.119 \pm 0.002$ & $1.06 \pm 0.02$ \\
			\hline
    \end{tabular}
    \caption{Figure of demerit results with no post-acceleration and nominal magnetic fields.}
    \label{tab:FDR_noPAE}
 \end{table*}

As discussed in Sec.~\ref{sec:apparatus:vacuum}, the post-acceleration electrode can boost signal electrons into an energy range with lower backgrounds. We took data at several post-acceleration voltages in order to examine background in these regions.  However, low-energy electrons emitted from the upstream blank flange in the standalone setups (Sec.~\ref{sec:performance:backgrounds}) were boosted along with the signal electrons, resulting in higher figures of demerit (Table~\ref{tab:FDR_PAE1}) than observed at the 0-kV setting.  This background source will be replaced by the main spectrometer in normal operation.

 \begin{table*}[tbp]
    \centering
    \begin{tabular}{cccccc} 
             \hline

       Location & PAE  & ROI  & $E_U - E_L$ & $F_{min}$ & $b_{det}$\\
       & (kV) & (keV) & (keV) & &  (mcps/keV) \\
       \hline \hline
         Seattle & 7.7   &  $23.7 \pm 0.1 -28.0 \pm 0.1 $ & 4.3 & $1.141 \pm 0.007$ & $1.30 \pm 0.12$ \\
         & 10   &  $25.6 \pm 0.1 -31.3 \pm 0.1 $ & 5.7 & $1.156 \pm 0.006$ & $1.63 \pm 0.11$ \\
         & 12   & $28.0 \pm 0.1 -32.2 \pm 0.1 $  & 4.2 & $1.137 \pm 0.006$ & $1.27 \pm 0.11$ \\
          \hline
         Karlsruhe & 7.8 & $23.0 \pm 0.1 -28.0 \pm 0.1 $ & 5.0 & $1.120 \pm 0.005$ & $1.13 \pm 0.07$ \\
          & 11 & $26.9 \pm 0.1  - 31.1 \pm 0.1 $ & 4.2 & $1.123 \pm 0.002$ & $1.23 \pm 0.04$ \\
          \hline
    \end{tabular}
    \caption{Figure of demerit results with post-acceleration and nominal magnetic fields. Background-reduction cuts have been applied.}
    \label{tab:FDR_PAE1}
 \end{table*}
 
\section{Conclusion}
\label{sec:conclusion}

The focal-plane detector system for the KATRIN neutrino-mass experiment was constructed, tested at the University of Washington in Seattle, and found to meet its design criteria, after which it was installed at the Karlsruhe Institute of Technology. The system has undergone several hardware, firmware, and software upgrades before re-commissioning in Karlsruhe in the winter of 2013.  It is designed for efficient detection of low-energy electrons in the $10-100$~keV energy range. The reflection of backscattered electrons from magnetic and electrostatic fields improves the average per-pixel efficiency beyond what can be achieved for a silicon detector in a field-free environment.

The silicon detector is a pixelated, 90-mm-diameter p-i-n diode with an independent readout for each of its 148~pixels, 146 of which were fully functional during commissioning.  The analyzing plane of the KATRIN main spectrometer maps directly onto this array; the segmentation of the detector reduces uncertainties due to variations in the analyzing potential. The requirements of ultra-high-vacuum compatibility, reliable electrical connections, and low background are jointly met through the use of a novel pogo-pin connection arrangement that isolates the electronics in a separate vacuum system and requires no ceramic carrier for the detector.   The resulting detector-side vacuum is compatible with direct connection to the KATRIN main spectrometer. A custom thermosiphon facilitates cooling of the detector and front-end electronics. 

The detector data are recorded in a multi-channel digitizer array, and several modes of recording are available to accommodate a wide range of data rates.  Control of the data-acquisition system is achieved with ORCA software, which includes an intuitive and flexible graphical user interface.  The setting and monitoring of apparatus parameters is carried out in a comprehensive slow-controls system based on industrial field-point controllers, while the web-based interface for data management provides near-real-time worldwide access to the data. 

Background contributions from cosmic rays, internal radioactivity, and the ambient $\gamma$ flux are minimized through the use of a scintillator veto; a layered Pb-Cu radiation shield; components selected and assayed for low radioactivity; post-acceleration to move the signal to an energy where the background is lower; magnetic-field adjustability to optimize the beam image size; and discrimination against multi-pixel events during analysis.  The total background in the detector in an optimized 5.5-keV energy window, as measured in Karlsruhe without post-acceleration and after offline background-reduction cuts, is $5.8 \pm 0.1$~mcps. Even without post-acceleration, this background level results in a figure of demerit that meets the system's commissioning goals, and is nearly low enough for the final KATRIN measurement. 

In late spring of 2013, the FPD system was successfully connected to the main spectrometer for the first commissioning phase of the KATRIN spectrometer-detector section. The FPD system was subsequently operated continuously for 118~days until the scheduled end of commissioning, demonstrating its reliability for the KATRIN experiment. 

This material is based upon work supported by the U.S. Department of Energy Office of Science, Office of Nuclear Physics under Award Number DE-FG02-97ER41020, which provided primary support for construction, and under Award Numbers DE-FG02-97ER41041 and DE-FG02-97ER41033. The electronics subsystem has been provided through grants from the German Helmholtz Gemeinschaft and Bundesministerium f\"ur Bildung und Forschung. Some travel support was provided by the Karlsruhe House of Young Scientists. The active interest of Canberra Industries in meeting our specialized requirements is appreciated. We gratefully acknowledge the participation, interest, and strong support of the entire KATRIN collaboration. 













\end{document}